%% file: paper.tex
\pdfoutput=1
\documentclass[letterpaper,twocolumn,10pt]{article}
\usepackage{usenix2019_v3}
\usepackage{amsmath,amssymb,amsthm}
\usepackage{xspace}
\usepackage{aliascnt}
\usepackage{booktabs}
\usepackage{colortbl}
\usepackage{tabularx}
\usepackage{float}
\usepackage{listings}
\usepackage{caption}
\usepackage{subcaption}
\usepackage{enumitem}
\setlist{topsep=3pt,partopsep=0pt,parsep=0pt,itemsep=1pt}
\usepackage{tikz}
\usetikzlibrary{arrows.meta,backgrounds,calc,fit,positioning}
\usepackage[varqu]{zi4}

\input{macros}
\begin{document}

\date{}
\title{\Large\bf \sys{}: Formal Verification of Warp-Specialized GPU Kernels}
\author{\rm
Mengdi Wu\textsuperscript{1}\quad
Sreeram Vennam\textsuperscript{1}\quad
Stephen Mell\textsuperscript{1}\quad
Ziwei Zhou\textsuperscript{1}\quad
Advay Kadam\textsuperscript{2}\\
\rm Xinhao Cheng\textsuperscript{1}\quad
Francois Costa\textsuperscript{3}\quad
Alex Aiken\textsuperscript{3}\quad
Oded Padon\textsuperscript{4}\quad
Zhihao Jia\textsuperscript{1}\\[4pt]
\rm\normalsize\textsuperscript{1}Carnegie Mellon University\quad
\normalsize\textsuperscript{2}University of Illinois Urbana-Champaign\\
\rm\normalsize\textsuperscript{3}Stanford University\quad
\normalsize\textsuperscript{4}Weizmann Institute of Science}

\maketitle

\input{abstract}
\input{sections/introduction}
\input{sections/background}

\input{sections/overview}
\input{sections/representation}
\input{sections/synchronization}
\input{sections/races}

\input{sections/scaling}
\input{sections/reference}
\input{sections/evaluation}
\input{sections/related}

\bibliographystyle{plain}
\bibliography{references}

\clearpage
\appendix

\input{sections/proofs-sync}
\input{sections/proofs-memory}

\end{document}

%% file: macros.tex
\newcommand{\sys}{SentryWard\xspace}

\newcommand{\code}[1]{\texttt{#1}}

\newcommand{\step}[1]{\tikz[baseline=(c.base)]{\node[circle,draw=black!70,fill=white,inner sep=0.5pt,minimum size=3.2mm,font=\scriptsize\sffamily\bfseries](c){#1};}}

\newcommand{\MissingWait}{\textsf{MissingWait}}
\newcommand{\SplitTx}{\textsf{SplitTx}}
\newcommand{\evl}[2]{\mathsf{E}_{#1,#2}}
\newcommand{\Sync}{\mathsf{E}_{\mathrm{sync}}}
\newcommand{\AEall}{\evl{0}{6}}
\newcommand{\QA}{\evl{0}{7}}
\newcommand{\LA}{\evl{0}{8}}
\newcommand{\CA}{\evl{0}{9}}

\newcommand{\LB}{\evl{0}{11}}
\newcommand{\CB}{\evl{0}{12}}
\newcommand{\MX}{\evl{32}{4}}
\newcommand{\Rt}{\evl{t}{4}}
\newcommand{\St}{\evl{t}{6}}
\newcommand{\AEA}{\mathsf{V}_1}
\newcommand{\QAv}{\mathsf{V}_2}
\newcommand{\CAv}{\mathsf{V}_3}
\newcommand{\EB}{\mathsf{V}_4}
\newcommand{\QBv}{\mathsf{V}_5}
\newcommand{\CBv}{\mathsf{V}_6}
\tikzset{evlab/.style={anchor=west,fill=black!13,text=black!80,font=\fontsize{5.5}{6}\selectfont,inner xsep=0pt,inner ysep=0pt,minimum width=5.4mm,minimum height=3.4mm,rounded corners=1.5pt}}

\newcommand{\eventof}{\mathit{eventOf}}

\newcommand{\reach}{\mathrel{\rightarrow_{\tau}}}
\newcommand{\hb}{\mathrel{\rightarrow_{\mathsf{hb}}}}
\newcommand{\labelof}{\operatorname{label}}
\newcommand{\stateof}{\operatorname{state}}
\newcommand{\contr}[2]{C_{#1}^{(#2)}}
\newcommand{\cmp}[2]{\mathit{end}_{#1}^{(#2)}}
\newcommand{\barname}[1]{\mathit{#1}}
\newcommand{\full}{\barname{full}}
\newcommand{\mmadone}{\barname{mma\_done}}
\newcommand{\lexge}{\mathrel{\ge_{\mathrm{lex}}}}
\newcommand{\lexgt}{\mathrel{>_{\mathrm{lex}}}}
\newcommand{\SyncCheck}{\mathrm{SyncCheck}}
\newcommand{\lexle}{\mathrel{\le_{\mathrm{lex}}}}
\newcommand{\nreach}{\mathrel{\nrightarrow_{\tau}}}
\newcounter{algoline}
\newenvironment{pseudocode}{%
  \footnotesize\raggedright\hyphenpenalty=10000
  \begin{list}{\tiny\arabic{algoline}}{\usecounter{algoline}%
    \setlength{\leftmargin}{1.4em}\setlength{\labelwidth}{1.0em}\setlength{\labelsep}{0.4em}%
    \setlength{\itemsep}{0pt}\setlength{\parsep}{0pt}\setlength{\topsep}{2pt}}}%
  {\end{list}}
\newcommand{\Ind}[1][1]{\hspace*{#1em}}
\newcommand{\Kw}[1]{\textbf{#1}}
\newcommand{\Cm}[1]{{\color{black!55}$\triangleright$~#1}}

\newcommand{\evs}{\operatorname{events}}
\newcommand{\ph}[2]{#1^{(#2)}}
\newcommand{\phaseof}{\Phi}
\newcommand{\initof}{\mathit{init}}
\newcommand{\issue}{\mathit{issue}}
\newcommand{\ret}{\mathit{return}}

\theoremstyle{plain}
\newtheorem{theorem}{Theorem}[section]
\newaliascnt{lemma}{theorem}
\newtheorem{lemma}[lemma]{Lemma}
\aliascntresetthe{lemma}
\theoremstyle{definition}
\newaliascnt{definition}{theorem}
\newtheorem{definition}[definition]{Definition}
\aliascntresetthe{definition}

\definecolor{swHB}{RGB}{30,92,155}
\definecolor{swBad}{RGB}{175,35,45}
\definecolor{swGood}{RGB}{40,110,50}
\definecolor{swKeep}{RGB}{222,240,224}
\definecolor{swChoice}{RGB}{130,65,150}

\lstdefinestyle{kernel}{
  basicstyle=\ttfamily\footnotesize,
  columns=fullflexible,
  keepspaces=true,
  showstringspaces=false,
  morecomment=[l]{//},
  commentstyle=\color{black!55},
  escapeinside={(*@}{@*)},
  numbers=left,
  numberstyle=\tiny\color{black!45},
  numbersep=6pt,
  xleftmargin=14pt,
  aboveskip=2pt,
  belowskip=2pt
}

\newfloat{algorithm}{tbp}{loa}
\floatname{algorithm}{Algorithm}

\tikzset{
  swbox/.style={draw=black!55,rounded corners=2pt,fill=white,align=center,
    font=\scriptsize\sffamily,minimum height=5.5mm,inner xsep=3pt,inner ysep=1.5pt},
  swphase/.style={swbox,draw=swHB,fill=blue!6},
  swdep/.style={-{Stealth[length=1.6mm]},line width=0.55pt,draw=black!70},
  swhb/.style={-{Stealth[length=1.6mm]},line width=0.7pt,draw=swHB},
  swrole/.style={font=\scriptsize\itshape,text=black!60},
  swkeep/.style={swbox,fill=swKeep},
  swomit/.style={swbox,fill=black!5,draw=black!35,text=black!55,dashed}
}

%% file: abstract.tex
\begin{abstract}
Modern GPU kernels overlap asynchronous data movement,
computation on general-purpose cores (e.g., CUDA cores), and computation on specialized cores (e.g., tensor cores) using \emph{warp specialization}.
These asynchronous-overlap optimizations introduce complex synchronization requirements, where incorrect ordering of asynchronous operations can lead to data races, premature resource reuse, and incorrect results.
Testing alone cannot guarantee correctness because such errors are nondeterministic and may only manifest in rare execution schedules, and
 existing GPU formal verification techniques have limited support for warp-specialized kernels.

We present \sys, a framework for formally verifying warp-specialized GPU kernels. \sys integrates synchronization verification, data-race and resource-lifetime analysis, and functional equivalence checking. The key idea is to establish that synchronization induces the same ordering constraints across all possible execution schedules. \sys extends this \emph{well-synchronization analysis} to modern GPU barriers with asynchronous transaction tracking and derives a schedule-independent happens-before graph. It then uses this graph to verify memory safety and construct symbolic output expressions for functional equivalence checking. 
While \sys generally assumes tensor shapes and kernel launch configurations are given and unrolls loops, it uses compact representations of repeated structures across threads and tensor computations to improve scalability.
\sys{} is publicly available at \url{https://github.com/gpu-verifier/sentryward}.
\end{abstract}

%% file: sections/introduction.tex
\section{Introduction}

The performance of machine-learning (ML) systems depends heavily on highly-optimized GPU kernels.
These kernels are developed by experts, generated by optimizing compilers~\cite{chen2018tvm,tillet2019triton,zheng2020ansor,wu2025mirage}, and increasingly produced by AI agents~\cite{ouyang2025kernelbench,baronio2026kevin,dai2026cudaagent,saba2026cutegen}.
To achieve high performance, GPU kernels employ sophisticated techniques such as tiling, software pipelining, and warp specialization to overlap computation with data movement. 
Although these optimizations can substantially improve performance, they also introduce complex synchronization requirements that make it difficult to establish correctness.

Modern GPUs have specialized units for asynchronous data movement and computation. For example, NVIDIA GPUs have tensor memory accelerators (TMAs) for transferring data and tensor cores for performing matrix operations. These operations execute asynchronously, allowing the issuing threads to continue before their memory accesses have completed.
Warp-specialized kernels exploit this capability by assigning data movement and computation to different {\em warps}, each consisting of 32 threads on NVIDIA GPUs.
A producer warp may load the next tile of data while a consumer warp performs matrix multiplication on the current tile. Correctness requires synchronization between these operations to ensure that data is available before it is consumed and that memory is not reused before outstanding operations have completed.

Consider a warp-specialized matrix multiplication kernel in which a producer warp loads input tiles into shared memory using a TMA, while a consumer warp issues asynchronous tensor-core operations on these tiles. After issuing tensor-core operations, the consumer warp reads the accumulated results from tensor memory and writes them to global memory. If the consumer warp omits a wait for the tensor-core computation to complete before reading the results, the output may depend on the execution schedule. The kernel produces correct results when the tensor-core operation finishes before the output is read, but may produce incorrect results otherwise. Similarly, if the producer warp reuses a shared-memory buffer before an outstanding tensor-core operation has finished reading it, the kernel may encounter data races that are difficult to detect and reproduce in testing.

A common approach to validating GPU kernels is to compare their outputs for random inputs against reference implementations~\cite{ouyang2025kernelbench}. Although such random testing can detect many bugs, it cannot guarantee correctness.
In particular, synchronization errors may remain hidden across repeated tests because they only manifest under specific interleaving schedules of asynchronous operations. Dynamic race detectors and sanitizers~\cite{vu2026racer,biberstein2026dirigo,vedadigargary2026resolve} provide additional checks, but their guarantees are limited to the executions they observe.

Formal verification provides a way to establish stronger correctness guarantees. Existing systems such as GPUVerify~\cite{betts2012gpuverify} and Faial~\cite{cogumbreiro2021faial} verify memory safety properties of GPU kernels, while WEFT~\cite{sharma2015weft} reasons about synchronization between producer and consumer warps. More recent systems, such as Volta~\cite{driscoll2025volta}, combine symbolic execution with equivalence checking to verify that GPU kernels produce the same results as their references. Although these approaches have demonstrated the feasibility of formally verifying GPU kernels, they face several challenges when applied to modern warp-specialized implementations.

Verifying warp-specialized GPU kernels presents three fundamental challenges. First, modern GPUs provide asynchronous transaction barriers, such as NVIDIA's \emph{mbarrier} objects.
Unlike traditional barriers that monotonically progress (as more threads reach the barrier), mbarriers have two counters that might be updated non-monotonically, making the structure of synchronization rounds or \emph{phases} more complex.
Because transaction counts can increase when new transactions are announced and decrease when they complete, a phase may complete prematurely depending on the order of these updates. Consequently, the assignment of operations to barrier phases may depend on the execution schedule, making synchronization correctness difficult to establish without enumerating possible interleavings. Second, asynchronous instructions may continue accessing memory after their issuing threads have proceeded to subsequent instructions. Verifying memory safety therefore requires reasoning about the actual execution and completion of asynchronous operations, while functional verification additionally requires establishing that the kernel produces consistent results across different schedules. Third, modern GPU kernels involve many concurrent threads and large tensor computations. Explicitly representing every thread-level event and scalar computation can result in substantial verification overhead, particularly when the same computational and synchronization patterns are repeated across threads and tensor elements.

We present \sys{}, a framework for formally verifying warp-specialized GPU kernels. \sys takes as input a GPU kernel in NVIDIA's Parallel Thread Execution (PTX) instruction set~\cite{nvidia2026ptx}, together with fixed tensor shapes, a launch configuration specifying the grid and thread-block dimensions, and input constraints. 
\sys verifies four properties: synchronization correctness, data-race freedom, resource-lifetime safety, and functional equivalence against a reference implementation. Unlike testing-based approaches that check sampled executions, \sys reasons about {\em all} possible execution schedules under its modeled PTX execution semantics.

\sys's analysis separates reasoning about synchronization from reasoning about computation,
targeting kernels where the synchronization behavior is not dependent on the scheduling or the input data, similarly to WEFT~\cite{sharma2015weft} and Volta~\cite{driscoll2025volta}.
Once \sys checks that synchronization behavior is independent of execution schedule and conflicting memory accesses are properly ordered, the kernel's computation can be analyzed as a deterministic function of its inputs. 
Instead of enumerating all possible interleavings of asynchronous operations, \sys first proves that all executions assign the same synchronization phases to barrier updates and waits. This property, which we call {\em well-synchronization}, allows \sys to derive a single happens-before graph that captures the ordering constraints imposed by synchronization across {\em all} executions.

Once the happens-before graph is established, \sys uses it to check whether potentially conflicting memory accesses are appropriately ordered. If the kernel is free of data races and resource-lifetime violations, every memory read obtains the same value across all executions under the assumed memory model. \sys can therefore eliminate intermediate memory accesses through symbolic execution and represent the kernel's output as a deterministic function of its inputs. Finally, checking equivalence of this function against a reference implementation yields functional verification.

We next outline the main components of \sys.

\paragraph{Program representation.} \sys derives two linked representations of GPU kernels that capture their scheduling and computation directly from PTX. The {\em event view} captures instruction issue, asynchronous execution, barrier operations, and completion events, while the {\em value view} captures memory accesses and symbolic computations. A correspondence between the two views associates each memory access with the event that performs it. \sys supports kernels with data-independent control flow, whose synchronization structure can be determined without knowing the values of input tensors. This allows \sys to analyze synchronization largely independently of symbolic computations, while preserving the correspondence between the two views needed for memory safety and functional verification.

\paragraph{Synchronization verification.} \sys extends well-synchronization analysis~\cite{sharma2015weft} to support the asynchronous transaction barriers used by modern GPUs. The main challenge is that mbarriers maintain arrival and transaction counters, with transaction counts potentially increasing and decreasing within the same round or phase.
A phase in an mbarrier completes when both counters reach zero, but this condition might be unintentionally satisfied before all updates intended for that phase have occurred. Such premature completion can cause subsequent updates to be assigned to a different phase depending on the execution schedule. \sys constructs a candidate happens-before graph from a single reference execution, recording the phase associated with each barrier update and wait. It then checks whether these phase assignments remain valid across all executions. In particular, our analysis introduces a completion condition that prevents any barrier phase from finishing before all updates assigned to that phase have occurred. We show that these conditions characterize well-synchronization under our execution model and develop an efficient checking procedure using graph-reachability checks and a minimum-cut algorithm.

\paragraph{Memory safety and functional verification.} Given a well-synchronized kernel, \sys combines the happens-before graph with symbolic memory addresses to rule out (or identify) data races and verify resource lifetimes. For memory accesses whose addresses depend on runtime values, \sys uses an SMT solver~\cite{demoura2008z3} to determine whether they may overlap under the specified input constraints. Because tensor shapes and loop bounds are fixed, \sys can specialize control flow and unroll loops before analysis. Once memory safety has been established, \sys symbolically executes the kernel, eliminating intermediate memory accesses and deriving symbolic expressions for the outputs in terms of the inputs. 
These expressions are compared against a reference implementation using algebraic reasoning, techniques that match the index sets of reductions in the kernel and reference, and template-based inductive invariants for supported stateful computations. 
Together, these analyses establish that the kernel produces the reference outputs across all execution schedules covered by our PTX execution model. Functional equivalence is checked using real arithmetic for floating-point operations.

\paragraph{Representing structured computations.} To scale verification to practical GPU kernels, we develop indexed representations that exploit repeated structure across threads and tensor computations in both the event and value views. \sys represents groups of similar events using templates parameterized by thread or tensor indices, and expresses their dependencies as relations between index domains. It also uses symbolic dataflow functions and folds to represent repeated scalar computations without explicitly expanding every operation. These representations allow synchronization analysis, data-race and resource-lifetime checking, and functional verification to operator on compact representations of recurring structures, reducing the need to enumerate individual thread events and scalar operations even for kernels with many threads and large tensors.

\paragraph{Evaluation.}
We evaluate \sys on 189 kernel configurations from 12 widely used GPU libraries and compilers (\autoref{sec:eval}), covering kernels developed by experts, generated by optimizing compilers, and produced by AI agents.
Among them, 105 are verified and 64 are flagged for potential correctness issues; successful verification takes a median of 14.9 seconds.
Our analysis identified a historical bug in PyTorch and synchronization issues that led to upstream fixes in CUTLASS, FlashMLA, FlashInfer, and vLLM.
We further evaluate verification-guided synchronization removal on 95 kernels, eliminating 37.7\% of executed synchronization points in aggregate.
Synchronization points are successfully removed from 83 of these kernels, reducing kernel run time by up to 15\% on an NVIDIA B200 GPU.

%% file: sections/background.tex
\section{Background}

\subsection{GPU Execution Model}
\label{sec:background:gpu}
GPUs achieve high throughput by running thousands of threads in parallel.
When a kernel is launched, all of its threads execute the same code, and each thread uses its index to determine the data it processes and the path it takes.
These threads are organized hierarchically.
Each group of 32 consecutive threads forms a warp, and several warps form a thread block (CTA).
On recent NVIDIA GPUs, several blocks can additionally form a cluster.
Threads within a block communicate through on-chip shared memory, and threads within a cluster can also access the shared memory of the other blocks in their cluster.
The warps of a block need not execute in lockstep.
They can perform different tasks, and even the threads within a warp can progress independently.

Traditional GPU kernels follow a bulk-synchronous structure.
The threads of a block cooperatively load a tile into shared memory, synchronize at a block-wide barrier, and then compute on the loaded tile.
A thread passes the barrier only after all threads of its block have reached it.
As a result, all accesses before the barrier are ordered before all accesses after it.

Recent GPUs add specialized units for asynchronous data movement and computation.
For example, NVIDIA GPUs have the Tensor Memory Accelerator (TMA) for transferring data and tensor cores for matrix multiply-accumulate operations (MMAs).
These operations execute asynchronously, allowing the issuing threads to continue before their memory accesses have completed.
Blackwell GPUs additionally introduce \emph{tensor memory}, a dedicated on-chip memory for MMA accumulators.
Kernels allocate and deallocate tensor memory explicitly.
These units change how high-performance kernels are structured.
Warp-specialized kernels assign data movement and computation to different warps.
A producer warp may load the next tile of data while a consumer warp performs matrix multiplication on the current tile.
This overlap requires synchronization in both directions.
Consumers must wait until their data is available, and producers must not reuse a buffer before outstanding operations have finished with it.

To support this pattern, NVIDIA GPUs provide mbarriers, barrier objects that reside in shared memory.
Both threads and asynchronous operations can update an mbarrier, and threads can wait on it.
An mbarrier is reusable, and each round of synchronization on it is called a \emph{phase}.
To determine when a phase completes, an mbarrier maintains two counters.
Its pending-arrival and pending-byte counters, $a$ and~$x$, begin each phase at $n$ and~$0$, where $n$ is the expected arrival count set at initialization.
An arrival decreases~$a$; announcing an expected transfer increases~$x$ by its byte count, and reporting completion decreases~$x$ by the completed byte count.
Each update applies atomically to the current phase, including an arrival that also announces a transfer.
When both counters reach zero, the phase completes and the counters reset to $(n,0)$ in the same atomic step.
Subsequent updates affect the next phase, so their execution order determines phase boundaries.
A wait lets a thread continue once a designated phase has completed, without changing the counters.

\subsection{Running Example}

\input{figures/running-example}
\input{figures/running-example-operations}

We use a matrix multiplication implemented with warp specialization as a running example.
\autoref{fig:running} shows its pseudocode.
For simplicity, we consider a single thread block multiplying a $128\times K$ matrix~$A$ by a $K\times128$ matrix~$B$.
For larger matrix multiplications, each thread block can compute one output tile, with multiple blocks computing different tiles in parallel.

The thread block has 128~threads, organized into four warps; thread~$t$ belongs to warp~$\lfloor t/32 \rfloor$.
To overlap data loading and computation, different warps have different roles.
Warp~0, the \emph{producer} warp, loads data from global memory into shared memory.
Warp~1, the \emph{consumer} warp, uses this data to compute the output tile with matrix multiply-accumulate (MMA) operations.
Note that data loading and computation are performed asynchronously by specialized hardware components (the TMA unit and tensor cores in this example), and only one thread in each participating warp (thread~0 in warp~0 and thread~32 in warp~1) needs to issue the operations.\footnote{A common implementation uses \code{elect.sync} to select one issuing thread in each participating warp.
For simplicity, this example fixes the issuing threads to threads 0 and 32.}
Once the output tile is complete, all four warps store it to global memory in the \emph{epilogue}.

This example uses three mbarriers: \code{full} signals that both input tiles are in shared memory, \code{empty} signals that the MMA has finished reading them, and \code{mma\_done} signals that the output tile is complete.

The producer first waits on \code{empty} to avoid overwriting the buffers (\code{sA} and \code{sB}) while the previous MMA still reads them~\step{1}.
It then arrives on \code{full}, announces $2\cdot128\cdot64$ elements for both tiles~\step{2}, and starts two asynchronous loads into \code{sA} and~\code{sB}~\step{3}.\footnote{PTX specifies transfer sizes in bytes.
Here we use element counts for clarity.}
Each load's completion subtracts $128\cdot64$ from \code{full}'s pending transaction count~\step{4}; once both finish, the phase completes and releases the consumer~\step{5}.
The consumer issues the MMA~\step{6}, which multiplies \code{sA} by \code{sB} and accumulates the product into \code{acc}.
The first MMA initializes \code{acc}; subsequent MMAs add to it.
After this thread's MMAs issued so far have completed, the operation triggers one arrival on \code{empty} for an intermediate iteration or \code{mma\_done} for the last iteration~\step{7}.
The final arrival completes phase~0 of \code{mma\_done} and releases the epilogue~\step{8}.
Each thread~$t$ then copies row~$t$ of \code{acc} to~$C$~\step{9}.

The barriers enforce three orders: each iteration's copies precede its MMA's reads; those reads precede the next iteration's buffer overwrites; and all MMA writes precede the epilogue's reads.

\input{figures/schedules-wait}

\paragraph{An incorrect variant.}
\label{sec:example:bug}
Now suppose the epilogue omits the wait~\step{8}; we call this variant program \MissingWait.
The barriers still complete their phases, but nothing orders the epilogue's loads of \code{acc} after the MMA's write.
The kernel has a data race, and whether the race corrupts~$C$ depends on the schedule (\autoref{fig:running:schedules}).
\autoref{sec:eval:cases} describes a closely related missing wait in CUTLASS.

%% file: figures/running-example.tex
\begin{figure}[t]
\lstset{
  style=kernel,
  basicstyle=\ttfamily\footnotesize,
  numbers=left,
  firstnumber=1,
  stepnumber=1,
  numberblanklines=false,
  xleftmargin=14pt,
  alsoletter={_},
  morekeywords=[1]{kernel,def,if,for,in,shared,tensor_memory,registers,barriers},
  columns=fixed,
  basewidth=0.5em,
  keywordstyle=[1]\color{swHB!85!black}\bfseries,
  morekeywords=[2]{gemm,thread_id,mbar_init,sync_threads,wait,arrive_expect,async_load_update_bar,async_mma_update_bar,async_load,wait_loads,store},
  keywordstyle=[2]\color{teal!75!black},
  morekeywords=[3]{bar,expected_arr,parity,tx,dst,src,lhs,rhs,accumulate,value},
  keywordstyle=[3]\color{swChoice!80!black},
  commentstyle=\color{black!55}
}
\begin{lstlisting}
kernel gemm(K, A[128,K], B[K,128], C[128,128])
t = thread_id()
shared sA[128,64], sB[64,128]
tensor_memory acc[128,128]
registers r[128]
shared barriers full, empty, mma_done

if t == 0: // initialize mbarriers
  mbar_init(bar=full, expected_arr=1)
  mbar_init(bar=empty, expected_arr=1)
  mbar_init(bar=mma_done, expected_arr=1)
sync_threads()

def producer_task(i):
  k = 64 * i
  wait(bar=empty, parity=(i + 1) % 2)        (*@\step{1}@*)
  arrive_expect(bar=full, tx=2*128*64)       (*@\label{line:running:copies:start}\step{2}@*)
  async_load_update_bar(                     (*@\step{3}@*)
      dst=sA, src=A[0:128, k:k+64], bar=full)
  // on completion, subtract 128*64 elems    (*@\step{4}@*)
  // from full's pending transaction count
  async_load_update_bar(                     (*@\step{3}@*)
      dst=sB, src=B[k:k+64, 0:128], bar=full)
  // on completion, subtract 128*64 elems    (*@\step{4}@*)
  // from full's pending transaction count   (*@\label{line:running:copies:end}@*)

def consumer_task(i):
  wait(bar=full, parity=i % 2)               (*@\step{5}@*)
  b = mma_done if i == K/64 - 1 else empty
  async_mma_update_bar(                      (*@\step{6}@*)
      dst=acc, lhs=sA, rhs=sB,
      accumulate=(i > 0), bar=b)
  // on completion, arrive once on b         (*@\step{7}@*)

for i in 0 .. K/64 - 1:     // tiles of K
  if t == 0:                // producer, in warp 0
    producer_task(i)
  if t == 32:               // consumer, in warp 1
    consumer_task(i)

// epilogue, all threads
wait(bar=mma_done, parity=0)                 (*@\step{8}@*)
async_load(dst=r, src=acc[t, 0:128])         (*@\step{9}@*)
wait_loads()
store(dst=C[t, 0:128], value=r)              (*@\step{9}@*)
\end{lstlisting}
\caption{Pseudocode of the running example.
We use abstract operations to make the code easier to understand.
\autoref{tab:example-operations} explains these operations and lists their corresponding PTX instructions.
}
\label{fig:running}
\end{figure}

%% file: figures/running-example-operations.tex
\begin{table*}[t]
\centering
\small
\setlength{\tabcolsep}{4pt}
\renewcommand{\arraystretch}{1.05}
\begin{tabularx}{\textwidth}{@{}>{\raggedright\arraybackslash}p{.22\textwidth}>{\raggedright\arraybackslash}X>{\raggedright\arraybackslash}p{.27\textwidth}@{}}
\toprule
Operation & Meaning & PTX correspondence \\
\midrule
\code{mbar\_init} &
Initialize \code{bar} with \code{expected\_arr} arrivals. &
\code{mbarrier.init} \\
\code{sync\_threads} &
Wait until all block threads reach this point. &
\code{bar.sync} \\
\code{wait} &
Wait until \code{bar}'s phase bit differs from \code{parity}. &
\code{mbarrier.try\_wait.parity} \\
\code{arrive\_expect} &
Arrive once and add \code{tx} to the transaction count. &
\code{mbarrier.arrive.expect\_tx} \\
\code{async\_load\_update\_bar} &
Copy \code{src} to \code{dst} asynchronously; subtract the transfer size from \code{bar}'s transaction count when done. &
\code{cp.async.bulk.tensor} \\
\code{async\_mma\_update\_bar} &
Async MMA with optional accumulation; arrive once on \code{bar} when all MMAs issued so far by this thread finish. &
\code{tcgen05.mma} + \code{tcgen05.commit} \\
\code{async\_load} &
Copy \code{src} to \code{dst} asynchronously. &
\code{cp.async}, \code{tcgen05.ld} \\
\code{wait\_loads} &
Wait for this thread's prior loads to finish. &
\code{cp.async.wait\_all}, \code{tcgen05.wait::ld} \\
\code{store} &
Write \code{value} to \code{dst}. &
\code{st.global} \\
\bottomrule
\end{tabularx}
\caption{Operations in the running example.}
\label{tab:example-operations}
\end{table*}

%% file: figures/schedules-wait.tex
\begin{figure}[t]
\centering
\scriptsize
\renewcommand{\arraystretch}{1.15}
\setlength{\tabcolsep}{2pt}
\begin{tabular}{@{}>{\raggedright\arraybackslash}p{\dimexpr0.5\columnwidth-5pt\relax}!{\hspace{6pt}}>{\raggedright\arraybackslash}p{\dimexpr0.5\columnwidth-5pt\relax}@{}}
\textbf{Schedule A:} the epilogue reads first & \textbf{Schedule B:} the tensor core writes first \\
\midrule
MMA and arrival request issued & MMA and arrival request issued \\
\cellcolor{red!8}epilogue reads \code{acc}: \textcolor{swBad}{\textbf{not yet written}} & tensor core writes \code{acc} \\
tensor core writes \code{acc} & \cellcolor{blue!8}arrival: \textbf{\code{mma\_done} completes} \\
\cellcolor{blue!8}arrival: \textbf{\code{mma\_done} completes} & \cellcolor{green!8}epilogue reads \code{acc}: \textcolor{swGood}{\textbf{written}} \\
\bottomrule
\end{tabular}
\caption{Two executions of \MissingWait{} (step~8 omitted), each listed top to bottom in execution order.}
\label{fig:running:schedules}
\end{figure}

%% file: sections/overview.tex
\section{Overview}
\label{sec:overview}

\input{figures/pipeline}

Given a kernel in PTX and a fixed configuration specifying grid and block dimensions, scalar kernel arguments, and tensor shapes and layouts, \sys performs three checks in order, each depending on the previous one.
\begin{itemize}
    \item \textbf{Well-synchronization.} We extend the notion of well-synchronization introduced by WEFT~\cite{sharma2015weft} to asynchronous transaction barriers, verifying that every barrier update and wait belongs to the same synchronization phase across {\em all} execution schedules.
    \item \textbf{Race freedom and resource safety.} Every two conflicting memory accesses are ordered by the happens-before relation, and every resource, such as tensor memory, remains allocated until all operations accessing it have completed.
    \item \textbf{Functional equivalence.} Given a specification, the kernel computes the same function as the reference.
\end{itemize}
\sys{} verifies one fixed configuration at a time. When verification succeeds, its guarantees hold for all execution schedules under the modeled semantics and all input tensor values consistent with the configuration and specified input constraints.

Well-synchronization guarantees that the ordering constraints imposed by barriers do not depend on the schedule, so one \emph{happens-before graph} describes the required ordering constraints across {\em all} executions.
Race freedom, checked against this graph, guarantees that every load reads the same write in every execution for fixed input values, so all executions compute the same values.
A kernel with both properties and valid resource lifetimes therefore computes a deterministic function of its inputs, independent of the schedule, and \sys{} extracts this function by symbolic execution and compares it with the reference.
This is how \sys{} covers every schedule without enumerating them: it establishes the required orders once, from the barriers, and then reasons about a pure function of the inputs.

\autoref{fig:pipeline} shows how \sys{} checks the three properties.
It first specializes the kernel for the launch configuration, resolving branches and unrolling loops, so that each thread executes a finite sequence of instructions (\autoref{sec:repr}).
From the resulting program it derives two views: the \emph{event view} records when instructions and their asynchronous effects happen, and the \emph{value view} records what each instruction computes, including the addresses it accesses.
The synchronization check verifies well-synchronization on the event view and constructs the happens-before graph used by subsequent analyses (\autoref{sec:sync}).
The memory-safety check combines the happens-before graph with the addresses from the value view (\autoref{sec:races:races}).
It requires every pair of conflicting accesses to be ordered and every resource to be deallocated after all accesses to it have completed.
Finally, \sys{} symbolically evaluates the race-free kernel, using the happens-before graph to determine the value observed by each load, yielding one expression per output element (\autoref{sec:races:determinate}).
It then proves these expressions equal to those of the reference program, written in PyTorch, by a reference-guided proof that follows the structure of the reference's computation (\autoref{sec:reference}).

Explicitly representing every thread-level event and scalar computation can produce very large representations for practical GPU kernels.
\sys{} therefore represents repeated structure across threads, data elements, and reduction steps by templates, and every stage above works on the templates directly (\autoref{sec:scaling}).

\sys{} relies on the following assumptions.
\begin{itemize}
\item \textbf{Static control flow.} The fixed launch configuration determines every branch output and finite loop bound. Branches and loop bounds cannot depend on runtime tensor values.
\item \textbf{Strong memory model.} Writes become visible to all threads and units when their modeled memory operations execute; fences and weak-memory effects are not checked.
\item \textbf{Real arithmetic.} Floating-point computation is interpreted over the reals, excluding rounding effects.
\end{itemize}

%% file: figures/pipeline.tex
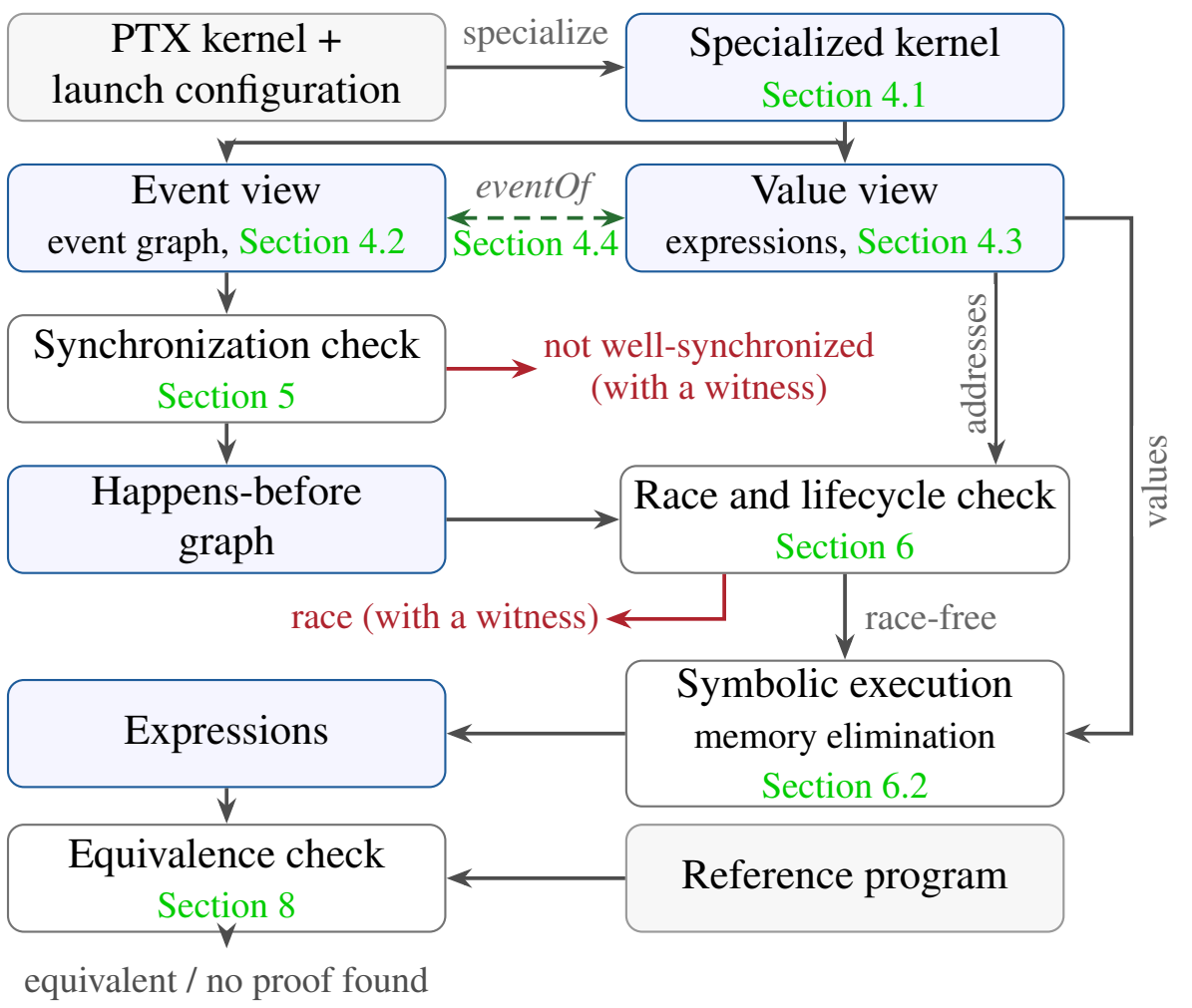
\begin{figure}[t]
\centering
\captionsetup{skip=3pt}
\begin{tikzpicture}[x=1cm,y=1cm,
  stage/.style={draw=black!55,rounded corners=3pt,fill=white,align=center,
    font=\footnotesize,minimum width=2.9cm,minimum height=7.1mm,
    inner xsep=2.5pt,inner ysep=1.5pt},
  artifact/.style={stage,draw=swHB,fill=blue!4},
  input/.style={stage,draw=black!40,fill=black!3},
  verdict/.style={font=\scriptsize,align=center,text=swBad,inner sep=1pt},
  outcome/.style={font=\scriptsize,align=center,text=black!70},
  flow/.style={-{Stealth[length=1.7mm]},line width=0.7pt,draw=black!70},
  link/.style={{Stealth[length=1.6mm]}-{Stealth[length=1.6mm]},densely dashed,
    line width=0.7pt,draw=swGood},
  lbl/.style={font=\scriptsize,text=black!60,fill=white,align=center,inner sep=1.5pt}]
  \node[input] (kernel) at (0,0) {PTX kernel +\\launch configuration};
  \node[artifact] (spec) at (4.1,0) {Specialized kernel\\{\scriptsize\autoref{sec:repr:program}}};
  \node[artifact] (events) at (0,-1.0) {Event view\\{\scriptsize event graph, \autoref{sec:repr:events}}};
  \node[artifact] (values) at (4.1,-1.0) {Value view\\{\scriptsize expressions, \autoref{sec:repr:values}}};
  \node[stage] (cert) at (0,-2.0) {Synchronization check\\{\scriptsize\autoref{sec:sync}}};
  \node[artifact] (hb) at (0,-3.0) {Happens-before\\graph};
  \node[stage] (race) at (4.1,-3.0) {Race and lifecycle check\\{\scriptsize\autoref{sec:races}}};
  \node[stage] (elim) at (4.1,-4.42) {Symbolic execution\\{\scriptsize memory elimination}\\{\scriptsize\autoref{sec:races:determinate}}};
  \node[artifact] (scalar) at (0,-4.42) {Expressions};
  \node[stage] (equiv) at (0,-5.38) {Equivalence check\\{\scriptsize\autoref{sec:reference:egraph}}};
  \node[input] (ref) at (4.1,-5.38) {Reference program};
  \draw[flow] (kernel) -- node[lbl,above=1pt] {specialize} (spec);
  \draw[flow] (spec.south) -- (4.1,-0.5) -| (events.north);
  \draw[flow] (spec) -- (values);
  \draw[link] (events) -- node[lbl,above=0.5pt,fill=none] {$\eventof$} node[lbl,below=0.5pt,fill=none] {\autoref{sec:repr:link}} (values);
  \draw[flow] (events) -- (cert);
  \draw[flow] (cert) -- (hb);
  \draw[flow] (hb) -- (race);
  \draw[flow] ([xshift=1.0cm]values.south) -- node[lbl,rotate=90,above=1pt] {addresses} ([xshift=1.0cm]race.north);
  \draw[flow] (race) -- node[lbl,right=2pt] {race-free} (elim);
  \draw[flow] (values.east) -- (6.0,-1.0) -- node[lbl,rotate=90,below=1pt] {values} (6.0,-4.42) -- (elim.east);
  \draw[flow] (elim) -- (scalar);
  \draw[flow] (scalar) -- (equiv);
  \draw[flow] (ref) -- (equiv);
  \node[verdict] (v1) at (3.2,-2.0) {not well-synchronized\\(with a witness)};
  \draw[flow,draw=swBad] (cert) -- (v1);
  \node[verdict] (v2) at (1.45,-3.66) {race (with a witness)};
  \draw[flow,draw=swBad] ([xshift=-0.8cm]race.south) |- (v2.east);
  \node[outcome] (v3) at (0,-6.08) {equivalent / no proof found};
  \draw[flow] (equiv) -- (v3);
\end{tikzpicture}
\caption{The pipeline: representations (blue), analyses (white), inputs (grey).}
\label{fig:pipeline}
\end{figure}

%% file: sections/representation.tex
\section{Program Representation}
\label{sec:repr}

\sys{} analyzes two views of one program: the \emph{event view} describes when operations happen and how their events may be scheduled, and the \emph{value view} describes what they compute, including the addresses they access.

\subsection{Program Specialization}
\label{sec:repr:program}

The launch configuration decides every branch and every loop bound of the kernel.
\sys{} therefore first specializes the kernel for it: it resolves every branch and unrolls every loop, so that the program of each thread becomes a finite sequence of \emph{instruction occurrences}, one per execution of a PTX instruction by that thread.

For simplicity, we specialize the running example with $K = 64$, for which each loop runs once.
Specialization resolves the role branches and unrolls the loops (\autoref{fig:specialized}).
Thread~0 executes the initializations, the producer's instructions, and the epilogue; thread~32 executes the consumer's instructions and the epilogue; every other thread executes only the epilogue.
All threads share the \code{syncthreads}.

\input{figures/representation}

\subsection{Event View}
\label{sec:repr:events}

\begin{table}[t]
\footnotesize
\renewcommand{\arraystretch}{1.1}
\setlength{\tabcolsep}{3pt}
\centering
\caption{Kinds of events.}
\label{tab:events}
\begin{tabularx}{\columnwidth}{@{}l>{\raggedright\arraybackslash}X@{}}
\toprule
Event & When it occurs \\
\midrule
\multicolumn{2}{@{}l}{\emph{Events of one thread's instruction}} \\
issue & the thread starts an asynchronous instruction or a wait \\
execute & the instruction's memory effects take place, by the thread or by the TMA or tensor core \\
return & a wait returns to its thread: an mbarrier wait, \code{barrier.cluster.wait}, \code{cp.async.wait\_group}, or \code{tcgen05.wait::ld} \\
\midrule
\multicolumn{2}{@{}l}{\emph{Events on a barrier}} \\
init & a thread initializes an mbarrier \\
update & the counters change: a thread arrives or announces bytes, a copy completes its bytes, a commit's arrival lands, or a thread arrives on a cluster barrier \\
\midrule
\multicolumn{2}{@{}l}{\emph{Events shared by all participants}} \\
collective & one event for all participating threads: \code{bar.sync} (\code{syncthreads} in our listings), \code{bar.warp.sync}, or a warp-wide data exchange such as \code{shfl.sync} or \code{elect.sync} \\
\bottomrule
\end{tabularx}
\end{table}

The event view records when the effects of each instruction occur.
\sys{} keeps only the instructions with memory or barrier effects; instructions that touch only registers have no events.
Some instructions are executed asynchronously.
For example, issuing a TMA copy does not block the issuing thread; the TMA unit performs the copy and records its completion on an mbarrier.
To express such executions, \sys{} splits each instruction occurrence into one or more \emph{events}, the moments that matter for scheduling, following the semantics of PTX~\cite{nvidia2026ptx}.
\autoref{tab:events} lists the kinds of events; the TMA copy above has three: its issue by the thread, its execution by the TMA unit, and its update of the mbarrier.
We write~$\mathcal{V}_P$ for the finite set of events of a program~$P$.

The event view is a directed acyclic graph whose vertices are $\mathcal{V}_P$ and whose edges are \emph{program-order} dependencies.
A program-order edge from~$e$ to~$f$ states that the program or the definition of the events requires~$f$ to take place after~$e$.
For instance, the events of one instruction are ordered from its issue to its execution and update; a thread's next instruction follows the issue of an asynchronous instruction, not its execution; and a collective event follows each participant's preceding event and precedes its next one.
A warp-wide data exchange, however, orders only execution, because PTX specifies memory ordering for \code{bar.warp.sync} but not for \code{shfl.sync} or \code{elect.sync}.
The race check therefore does not use the edges of such collectives.
Appendix~\ref{app:sync-model} derives these dependencies from the instruction-family semantics.

We label the events of thread~$t$ as $\evl{t}{1}, \evl{t}{2}, \ldots$, numbering them instruction by instruction, and the shared \code{syncthreads} as~$\Sync$.
In the running example, the copy of~A contributes three events: its issue~$\QA$ by the producer, its execute event~$\LA$, which writes \code{sA}, and its update~$\CA$ of \code{full}.
The commit contributes its issue and an arrival on \code{mma\_done}; the arrival follows the execute event~$\MX$ of the MMA.
Each wait contributes an issue and a return, and $\Rt$ is the execute event of thread~$t$'s \code{tcgen05.ld}.
\autoref{fig:event-graph} shows the whole event view.
Barriers add no edges to it, because which updates a wait observes can differ between executions; \autoref{sec:sync} derives these \emph{synchronization edges} (\autoref{fig:candidate-graph}) and checks that they hold in every execution.

\subsection{Value View}
\label{sec:repr:values}

\begin{table}[t]
\footnotesize
\renewcommand{\arraystretch}{1.15}
\centering
\caption{Value-view operators.}
\label{tab:values}
\begin{tabular}{@{}l@{\hspace{1em}}l@{}}
\toprule
Operator & Meaning \\
\midrule
constant & a fixed value \\
parameter & a kernel argument, e.g., \code{A} \\
coordinate & a thread or block index \\
\code{op(x1, ..., xn)} & a scalar operation \\
\code{load(a)} & a read at address \code{a} \\
\code{store(a, v)} & a write of \code{v} to \code{a} \\
\code{matmul(x, y)} & the product of tiles \\
\bottomrule
\end{tabular}
\end{table}

The value view records what each thread computes, as a set of expressions over the operators of \autoref{tab:values}.
It tracks dataflow through registers and within an instruction, but not through memory: a load records where it reads, but does not track which values it reads.

In the running example (\autoref{fig:views}), the copy of~\code{A} is a load of~\code{A} and a store to~\code{sA}, and the MMA is two loads, a \code{matmul}, and a store to~\code{acc}; every memory operation takes its address as its first operand.
Here \code{A}, \code{B}, and \code{C} are pointer parameters, \code{sA} and \code{sB} are buffers in shared memory, and \code{acc} is the accumulator in tensor memory.
The address of row~$t$ of~\code{C}, written in logical element offsets as $\code{C} + 128 \cdot t$, is a formula rather than a number, which allows parameterized descriptions of all threads in \autoref{sec:scaling}.
Nothing yet connects the store to \code{sA} with the load from \code{sA}: whether the load sees the stored tile depends on the order of the two operations.

\subsection{Correspondence Between Views}
\label{sec:repr:link}

The two views are linked by a mapping from memory operations to events.
For every load and store~$m$ of the value view, \sys{} maintains the event that performs it, written $\eventof(m)$.
In the running example (\autoref{fig:views}), the load from~\code{A} and the store to~\code{sA} both map to~$\LA$, and the MMA's loads and store map to~$\MX$.
The analyses of the event view establish orders between events (Sections~\ref{sec:sync} and~\ref{sec:races}), and the mapping transfers ordering information to memory operations: two memory operations are ordered if their events are.

%% file: figures/representation.tex
\begin{figure*}[t]
\centering
\captionsetup[subfigure]{skip=3pt}
\begin{subfigure}{\textwidth}
\centering
\begin{tikzpicture}[x=1cm,y=-0.38cm,
  ins/.style={swbox,minimum width=3.6cm,minimum height=3.3mm,inner ysep=0.6pt,font=\fontsize{6}{7}\selectfont\ttfamily},
  val/.style={font=\scriptsize,text=black!60,anchor=west,align=left,inner sep=1pt},
  po/.style={-{Stealth[length=1.3mm]},line width=0.5pt,draw=black!70}]
  \node[swrole] at (0,-0.9) {thread 0};
  \node[swrole] at (5.9,-0.9) {thread 32};
  \node[swrole] at (11.8,-0.9) {thread $t \notin \{0, 32\}$};
  \node[ins,minimum width=0pt,anchor=west] (a0) at (-1.8,0) {init(full, 1)};
  \node[ins,minimum width=0pt,anchor=west] (a1) at ([xshift=2.5mm]a0.east) {init(empty, 1)};
  \node[ins,minimum width=0pt,anchor=west] (a2) at ([xshift=2.5mm]a1.east) {init(mma\_done, 1)};
  \node[ins,minimum width=15.4cm] (sync) at (5.9,1.05) {syncthreads: one occurrence shared by all 128 threads};
  \node[ins] (a4) at (0,2.1) {wait(empty, 1)};
  \node[val] at (1.9,2.1) {$(i+1) \bmod 2 = 1$};
  \node[ins] (a5) at (0,3.1) {arrive\_expect(full, tx=16384)};
  \node[ins,minimum height=7.0mm,align=center] (a6) at (0,4.6) {tma\_load(sA <- A[0:128, 0:64], full)\\.complete\_tx(full, tx=8192)};
  \node[val] at (2.15,4.6) {store(sA, load(A))};
  \node[ins,minimum height=7.0mm,align=center] (a7) at (0,6.6) {tma\_load(sB <- B[0:64, 0:128], full)\\.complete\_tx(full, tx=8192)};
  \node[val] at (2.15,6.6) {store(sB, load(B))};
  \node[ins] (a8) at (0,8.1) {wait(mma\_done, 0)};
  \node[ins] (a9) at (0,9.1) {r = tcgen05.ld(acc[0, 0:128])};
  \node[val] at (1.9,9.1) {load(acc $+$ $128 \cdot 0$)};
  \node[ins] (a10) at (0,10.1) {tcgen05.wait::ld()};
  \node[ins] (a11) at (0,11.1) {st.global(C[0, 0:128] <- r)};
  \node[val] at (1.9,11.1) {store(C $+$ $128 \cdot 0$, r)};
  \node[ins] (b4) at (5.9,2.1) {wait(full, 0)};
  \node[val] at (7.8,2.1) {$i \bmod 2 = 0$};
  \node[ins] (b5) at (5.9,3.1) {tcgen05.mma(acc = sA * sB)};
  \node[val,text width=2.1cm] at (7.8,3.4) {store(acc,\\\ \ matmul(load(sA),\\\ \ \ load(sB)))};
  \node[ins] (b7) at (5.9,4.1) {tcgen05.commit(mma\_done)};
  \node[ins] (b8) at (5.9,8.1) {wait(mma\_done, 0)};
  \node[ins] (b9) at (5.9,9.1) {r = tcgen05.ld(acc[32, 0:128])};
  \node[val] at (7.8,9.1) {load(acc $+$ $128 \cdot 32$)};
  \node[ins] (b10) at (5.9,10.1) {tcgen05.wait::ld()};
  \node[ins] (b11) at (5.9,11.1) {st.global(C[32, 0:128] <- r)};
  \node[val] at (7.8,11.1) {store(C $+$ $128 \cdot 32$, r)};
  \node[ins] (c8) at (11.8,8.1) {wait(mma\_done, 0)};
  \node[ins] (c9) at (11.8,9.1) {r = tcgen05.ld(acc[t, 0:128])};
  \node[val] at (13.7,9.1) {load(acc $+$ $128 \cdot t$)};
  \node[ins] (c10) at (11.8,10.1) {tcgen05.wait::ld()};
  \node[ins] (c11) at (11.8,11.1) {st.global(C[t, 0:128] <- r)};
  \node[val] at (13.7,11.1) {store(C $+$ $128 \cdot t$, r)};
  \draw[po] (a0) -- (a1);
  \draw[po] (a1) -- (a2);
  \draw[po] (a2) -- (a2 |- sync.north);
  \draw[po] (a4 |- sync.south) -- (a4);
  \draw[po] (b4 |- sync.south) -- (b4);
  \draw[po] (c8 |- sync.south) -- (c8);
  \foreach \p/\q in {a4/a5,a5/a6,a6/a7,a7/a8,a8/a9,a9/a10,a10/a11,b4/b5,b5/b7,b7/b8,b8/b9,b9/b10,b10/b11,c8/c9,c9/c10,c10/c11} {
    \draw[po] (\p) -- (\q);
  }
\end{tikzpicture}
\caption{Specialized per-thread programs in program order.
Gray text shows memory operations and resolved wait parities.}
\label{fig:specialized}
\end{subfigure}

\smallskip

\begin{subfigure}{\textwidth}
\centering
\begin{tikzpicture}[x=1.2cm,y=-0.42cm,
  ev/.style={swbox,minimum width=#1,minimum height=3.4mm,inner ysep=0.6pt,execute at begin node={\hspace*{5.8mm}}},ev/.default=3.1cm,
  hw/.style={ev=#1,fill=black!7},hw/.default=2.75cm,
  swdep/.style={-{Stealth[length=1.3mm]},line width=0.5pt,draw=black!70}]
\def\evmode{0}
\input{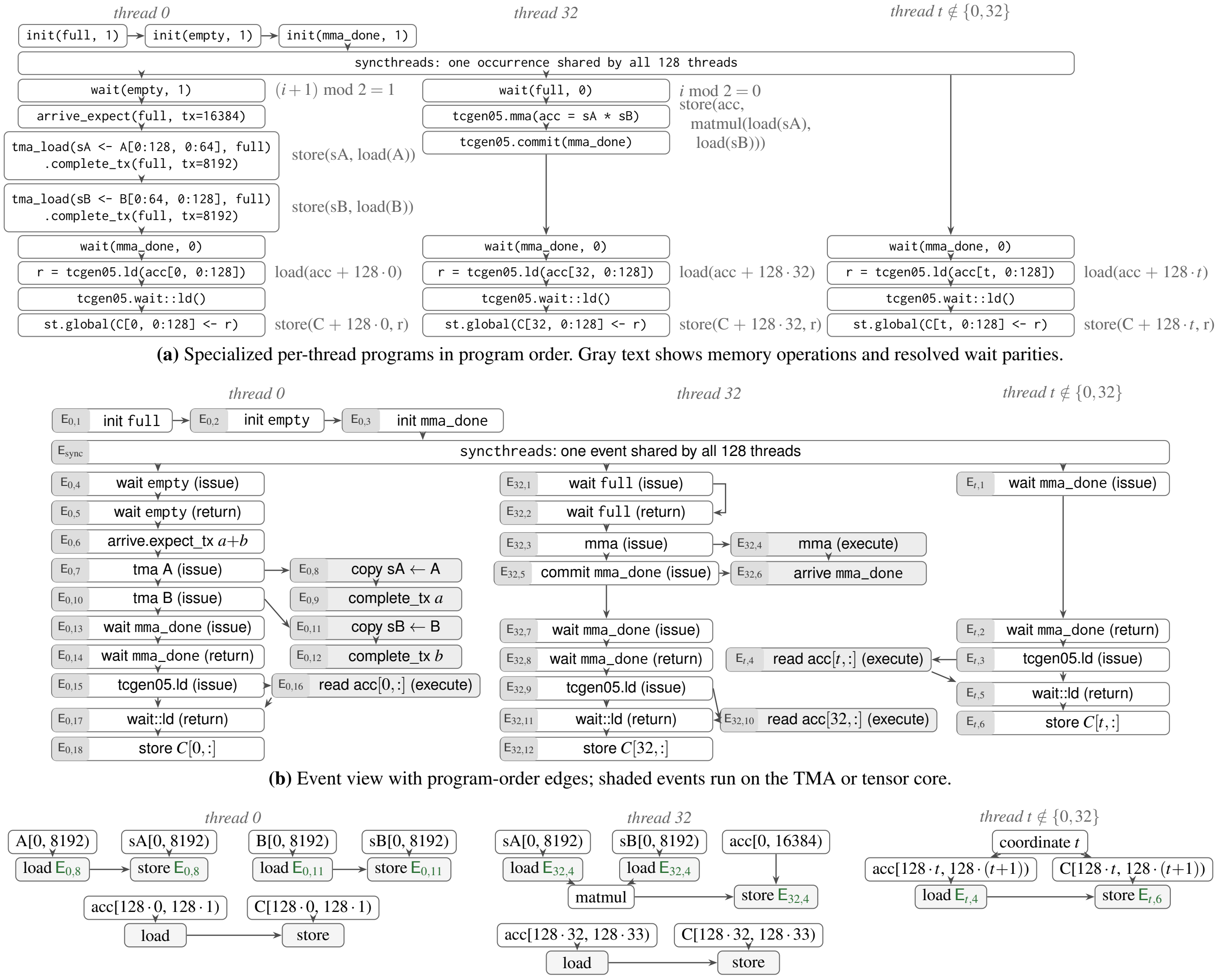}
\end{tikzpicture}
\caption{Event view with program-order edges; shaded events run on the TMA or tensor core.}
\label{fig:event-graph}
\end{subfigure}

\smallskip

\begin{subfigure}{\textwidth}
\centering
\begin{tikzpicture}[x=1cm,y=-0.40cm,
  memop/.style={draw=black!60,rounded corners=2pt,fill=black!4,minimum height=3.4mm,minimum width=9mm,inner ysep=0.6pt,inner xsep=3pt,font=\scriptsize},
  op/.style={memop,fill=white},
  val/.style={-{Stealth[length=1.3mm]},line width=0.5pt,draw=black!70}]
  \newcommand{\tg}[1]{\,\textcolor{swGood}{$#1$}}
  \node[swrole] at (-0.27,-0.9) {thread 0};
  \node[swrole] at (5.95,-0.9) {thread 32};
  \node[swrole] at (11.5,-0.9) {thread $t \notin \{0, 32\}$};
  \node[op] (a1) at (-2.9,0) {A[0, 8192)};       \node[op] (a2) at (-1.2,0) {sA[0, 8192)};
  \node[memop] (la) at (-2.9,1) {load\tg{\LA}};  \node[memop] (sa) at (-1.2,1) {store\tg{\LA}};
  \draw[val] (a1) -- (la); \draw[val] (a2) -- (sa); \draw[val] (la) -- (sa);
  \node[op] (b1) at (0.6,0) {B[0, 8192)};        \node[op] (b2) at (2.3,0) {sB[0, 8192)};
  \node[memop] (lb) at (0.6,1) {load\tg{\LB}};   \node[memop] (sb) at (2.3,1) {store\tg{\LB}};
  \draw[val] (b1) -- (lb); \draw[val] (b2) -- (sb); \draw[val] (lb) -- (sb);
  \node[op] (c1) at (-1.4,2.4) {acc[$128 \cdot 0$, $128 \cdot 1$)}; \node[op] (c2) at (0.9,2.4) {C[$128 \cdot 0$, $128 \cdot 1$)};
  \node[memop] (l0) at (-1.4,3.4) {load};
  \node[memop] (s0) at (0.9,3.4) {store};
  \draw[val] (c1) -- (l0); \draw[val] (c2) -- (s0); \draw[val] (l0) -- (s0);
  \node[op] (m1) at (4.25,0) {sA[0, 8192)};   \node[op] (m2) at (5.95,0) {sB[0, 8192)};   \node[op] (m3) at (7.65,0) {acc[0, 16384)};
  \node[memop] (ma) at (4.25,1) {load\tg{\MX}}; \node[memop] (mb) at (5.95,1) {load\tg{\MX}};
  \node[op] (mm) at (5.1,2) {matmul};
  \node[memop] (sacc) at (7.65,2) {store\tg{\MX}};
  \draw[val] (m1) -- (ma); \draw[val] (m2) -- (mb); \draw[val] (ma) -- (mm); \draw[val] (mb) -- (mm);
  \draw[val] (mm) -- (sacc); \draw[val] (m3) -- (sacc);
  \node[op] (d1) at (4.75,3.4) {acc[$128 \cdot 32$, $128 \cdot 33$)}; \node[op] (d2) at (7.25,3.4) {C[$128 \cdot 32$, $128 \cdot 33$)};
  \node[memop] (l32) at (4.75,4.4) {load};
  \node[memop] (s32) at (7.25,4.4) {store};
  \draw[val] (d1) -- (l32); \draw[val] (d2) -- (s32); \draw[val] (l32) -- (s32);
  \node[op] (ct) at (11.5,0) {coordinate $t$};
  \node[op] (t1) at (10.2,1) {acc[$128 \cdot t$, $128 \cdot (t{+}1)$)}; \node[op] (t2) at (12.85,1) {C[$128 \cdot t$, $128 \cdot (t{+}1)$)};
  \node[memop] (lt) at (10.2,2) {load\tg{\Rt}};  \node[memop] (st) at (12.85,2) {store\tg{\St}};
  \draw[val] (ct) -- (t1); \draw[val] (ct) -- (t2);
  \draw[val] (t1) -- (lt); \draw[val] (t2) -- (st); \draw[val] (lt) -- (st);
\end{tikzpicture}
\caption{Value view: memory operations (gray), dataflow arrows, and $\eventof$ tags (green).
\code{X[lo, hi)} denotes logical element offsets within \code{X}.}
\label{fig:views}
\end{subfigure}
\caption{Representations of the running example with $K = 64$.}
\label{fig:representation}
\end{figure*}

%% file: sections/synchronization.tex
\section{Synchronization Analysis}
\label{sec:sync}

The event view captures program order.
Barrier dependencies additionally depend on which updates belong to each phase and which phase each wait targets. These phase assignments may vary with the schedule, and analyzing the resulting dependencies separately for every possible execution is impractical.

This section checks \emph{well-synchronization}, allowing the race check to use the same barrier dependencies across all executions.
For kernels that pass, we combine barrier dependencies with program order to obtain a single happens-before graph for race checking.

\subsection{Well-Synchronization}

Following WEFT~\cite{sharma2015weft}, which introduced the notion for named barriers, a kernel is well-synchronized if all its executions agree on the phase of every update and every wait.

\paragraph{Traces.}
We describe executions by traces.
A \emph{trace} is a sequence of events that follows program order and the barrier semantics.
It may still be in progress, get stuck at a wait whose phase never completes, or stop at an error, such as using a barrier before its initialization or an update that makes a counter negative.
For a trace~$\pi$, we write $\evs(\pi)$ for the events it performs.
A trace is \emph{complete} if it performs all events, and \emph{safe} if it contains no error.
Appendix~\ref{app:sync-model} defines the semantics and the errors precisely.

\paragraph{Phase assignments.}
Within a trace~$\pi$, each update and each wait on a barrier is associated with a phase; we write $\ph{B}{k}$ for the $k$-th phase of barrier~$B$.
An update belongs to the phase of its barrier that is in progress when the update executes.
A wait is associated with the phase whose completion allows it to return.
We call this mapping the \emph{phase assignment} of~$\pi$ and write it as $\phaseof_\pi$.

Well-synchronization requires all traces to agree on their phase assignments.
It also requires every trace to be safe.

\begin{definition}[Well-synchronization]
A program~$P$ is \emph{well-synchronized} if
\begin{enumerate}[leftmargin=2em,itemsep=1pt]
\item every trace of~$P$ is safe, and
\item for any two traces $\pi$ and $\sigma$ of~$P$, $\phaseof_\pi(e) = \phaseof_\sigma(e)$ for every update and every wait~$e$ that both contain.
\end{enumerate}
\end{definition}

The running example in \autoref{fig:representation} is well-synchronized.
Every trace assigns the same phases to its updates and waits:
\begin{itemize}[leftmargin=1.5em]
\item $\evl{32}{1}$ waits for $\AEall$, $\CA$, and $\CB$, which form $\ph{\full}{0}$;
\item $\evl{0}{13}$, $\evl{32}{7}$, and $\evl{t}{1}$ wait for $\evl{32}{6}$, which forms $\ph{\mmadone}{0}$;
\item $\evl{0}{4}$ waits for no update and returns at once.
\end{itemize}

We next consider program \SplitTx, a variant of the running example in which the producer announces each copy separately.
The following snippet replaces lines~\ref{line:running:copies:start}--\ref{line:running:copies:end} of \autoref{fig:running}, where $a$ and $b$ denote the numbers of elements in the tiles of~$A$ and~$B$, respectively:
\begin{lstlisting}[style=kernel,numbers=none,xleftmargin=1.5em,literate={<-}{{$\leftarrow$}}2]
if t == 0:   // producer, separate announcements
    (*@\textcolor{swBad}{\ttfamily arrive.expect\_tx(full, a)}@*)                 (*@$\AEA$@*)
    tma_load(sA <- A, full)                   (*@$\QAv$@*)
      .complete_tx(full, a)                   (*@$\CAv$@*)
    (*@\textcolor{swBad}{\ttfamily expect\_tx(full, b)}@*)                        (*@$\EB$@*)
    tma_load(sB <- B, full)                   (*@$\QBv$@*)
      .complete_tx(full, b)                   (*@$\CBv$@*)
\end{lstlisting}
The first announcement also carries the single arrival on \code{full}.
We label the events of the listing in order.
$\QAv$ and $\QBv$ issue the copies.
The updates of \code{full} are the two announcements $\AEA$ and $\EB$ and the completions $\CAv$ and $\CBv$ that the TMA performs after each copy.
\autoref{fig:schedules} shows two of its traces.
In Schedule~B, $\EB$ occurs before $\CAv$, so all four updates of \code{full} belong to $\ph{\full}{0}$.
In Schedule~A, $\CAv$ occurs first.
The counters of \code{full} then reach zero after $\AEA$ and $\CAv$ alone, and $\ph{\full}{0}$ completes early.
$\EB$ and $\CBv$ belong to $\ph{\full}{1}$, and the MMA may read \code{sB} before the second copy writes it.

\input{figures/schedules}

\subsection{The Synchronization Check}

Well-synchronization quantifies over all executions.
Prior work~\cite{sharma2015weft} checks this property for named barriers by requiring consecutive phases of each barrier to be ordered.
This condition suffices because the counters of named barriers only decrease.
The mbarriers of modern GPU kernels lack this property: their counters, in particular the transaction counters, can both increase and decrease within a phase.
The checks of WEFT are then insufficient.
\sys{} extends the approach of WEFT to such barriers by adding a check that every phase completes exactly when all of its updates have occurred.

\paragraph{Reference trace.}
Specialization fixes barrier identities and the count and parity arguments of synchronization instructions.
The synchronization check (\autoref{alg:certify}) starts by simulating the kernel once to completion.
If the simulation fails, it already exhibits a problem: an error is a safety violation, and getting stuck is a deadlock.

If the simulation succeeds, the resulting safe complete trace is the \emph{reference trace}~$\tau$, so $\phaseof_\tau$ is defined on every update and every wait.
The check then asks whether every trace~$\pi$ is safe and agrees with~$\tau$, that is, $\phaseof_\pi(e) = \phaseof_\tau(e)$ for every update and every wait~$e$ in~$\pi$.
If so, the kernel is well-synchronized and free of deadlock, since $\tau$ completes (\autoref{thm:well-synchronization} and \autoref{lem:graph-execution-correspondence}).

\paragraph{The candidate graph.}
The reference trace also suggests the orders that synchronization creates.
We call the updates that $\tau$ assigns to phase $\ph{B}{k}$ its \emph{contributors} $\contr{B}{k}$.
In~$\tau$, each wait returns only after all contributors of its target phase have occurred. \sys{} records
these orders in a \emph{candidate graph}~$H$, built from the event graph and~$\tau$. For
every phase $\ph{B}{k}$ that completes in~$\tau$, it adds a completion vertex $\cmp{B}{k}$, an edge to it
from every contributor of $\ph{B}{k}$, and an edge from it to $\ret(w)$ for every wait~$w$
whose issue $\tau$ assigns to $\ph{B}{k}$.
We write $e \reach f$ if $H$ has a nonempty path from $e$ to~$f$.

\input{figures/hb-graph}

\autoref{fig:candidate-graph} shows the candidate graph of the running example: the event graph of \autoref{fig:event-graph} together with the completion vertices and the synchronization edges into and out of them.
The figure also overlays program \MissingWait{}, which lacks the green parts.

The synchronization check then examines~$H$ with four checks: separation, wait, initialization, and completion.
Each is a condition on the candidate graph alone, which \sys{} built from the single trace~$\tau$.
A kernel passes all four if and only if it is well-synchronized (\autoref{thm:well-synchronization}).
The pseudocode displays the mbarrier/parity-wait case without \code{.noComplete} updates, with counter upper bounds already checked; $\SyncCheck$ includes the variant-specific conditions below.

\paragraph{Separation check.}

As in WEFT, consecutive phases of each barrier must be ordered: for every phase $\ph{B}{k}$ with $k > 0$, \sys{} requires $d \reach e$ for every $d \in \contr{B}{k - 1}$ and every $e \in \contr{B}{k}$, so that no update falls into an earlier phase.
A path $\cmp{B}{k - 1} \reach e$ establishes all of these at once.

\paragraph{Wait check.}

A parity names a phase only modulo two, so a wait for $\ph{B}{k}$ is unambiguous only if it issues after $\ph{B}{k - 1}$ has completed and before $\ph{B}{k + 1}$ can complete.
For every parity wait~$w$ whose issue $\tau$ assigns to $\ph{B}{k}$, \sys{} requires:
\begin{itemize}[leftmargin=1.5em,itemsep=1pt]
\item if $k > 0$, then $\issue(w)$ follows every contributor of $\ph{B}{k - 1}$: $d \reach \issue(w)$ for every $d \in \contr{B}{k - 1}$;
\item if $\ph{B}{k + 1}$ completes in~$\tau$, then $\issue(w)$ precedes some contributor of $\ph{B}{k + 1}$: $\issue(w) \reach d$ for some $d \in \contr{B}{k + 1}$.
\end{itemize}

A state wait targets the phase captured at runtime by its statically identified token-producing arrival.
For target $\ph{B}{k}$, $\ret(w)$ must reach some contributor of $\ph{B}{k+1}$ in~$H$ if that phase completes in~$\tau$.
Named waits target the phase of their blocking registration.

\paragraph{Initialization check.}

For every mbarrier~$B$, with initialization event $\initof(B)$, and every update or wait issue~$e$ on~$B$, \sys{} requires $\initof(B) \reach e$.
It extends the separation and wait checks to the start of the barrier: the initialization plays the role of the completion of the phase before phase~0.

\paragraph{Completion check.}
The completion check guards against phases that complete too early.
The three checks above miss this case when counters can increase.
Program \SplitTx{} passes them, yet in Schedule~A of \autoref{fig:schedules} phase~0 of \code{full} completes after $\AEA$ and $\CAv$ alone.
The completion check must therefore consider every prefix of a phase.

\begin{definition}[Ideal, zero ideal]
An \emph{ideal} of phase $\ph{B}{k}$ is a set $I \subseteq \contr{B}{k}$ such that, for all
contributors $e$ and~$f$, $f \in I$ and $e \reach f$ imply $e \in I$. Its \emph{state} is
\[
  \stateof(I) \;=\; (n, 0) + \sum_{e \in I} \labelof(e),
\]
where $n$ is the phase's initial arrival count: the initialized count for an mbarrier, or the count declared by its first registration for a named barrier.
The \emph{label} $\labelof(e) = (\Delta a, \Delta x)$ is the change that update~$e$ makes to the pending arrivals and bytes.
An ideal is a \emph{zero ideal} if $\stateof(I) = (0, 0)$.
\end{definition}

For every phase $\ph{B}{k}$, the completion check requires
\[
  \stateof(I) \lexgt (0, 0) \quad \text{for every ideal } I \subsetneq \contr{B}{k}.
\]
Every ideal state must also respect the instruction's counter bounds.
An update marked \code{.noComplete} must not complete a phase; if its phase completes in~$\tau$, the update must reach another contributor of that phase in~$H$.

\input{figures/ideal-check}

\autoref{fig:ideal-check}(a) shows phase~0 of \code{full} for program \SplitTx,
as recorded by Schedule~B. The event graph orders its contributors as
$\AEA \reach \EB \reach \CBv$ and $\AEA \reach \CAv$. The ideal $\{\AEA, \CAv\}$ has state
$(1, 0) + (-1, +a) + (0, -a) = (0, 0)$, so the check fails. The ideal also explains the
failure: it is exactly the beginning of Schedule~A. For the running example
(\autoref{fig:ideal-check}b), every ideal containing $\AEall$ has state $(0, a + b - x)$,
where $x$ counts the completed elements in the ideal. Only the full phase reaches
$(0, 0)$, so the phase passes.

\sys{} does not enumerate the exponentially many ideals of a phase.
Instead, it finds the proper ideal of lexicographically minimal state with a few minimum cuts, as an instance of the classic \emph{closure problem}~\cite{picard1976closure}; Appendix~\ref{app:min-ideal} gives the construction.

\begin{algorithm}[t]
\footnotesize
\Kw{procedure} $\SyncCheck(P)$\\
\textbf{Input:} the event view of a kernel~$P$.\\
\textbf{Output:} the happens-before graph~$H$ of~$P$, or a failed check with its witness.
\begin{pseudocode}
\item $\tau \gets$ simulate $P$ once to completion
\item \Kw{reject if} the simulation hits an error or gets stuck
\item $H \gets$ candidate graph of $P$ and~$\tau$
\item \Kw{for each} barrier~$B$ \Kw{do}
\item \Ind \Cm{initialization check}
\item \Ind \Kw{reject if} $\initof(B) \nreach e$ for some update or wait issue~$e$ on~$B$
\item \Ind \Kw{for each} phase $\ph{B}{k}$ of~$B$ \Kw{do}
\item \Ind[2] \Cm{separation check}
\item \Ind[2] \Kw{reject if} $k > 0$ and $d \nreach e$ for some $d \in \contr{B}{k - 1}$, $e \in \contr{B}{k}$
\item \Ind[2] \Cm{wait check}
\item \Ind[2] \Kw{for each} wait~$w$ that $\tau$ assigns to $\ph{B}{k}$ \Kw{do}
\item \Ind[3] \Kw{reject if} $k > 0$ and $d \nreach \issue(w)$ for some $d \in \contr{B}{k - 1}$
\item \Ind[3] \Kw{reject if} $\cmp{B}{k + 1} \in H$ and $\issue(w) \nreach d$ for all $d \in \contr{B}{k + 1}$
\item \Ind[2] \Cm{completion check, by minimum cuts}
\item \Ind[2] $I \gets$ a proper ideal of $\contr{B}{k}$ with minimal state
\item \Ind[2] \Kw{reject if} $\stateof(I) \lexle (0, 0)$
\item \Kw{return} $H$
\end{pseudocode}
\caption{The synchronization check, $\SyncCheck$.}
\label{alg:certify}
\end{algorithm}

\paragraph{Soundness and completeness.}
The reference trace supplies one phase assignment, whereas well-synchronization concerns all traces.
The following theorem connects the two.
All proofs are given in the appendices.

\begin{theorem}[Characterization of well-synchronization]
\label{thm:well-synchronization}
Let $P$ be a specialized kernel within the scope of \autoref{sec:overview}, and suppose its simulation in $\SyncCheck(P)$ yields a safe complete trace~$\tau$.
Then $P$ is well-synchronized if and only if $\SyncCheck(P)$ accepts.
\end{theorem}

\paragraph{The happens-before graph.}
If $\SyncCheck(P)$ accepts, $H$ is the \emph{happens-before graph} of the kernel, and we write $e \hb f$ for $e \reach f$.
$H$ captures the ordering of every execution: each execution of~$P$ orders its events consistently with~$H$, and each ordering consistent with~$H$ is an execution of~$P$.

%% file: figures/schedules.tex
\begin{figure}[t]
\centering
\scriptsize
\renewcommand{\arraystretch}{1.2}
\setlength{\tabcolsep}{1.8pt}
\begin{tabular}{@{}l c!{\hspace{6pt}}l c@{}}
\multicolumn{2}{@{}l}{\textbf{Schedule A:} $\CAv$ before $\EB$} & \multicolumn{2}{l@{}}{\textbf{Schedule B:} $\EB$ before $\CAv$} \\
\cmidrule(r{3pt}){1-2}\cmidrule(l{3pt}){3-4}
Event & $(n, x)$ & Event & $(n, x)$ \\
\midrule
init(\code{full}, 1) & $(1, 0)$ & init(\code{full}, 1) & $(1, 0)$ \\
$\AEA$: arrive.expect\_tx $a$ & $(0, a)$ & $\AEA$: arrive.expect\_tx $a$ & $(0, a)$ \\
\cellcolor{blue!8}$\CAv$: complete\_tx $a$ & \cellcolor{blue!8}$(0, 0) \mapsto (1, 0)$ & $\EB$: expect\_tx $b$ & $(0, a{+}b)$ \\
\multicolumn{2}{@{}>{\columncolor{red!8}}l}{MMA reads \code{sA}, \code{sB}: \textcolor{swBad}{\textbf{\code{sB} not written}}} & $\CAv$: complete\_tx $a$ & $(0, b)$ \\
$\EB$: expect\_tx $b$ & $(1, b)$ & \cellcolor{blue!8}$\CBv$: complete\_tx $b$ & \cellcolor{blue!8}$(0, 0) \mapsto (1, 0)$ \\
$\CBv$: complete\_tx $b$ & $(1, 0)$ & \multicolumn{2}{>{\columncolor{green!8}}l@{}}{MMA reads \code{sA}, \code{sB}: \textcolor{swGood}{\textbf{both written}}} \\
\bottomrule
\end{tabular}
\caption{Two traces of \SplitTx.
$(n, x)$: pending arrivals and pending elements of \code{full}.
In a blue update, the state reaches $(0, 0)$, the phase completes, and the same update resets the state to $(1, 0)$.}
\label{fig:schedules}
\end{figure}

%% file: figures/hb-graph.tex
\definecolor{swOnlyBug}{RGB}{214,118,0}
\begin{figure*}[t]
\centering
\begin{tikzpicture}[x=1.2cm,y=-0.42cm,
  ev/.style={swbox,minimum width=#1,minimum height=3.4mm,inner ysep=0.6pt,execute at begin node={\hspace*{5.8mm}}},ev/.default=3.1cm,
  hw/.style={ev=#1,fill=black!7},hw/.default=2.75cm,
  cwait/.style={draw=swGood,fill=green!5,line width=0.6pt},
  endn/.style={swphase,minimum width=#1,minimum height=3.4mm,inner ysep=0.6pt},endn/.default=2.6cm,
  swdep/.style={-{Stealth[length=1.3mm]},line width=0.5pt,draw=black!30},
  cwdep/.style={-{Stealth[length=1.3mm]},line width=0.55pt,draw=swGood},
  mwdep/.style={-{Stealth[length=1.3mm]},line width=0.6pt,draw=swOnlyBug,densely dashed},
  swhb/.style={-{Stealth[length=1.4mm]},line width=0.7pt,draw=swHB},
  cwhb/.style={-{Stealth[length=1.4mm]},line width=0.7pt,draw=swGood},
  race/.style={draw=swBad,line width=0.9pt,rounded corners=2pt,inner sep=0.6pt}]
\def\evmode{1}
\input{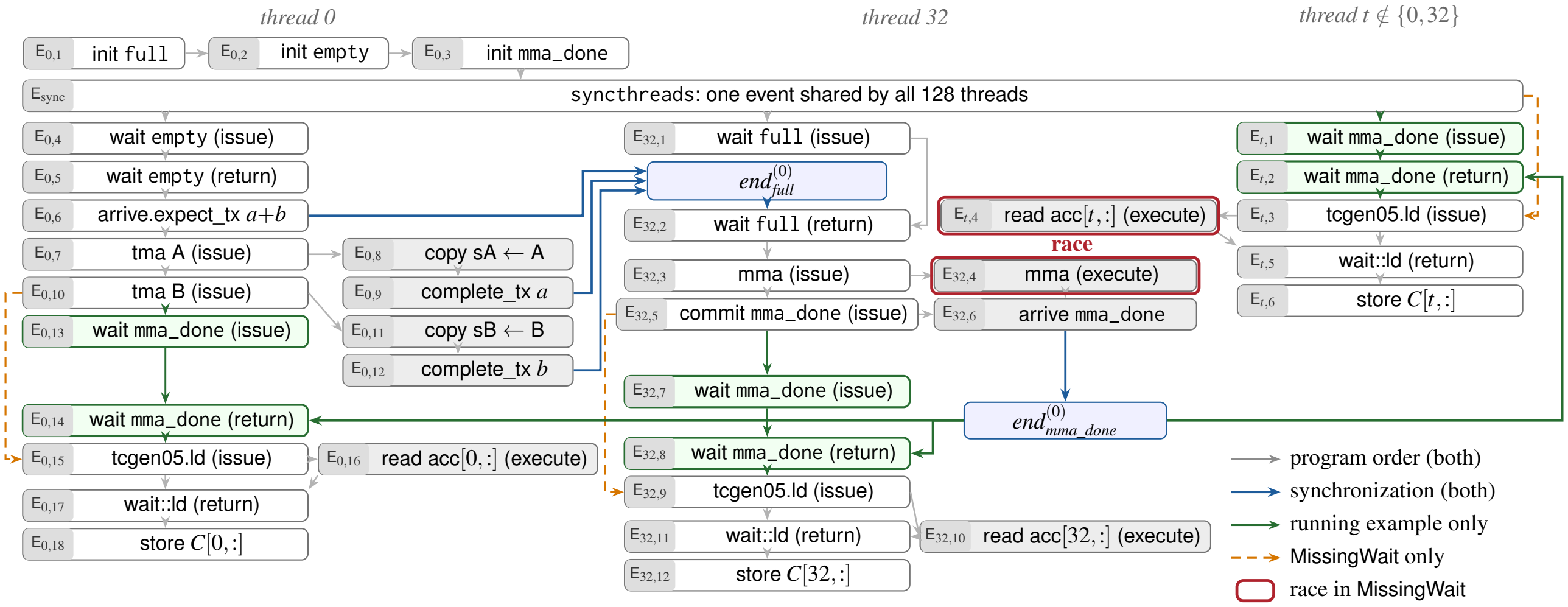}
  \node[endn] (ef) at (5.45,3.3) {$\cmp{\full}{0}$};
  \node[endn=2.2cm] (ed) at (8.15,9.5) {$\cmp{\mmadone}{0}$};
  \draw[swhb] (ae.east) -- (3.78,\yAE) -- (3.78,3.05) -- (ef.west |- 0,3.05);
  \draw[swhb] (ca.east) -- (3.86,\yQB) -- (3.86,3.3) -- (ef.west);
  \draw[swhb] (cb.east) -- (3.94,\yCB) -- (3.94,3.55) -- (ef.west |- 0,3.55);
  \draw[swhb] (ef) -- (wr);
  \draw[swhb] (ka) -- (ed);
  \draw[cwhb] (ed.west) -- (vr0.east);
  \draw[cwhb] (ed.west) -- (6.95,9.5) |- (vr32.east);
  \draw[cwhb] (ed.east) -- (12.65,9.5) |- (vrt.east);
  \node[race,fit=(mx)] {};
  \node[race,fit=(rct)] {};
  \node[font=\scriptsize\bfseries,text=swBad,inner sep=0pt] at ($(rct.south)!0.5!(mx.north)$) {race};
  \begin{scope}[shift={(9.65,10.5)},every node/.style={anchor=west,font=\scriptsize,inner sep=1pt}]
    \draw[swdep,draw=black!45] (0,0) -- (0.45,0);
    \node at (0.5,0) {program order (both)};
    \draw[swhb] (0,0.85) -- (0.45,0.85);
    \node at (0.5,0.85) {synchronization (both)};
    \draw[cwdep,line width=0.7pt] (0,1.7) -- (0.45,1.7);
    \node at (0.5,1.7) {running example only};
    \draw[mwdep] (0,2.55) -- (0.45,2.55);
    \node at (0.5,2.55) {\MissingWait{} only};
    \node[race,anchor=center,minimum width=4mm,minimum height=2.2mm] at (0.22,3.4) {};
    \node at (0.5,3.4) {race in \MissingWait{}};
  \end{scope}
\end{tikzpicture}
\caption{The running example's candidate graph, accepted as its happens-before graph, overlaid with the happens-before graph of \MissingWait{}.
Blue vertices complete phases.
Event labels follow the running example.}
\label{fig:candidate-graph}
\label{fig:race-graph}
\end{figure*}

%% file: figures/ideal-check.tex
\begin{figure}[t]
\centering
\begin{subfigure}[t]{\linewidth}
\centering
\begin{tikzpicture}[x=1cm,y=-1cm,
  n/.style={minimum width=3.0cm,minimum height=6.6mm,inner ysep=1pt,font=\scriptsize\sffamily}]
  \node[swkeep,n] (ae) at (0,0) {$\AEA$: arrive.expect\_tx $a$\\label $(-1, +a)$};
  \node[swkeep,n] (ca) at (3.4,0) {$\CAv$: complete\_tx $a$\\label $(0, -a)$};
  \node[swomit,n] (eb) at (0,0.95) {$\EB$: expect\_tx $b$\\label $(0, +b)$};
  \node[swomit,n] (cb) at (3.4,0.95) {$\CBv$: complete\_tx $b$\\label $(0, -b)$};
  \draw[swdep] (ae) -- (ca);
  \draw[swdep] (ae) -- (eb);
  \draw[swdep] (eb) -- (cb);
  \node[font=\scriptsize,anchor=north] at (1.7,1.35)
    {$I = \{\AEA, \CAv\}$, \quad $\stateof(I) = (1,0) + (-1,+a) + (0,-a) = (0, 0)$};
\end{tikzpicture}
\caption{\SplitTx: a proper zero ideal causes rejection.}
\end{subfigure}

\medskip

\begin{subfigure}[t]{\linewidth}
\centering
\begin{tikzpicture}[x=1cm,y=-1cm,
  n/.style={minimum width=3.0cm,minimum height=6.6mm,inner ysep=1pt,font=\scriptsize\sffamily}]
  \node[swbox,n] (ae) at (0,0.475) {$\AEall$: arrive.expect\_tx $a{+}b$\\label $(-1, +a{+}b)$};
  \node[swbox,n] (ca) at (3.4,0) {$\CA$: complete\_tx $a$\\label $(0, -a)$};
  \node[swbox,n] (cb) at (3.4,0.95) {$\CB$: complete\_tx $b$\\label $(0, -b)$};
  \draw[swdep] (ae.east) -- ++(0.2,0) |- (ca.west);
  \draw[swdep] (ae.east) -- ++(0.2,0) |- (cb.west);
  \node[font=\scriptsize,align=center,anchor=north] at (1.7,1.35)
    {Every ideal that contains $\AEall$ has state $(0,\ a + b - \text{elements already completed})$,\\
     which is $(0,0)$ only for the full phase.};
\end{tikzpicture}
\caption{Running example: no proper zero ideal; the check passes.}
\end{subfigure}
\caption{Completion checks for phase~0 of \code{full}.
Arrows show event-graph order; green nodes form ideal~$I$.}
\label{fig:ideal-check}
\end{figure}

%% file: sections/races.tex
\section{Race Freedom and Determinate Reads}
\label{sec:races}

The happens-before graph records which events are ordered in every execution.
\sys{} uses this order, together with the addresses from the value view, to check for data races and lifetime errors.
In a kernel that passes both checks, every read obtains the same value in every execution, and functional verification builds on this property.

\subsection{Race and Lifetime Checking}
\label{sec:races:races}

A data race occurs when two events access the same memory, at least one of them writes, and nothing orders them.
In a race, the value that a read obtains, or the value that memory finally holds, can depend on the schedule.

\begin{definition}[Race]
Two events \emph{race} if they conflict and neither happens before the other in~$\hb$.
A kernel is \emph{race-free} if no two of its events race.
\end{definition}

Events of one thread are not exempt: an asynchronous MMA can race with a later load by the thread that issued it.

\paragraph{Conflicts.}
Two events \emph{conflict} if their memory accesses overlap and at least one of them writes.
The accesses of an event are the memory operations of the value view that it performs (\autoref{sec:repr:link}).
Here we assume accesses to different allocations never overlap.
Within one allocation, most offsets are constants under a fixed launch configuration, so deciding overlap is straightforward.
Some offsets, however, depend on values loaded from memory, such as a gather index or a page table entry, which the launch configuration does not fix.
For these, \sys{} lets the user provide properties of the loaded values, such as bounds on an index or that the entries of an index tensor are pairwise distinct.
\sys{} then asks an SMT solver whether the two accesses can overlap under these properties.

\begin{algorithm}[b]
\footnotesize
\Kw{procedure} $\mathrm{RaceCheck}(P, H)$\\
\textbf{Input:} a kernel~$P$ and its happens-before graph~$H$.\\
\textbf{Output:} race-free, a race with its two events, or unresolved.
\begin{pseudocode}
\item $\mathit{unknown} \gets \mathit{false}$
\item \Kw{for each} pair of events $e, f$ of~$P$ with memory accesses \Kw{do}
\item \Ind \Kw{if} neither writes, $e \hb f$, or $f \hb e$ \Kw{then continue}
\item \Ind $r \gets \mathrm{Overlap}(e, f)$ \Cm{SMT: disjoint, overlap, or unknown}
\item \Ind \Kw{if} $r = \text{overlap}$ \Kw{then return} race$(e, f)$
\item \Ind \Kw{if} $r = \text{unknown}$ \Kw{then} $\mathit{unknown} \gets \mathit{true}$
\item \Kw{return} unresolved \Kw{if} $\mathit{unknown}$, \Kw{else} race-free
\end{pseudocode}
\caption{Pairwise race checking; \autoref{sec:scaling} avoids explicit pair enumeration.}
\label{alg:races}
\end{algorithm}

\paragraph{Example.}
Program \MissingWait{} (\autoref{sec:example:bug}) omits the epilogue's \code{wait(mma\_done, 0)}.
It is well-synchronized, and \autoref{fig:race-graph} overlays its happens-before graph on that of the running example.
In the running example, the only path from the MMA's execute event~$\MX$ to the read~$\Rt$ of \code{acc} goes through $\cmp{\mmadone}{0}$ and the green wait, which \MissingWait{} lacks.
Without the wait, nothing orders the two accesses.
Both touch row~$t$ of \code{acc}, so the two events race.

\paragraph{Lifetimes.}
Some resources are allocated and deallocated during a kernel, such as tensor memory, which \code{tcgen05.alloc} allocates and \code{tcgen05.dealloc} deallocates.
A kernel \emph{respects the lifetime} of such a resource if happens-before orders every access to it after its allocation and before its deallocation.

\subsection{Determinate Reads}
\label{sec:races:determinate}
\label{sec:reference:elimination}

Once the synchronization, race, and lifetime checks pass, happens-before orders every pair of distinct conflicting events.
Each read byte therefore has the same supplying write, or initial-memory cell, across executions.
For each admissible input, all complete executions give each read the same value and leave the same final memory.

In the running example, the MMA thus reads the copies' writes, and each load of \code{acc} reads the MMA's write.

\paragraph{Memory elimination.}
\sys{} executes the events symbolically, in an order that respects~$\hb$, and replaces every read by the value that its supplying write stored, or by the initial contents of an input.
Since read values are fixed, the result does not depend on the order.
At the end, each output element is an expression over the inputs alone, in which memory has been eliminated (\autoref{fig:mem-elim}).

\input{figures/mem-elim}

%% file: figures/mem-elim.tex
\definecolor{swMemA}{HTML}{1E5C9B}
\definecolor{swMemB}{HTML}{6D28D9}
\definecolor{swMemC}{HTML}{0F766E}
\begin{figure}[t]
\centering
\footnotesize
\renewcommand{\arraystretch}{1.2}
\setlength{\tabcolsep}{3.5pt}
\begin{tabular}{@{}lll@{}}
\toprule
Event & Reads & Write $\leftarrow$ value after memory elimination \\
\midrule
$\LA$ & \code{A}[$i$, $k$] & \textcolor{swMemA}{\code{sA}[$i$, $k$]} $\leftarrow A[i, k]$ \\
$\LB$ & \code{B}[$k$, $j$] & \textcolor{swMemB}{\code{sB}[$k$, $j$]} $\leftarrow B[k, j]$ \\
$\MX$ & \textcolor{swMemA}{\code{sA}[$i$, $k$]}, \textcolor{swMemB}{\code{sB}[$k$, $j$]} & \textcolor{swMemC}{\code{acc}[$i$, $j$]} $\leftarrow \sum_k A[i, k] \cdot B[k, j]$ \\
$\Rt$, $\St$ & \textcolor{swMemC}{\code{acc}[$t$, $j$]} & \code{C}[$t$, $j$] $\leftarrow \sum_k A[t, k] \cdot B[k, j]$ \\
\bottomrule
\end{tabular}
\caption{Memory elimination for the running example; matching colors denote the same location.}
\label{fig:mem-elim}
\end{figure}

%% file: sections/scaling.tex
\section{Templated Representations}
\label{sec:scaling}
Explicitly representing every thread and every scalar operation produces large graphs and expressions.
Much of this size comes from repeated structure across threads and within multi-element instructions and reductions.

We capture this repetition using templates, which describe shared structure parameterized by indices.
Dependencies and dataflow are expressed as relations and functions over these indices.
The corresponding analysis operations are lifted from individual instances to templates, manipulating these relations and functions directly.
This reduces explicit enumeration.

\input{figures/templated-representations}

\subsection{Indexed Templates}
\label{sec:scaling:templates}

A template consists of a shared body and a finite index domain.
For an event template~$E$ with domain~$D_E$, $E(\vec{i})$ denotes the event instance at index~$\vec{i}$; its operands and addresses are functions of~$\vec{i}$.
Instantiating the body for every index recovers the individual events, with their distinct identities and dependencies.

Threads that execute the same instruction sequence share its templates, indexed by thread ID.
Where threads follow different branches or participate in different operations, their domains are partitioned accordingly.
In \autoref{fig:representation}, for example, threads $t\notin\{0,32\}$ share the rightmost program.

In the epilogue of the running example (\autoref{fig:templated-representations}), each of the $T$ threads waits on \code{mma\_done} and then loads its $N$-element row of \code{acc}.
The waits share one template~$W(t)$, and the loads share one template~$L(t)$, both indexed by thread~$t$.

\subsection{Dependency and Conflict Relations}

Dependencies in GPU kernels often follow a few regular patterns across many threads.
Representing these patterns as relations lets the analysis operate on whole index domains without splitting templates into individual events.

An edge between templates $E$ and~$F$ carries a relation $R_{EF}\subseteq D_E\times D_F$.
A pair $(\vec{i},\vec{j})\in R_{EF}$ denotes a dependency from $E(\vec{i})$ to~$F(\vec{j})$.
Common patterns include one-to-one dependencies between corresponding thread events, all-to-one dependencies into a synchronization event, one-to-all dependencies out of it, and all-to-all dependencies between participating groups.
In \autoref{fig:templated-representations}, the edge from the MMA~$M$ to the waits carries the one-to-all relation $D_M\times D_W$, and program order gives the one-to-one relation $\{(t,t)\}$ from $W(t)$ to~$L(t)$.

Reachability is computed by composing these relations along paths and taking unions across alternative paths.
This allows dependencies to be propagated over whole index domains, splitting templates only where different subsets require different treatment.

After the synchronization check succeeds, the resulting relation $H_{EF}$ records the pairs $(\vec{i},\vec{j})$ for which $E(\vec{i})\hb F(\vec{j})$.
Let $C_{EF}$ records index pairs identifying distinct events whose accesses can conflict.
Address overlaps are expressed as constraints on the indexed addresses.
Race checking is to test whether $C_{EF}\setminus(H_{EF}\cup H_{FE}^{-1})$ is empty.

When a relation or an operation result cannot be handled compactly, the analysis enumerates the affected instance pairs exactly.

\subsection{Indexed Dataflow and Folds}

\paragraph{Read substitution.}
With the compact representation, memory elimination (\autoref{sec:reference:elimination}) can substitute a shared expression for many read values at once.
In \autoref{fig:templated-representations}, the MMA~$M$ writes all of \code{acc}, and each load~$L(t)$ reads row~$t$.
Memory elimination substitutes the shared expression $\sum_{k=0}^{K-1} A[t,k]B[k,j]$, indexed by thread~$t$ and column~$j$, for all $TN$ read values.
If different regions have different supplying expressions, the result is represented piecewise over disjoint index domains.

\paragraph{Folds.}
A \emph{Fold} compactly represents a state update $s_{k+1}=f(k,s_k)$ by its finite iteration bounds, initial state, and update function~$f$.
\sys{} supports common cases such as sums, products, and maxima, as well as coupled updates such as online softmax.
Online softmax updates a running maximum, normalizer, and weighted sum together; equivalence is proved using predefined inductive invariants (\autoref{sec:reference:egraph}).

In \autoref{fig:templated-representations}, the value of \code{acc}$[t,j]$ is an additive Fold with update $f(k,s)=s+A[t,k]B[k,j]$, initial value~$0$, and bounds $0\le k<K$.
Selecting an output element specializes $t$ and~$j$ while retaining the Fold over~$k$.
Folds are obtained from the known semantics of tile operations or from repeated updates that instantiate a common update function.
Subsequent analysis operates on the bounds and body, with equivalence checking described in \autoref{sec:reference:egraph}.

Expansion of these representations preserves the event dependencies, accessed bytes, and scalar computations.

%% file: figures/templated-representations.tex
\providecommand{\tplshadow}[1]{%
  \begin{scope}[on background layer]
    \foreach \d in {2,1} {
      \draw[rounded corners=2pt,draw=black!40,fill=white]
        ([shift={(\d*1.6pt,-\d*1.6pt)}]#1.south west) rectangle
        ([shift={(\d*1.6pt,-\d*1.6pt)}]#1.north east);
    }
  \end{scope}}
\begin{figure*}[t]
\centering
\captionsetup{skip=3pt}
\begin{tikzpicture}[x=1cm,y=1cm,
  ev/.style={swbox,minimum width=2.8cm,minimum height=4mm,inner ysep=0.8pt},
  head/.style={font=\scriptsize\bfseries,inner sep=0.5pt},
  role/.style={swrole,inner sep=0.5pt},
  prog/.style={font=\scriptsize\ttfamily,anchor=west,inner sep=0pt,align=left},
  cmt/.style={font=\scriptsize\ttfamily,text=black!55,anchor=west,inner sep=0pt},
  tag/.style={font=\scriptsize,text=black!55,anchor=east,inner sep=0pt},
  note/.style={font=\scriptsize,anchor=west,inner sep=0pt},
  gray/.style={font=\scriptsize,text=black!60,anchor=west,inner sep=0pt},
  dep/.style={swhb}]

  \node[head] at (2.25,0.70) {Program: epilogue of \autoref{fig:running}, $K = 64$};
  \node[prog] at (0,0.3) {if t == 32:};
  \node[prog] at (0,0) {\phantom{xx}tcgen05.mma(acc = sA * sB)};
  \node[tag] at (4.55,0) {\code{//} $M$};
  \node[prog] at (0,-0.3) {\phantom{xx}tcgen05.commit(mma\_done)};
  \node[cmt] at (0,-0.69) {// every thread t, 0 <= t < T};
  \node[prog] at (0,-1.05) {wait(mma\_done, 0)};
  \node[tag] at (4.55,-1.05) {\code{//} $W(t)$};
  \node[prog] at (0,-2.10) {r = tcgen05.ld(acc[t, 0:N])};
  \node[tag] at (4.55,-2.10) {\code{//} $L(t)$};
  \node[gray] at (-0.05,-2.63) {\code{sA}, \code{sB}: tiles of $A$, $B$ in shared memory};

  \draw[black!20] (4.75,0.86) -- (4.75,-3.10);

  \node[head] at (8.6,0.70) {Expanded: one event per thread, one sum per value};
  \node[role] at (8.6,0.34) {thread 32};
  \node[ev] (m) at (8.6,0) {$M$: \code{tcgen05.mma}, \ \code{acc = sA} $\cdot$ \code{sB}};
  \node[role] at (5.95,-0.72) {thread 0};
  \node[role] at (11.2,-0.72) {thread $T{-}1$};
  \node[ev] (w0) at (6.4,-1.05) {$W(0)$: \code{wait(mma\_done)}};
  \node[ev] (w2) at (10.55,-1.05) {$W(T{-}1)$: \code{wait(mma\_done)}};
  \node[ev] (l0) at (6.4,-2.10) {$L(0)$: \code{r} $\leftarrow$ \code{acc}$[0,{:}]$};
  \node[ev] (l2) at (10.55,-2.10) {$L(T{-}1)$: \code{r} $\leftarrow$ \code{acc}$[T{-}1,{:}]$};
  \node[font=\scriptsize] at (8.45,-1.05) {$\cdots$};
  \node[font=\scriptsize] at (8.45,-2.10) {$\cdots$};
  \draw[dep] ([xshift=5mm]m.south west) -- ([xshift=-3.5mm]w0.north east);
  \draw[dep] ([xshift=-5mm]m.south east) -- ([xshift=3.5mm]w2.north west);
  \node[font=\scriptsize,text=black!60] at (8.6,-0.50) {via \code{mma\_done}};
  \draw[dep] (w0) -- (l0);
  \draw[dep] (w2) -- (l2);
  \node[note] at (4.95,-2.63)
    {\code{acc}$[0,0] = A[0,0]\,B[0,0] + A[0,1]\,B[1,0] + \cdots + A[0,K{-}1]\,B[K{-}1,0]$};
  \node[gray] at (4.95,-2.94) {one such sum for each of the $TN$ values that the loads read};

  \draw[black!20] (12.45,0.86) -- (12.45,-3.10);

  \node[head] at (15.1,0.70) {Templated: indexed by thread $t$};
  \node[ev,minimum width=2.1cm] (tm) at (14.0,0) {$M$: \code{tcgen05.mma}};
  \node[ev] (tw) at (14.0,-1.05) {$W(t)$: \code{wait(mma\_done)}};
  \tplshadow{tw}
  \node[ev] (tl) at (14.0,-2.10) {$L(t)$: \code{r} $\leftarrow$ \code{acc}$[t,{:}]$};
  \tplshadow{tl}
  \draw[dep] (tm) -- (tw);
  \draw[dep] ([yshift=-3.2pt]tw.south) -- (tl.north);
  \node[note] at (14.2,-0.40) {$R_{MW} = D_M \times D_W$, \ one-to-all};
  \node[gray] at (14.2,-0.66) {$M$ precedes every $W(t)$};
  \node[note] at (14.2,-1.48) {$R_{WL} = \{(t,t) \mid 0 \le t < T\}$};
  \node[gray] at (14.2,-1.74) {$W(t)$ precedes $L(t)$};
  \node[note] at (12.6,-2.63)
    {\code{acc}$[t,j] = \mathit{Fold}_{+}\bigl(k \in [0,K),\ 0,\ A[t,k]\,B[k,j]\bigr)$};
  \node[gray] at (12.6,-2.94) {start at $0$, \ add $A[t,k]\,B[k,j]$ for $k = 0, \ldots, K{-}1$};
\end{tikzpicture}
\caption{Expanded and templated views of the running example's epilogue ($K = 64$, $T = N = 128$).
Blue arrows are dependencies; the bottom row shows read values after memory elimination.}
\label{fig:templated-representations}
\end{figure*}
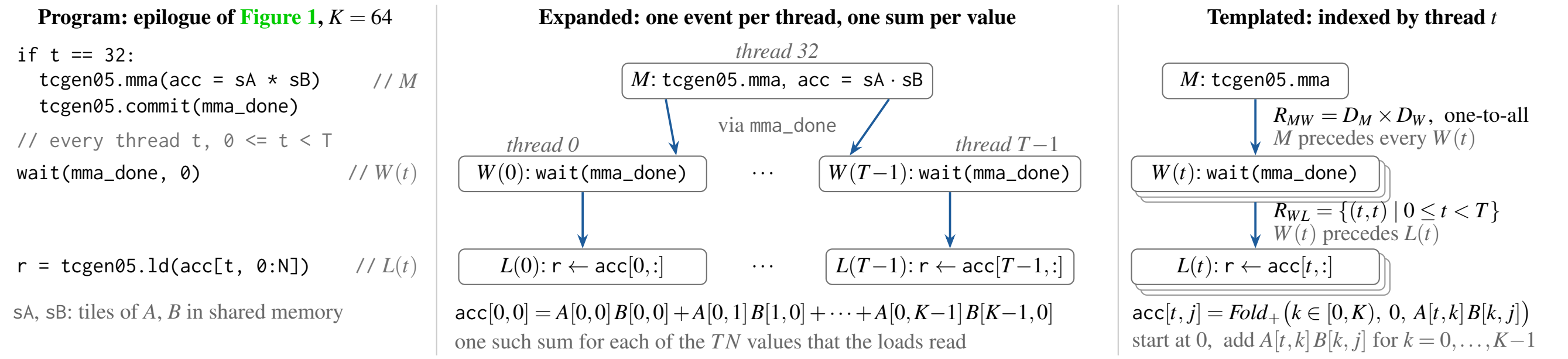

%% file: sections/reference.tex
\section{Functional Verification}
\label{sec:reference}
\label{sec:reference:egraph}

\sys{} checks functional correctness by establishing that the kernel and a reference program compute the same outputs over real arithmetic.

\input{figures/gemm-proof}

The kernel and its reference program may compute the same result through different arithmetic expressions and different decompositions of accumulation.
We check algebraic equalities using properties of exponential polynomials~\cite{li2026identity}, with operator-specific handling for other operations.

We aim to make verification cost depend on the compact representations rather than the number of operations obtained by expanding the indexed templates or unrolling folds.
For sums, we compare the summation bodies symbolically and use generating functions to check that both computations include each term the same number of times.

In \autoref{fig:gemm-proof}, both computations accumulate products of the form $A[t,k]B[k,j]$.
The reference ranges directly over $0\le k<K$, whereas the kernel uses $k=64i+k'$, with $0\le i<K/64$ and $0\le k'<64$.
After matching the bodies, we compare the generating functions for these index domains.
Their equality shows that the kernel includes each reference term exactly once.

For more complex folds
such as online softmax, we use predefined inductive invariants to relate the accumulated state to the reference computation.
Our support is therefore limited to recurrence patterns covered by these invariants, rather than arbitrary recurrences.

Appendix~\ref{app:memory-equivalence} presents the soundness argument for our approach.

%% file: figures/gemm-proof.tex
\providecommand{\rf}[1]{\textcolor{swHB}{#1}}
\providecommand{\kn}[1]{\textcolor{swChoice}{#1}}
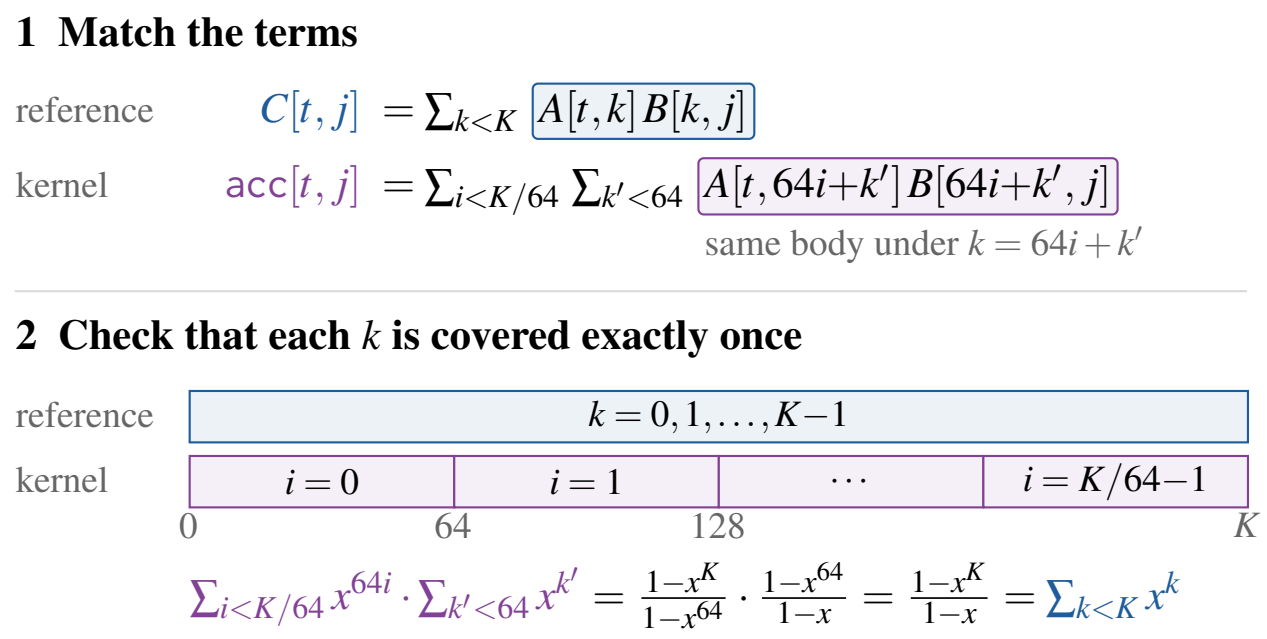
\begin{figure}[t]
\centering
\captionsetup{skip=4pt}
\begin{tikzpicture}[x=1cm,y=1cm,
  stp/.style={font=\footnotesize\bfseries,anchor=base west,inner sep=0pt},
  lab/.style={font=\scriptsize,text=black!60,anchor=base west,inner sep=0pt},
  txt/.style={font=\footnotesize,anchor=base west,inner sep=0pt},
  rtxt/.style={font=\footnotesize,anchor=base east,inner sep=0pt},
  note/.style={font=\scriptsize,text=black!60,anchor=base west,inner sep=0pt},
  body/.style={font=\footnotesize,anchor=base west,inner sep=1.2pt,rounded corners=1pt},
  bar/.style={draw=swHB,fill=swHB!8,minimum height=3.4mm,inner sep=0pt,font=\scriptsize},
  tile/.style={draw=swChoice,fill=swChoice!8,minimum height=3.4mm,inner sep=0pt,font=\scriptsize},
  tick/.style={font=\scriptsize,text=black!60,anchor=north,inner sep=0.5pt}]

  \node[stp] at (0,0) {1\enspace Match the terms};
  \node[lab] at (0,-0.5) {reference};
  \node[rtxt] at (2.3,-0.5) {$\rf{C[t,j]}$};
  \node[txt] (rs) at (2.36,-0.5) {${}= \sum_{k<K}$};
  \node[body,draw=swHB,fill=swHB!8] at ([xshift=2pt]rs.base east) {$A[t,k]\,B[k,j]$};
  \node[lab] at (0,-1.0) {kernel};
  \node[rtxt] at (2.3,-1.0) {$\kn{\code{acc}[t,j]}$};
  \node[txt] (ks) at (2.36,-1.0) {${}= \sum_{i<K/64}\, \sum_{k'<64}$};
  \node[body,draw=swChoice,fill=swChoice!8] (kb) at ([xshift=2pt]ks.base east) {$A[t,64i{+}k']\,B[64i{+}k',j]$};
  \coordinate (n1) at (0,-1.38);
  \node[note] at ([xshift=1.2pt]kb.base west |- n1) {same body under $k = 64i + k'$};

  \draw[black!15,line width=0.4pt] (0,-1.62) -- (8.15,-1.62);

  \node[stp] at (0,-2.0) {2\enspace Check that each $k$ is covered exactly once};
  \node[lab] at (0,-2.52) {reference};
  \node[bar,minimum width=7.0cm,anchor=west] at (1.15,-2.45) {$k = 0, 1, \ldots, K{-}1$};
  \node[lab] at (0,-2.95) {kernel};
  \foreach \x/\l in {0/{$i = 0$},1/{$i = 1$},2/{$\cdots$},3/{$i = K/64{-}1$}}
    \node[tile,minimum width=1.75cm,anchor=west] at (1.15+1.75*\x,-2.88) {\l};
  \foreach \x/\l in {0/{$0$},1/{$64$},2/{$128$},4/{$K$}}
    \node[tick] at (1.15+1.75*\x,-3.06) {\l};
  \node[txt] at (1.15,-3.72)
    {$\kn{\sum_{i<K/64} x^{64i} \cdot \sum_{k'<64} x^{k'}}
      = \tfrac{1-x^{K}}{1-x^{64}} \cdot \tfrac{1-x^{64}}{1-x}
      = \tfrac{1-x^{K}}{1-x}
      = \rf{\sum_{k<K} x^{k}}$};
\end{tikzpicture}
\caption{Equivalence with $C = AB$ by matching summation bodies and checking index coverage ($64 \mid K$).}
\label{fig:gemm-proof}
\end{figure}

%% file: sections/evaluation.tex
\section{Evaluation}
\label{sec:eval}

\subsection{Experimental Setup}

\input{sections/confirmed-findings}

We analyze kernels from widely used GPU libraries and compilers.
Each analysis fixes a kernel's input shapes and launch configuration.
Our benchmark contains 189 configurations drawn from 12 GPU libraries and compilers (\autoref{tab:eval:overall}).
By operator, 77 are matrix multiplications, 56 are attention kernels, and 27 are linear-attention or state-space kernels.
The other 29 include normalization, softmax, rotary-embedding, quantization, top-k, and copy kernels.
All target Blackwell GPUs.
The configurations include 142 that use mbarriers, 130 that use TMA, 121 that use \code{tcgen05} instructions, and 38 that span multiple CTAs within a cluster.

We compile library kernels from fixed source revisions.
Most kernels are compiled to PTX with \code{nvcc} from CUDA 13.0--13.4; Triton and Triton Gluon kernels use the Triton compiler.
We select input shapes and launch parameters from the kernels' accompanying examples or tests where available.
We use PyTorch implementations as functional specifications for 180 configurations (\autoref{sec:reference}).

We run experiments on an AWS \code{m8azn.6xlarge} instance with a 24-core AMD EPYC 9R05 processor and 92~GiB of memory.
Each run has a 2,000-second end-to-end timeout.

\subsection{Overall Results}

\begin{table}[t]
\centering
\small
\setlength{\tabcolsep}{2.4pt}
\begin{tabular}{@{}lrrrrrr@{}}
\toprule
Source & Total & Verified & Findings & Unsup. & \shortstack{Not\\proven} & Budget \\
\midrule
CUTLASS & 30 & 18 & 9 & 3 & 0 & 0 \\
DeepGEMM & 2 & 0 & 2 & 0 & 0 & 0 \\
FlashAttention & 21 & 2 & 16 & 3 & 0 & 0 \\
FlashInfer & 54 & 32 & 15 & 5 & 1 & 1 \\
FlashMLA & 12 & 7 & 5 & 0 & 0 & 0 \\
PyTorch & 11 & 8 & 3 & 0 & 0 & 0 \\
TensorRT-LLM & 13 & 9 & 2 & 1 & 1 & 0 \\
ThunderKittens & 1 & 0 & 1 & 0 & 0 & 0 \\
TileLang & 7 & 5 & 2 & 0 & 0 & 0 \\
TIRx & 10 & 2 & 3 & 5 & 0 & 0 \\
Triton & 13 & 11 & 2 & 0 & 0 & 0 \\
vLLM & 15 & 11 & 4 & 0 & 0 & 0 \\
\midrule
Total & 189 & 105 & 64 & 17 & 2 & 1 \\
\bottomrule
\end{tabular}
\caption{Outcomes by source.
Findings: potential correctness issues.
Unsup.: dynamic control flow; Not proven: unproved equivalence; Budget: analysis budget exhausted.}
\label{tab:eval:overall}
\end{table}

\begin{figure}[t]
\centering
\input{figures/fig-verification-time}
\caption{End-to-end analysis time for 189 configurations (log scale); dashed line: median.}
\label{fig:eval:time}
\end{figure}
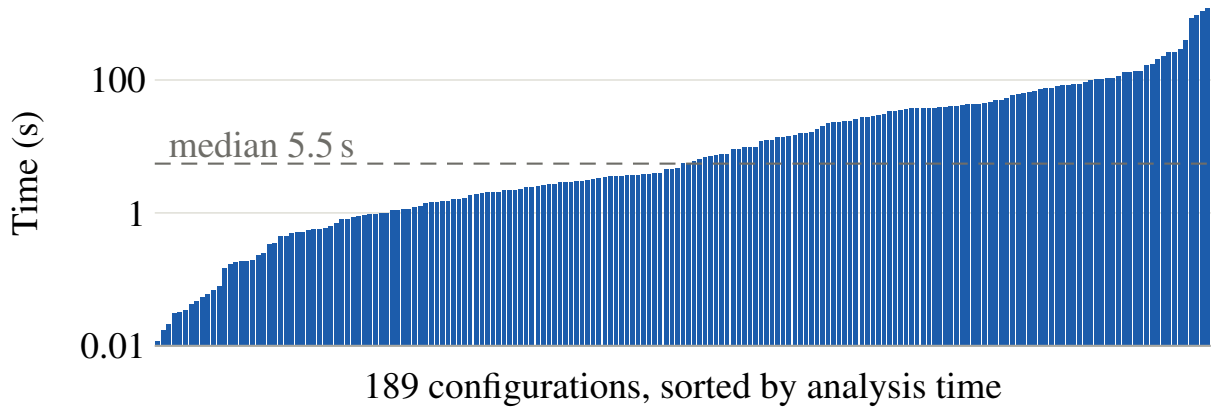

\autoref{tab:eval:overall} summarizes the results.
\autoref{fig:eval:time} shows the end-to-end analysis time for each configuration.
\autoref{tab:confirmed-findings} summarizes selected findings and their confirmation and repair status.

Among the 189 configurations, 105 (55.6\%) are verified and 64 (33.9\%) are flagged for potential correctness issues.
The findings comprise 8 well-synchronization violations, 18 potential races, 33 lifecycle findings, and 5 reports of undefined behavior.
In 22 of the lifecycle findings, an access is not proved to complete before deallocation.
Reported issues have been confirmed and fixed in libraries including CUTLASS, FlashMLA, and FlashInfer; other findings correspond to independent fixes in PyTorch and TensorRT-LLM.

Two configurations are not proved equivalent to their references.
Manual inspection identified functional errors in these two cases: the kernels differ from their references on some inputs permitted by the analyzed configurations.
Seventeen configurations are flagged as unsupported; inspection shows that all involve dynamic control flow.
One configuration exhausts the analysis budget.

\subsection{Scalability}

\begin{figure}[t]
\centering
\input{figures/fig-scaling-size}
\caption{Width scales batch/output dimensions; depth scales reduction/sequence dimensions.
Input size and verification time are relative to each curve's smallest tested configuration (log scales).}
\label{fig:eval:size}
\end{figure}
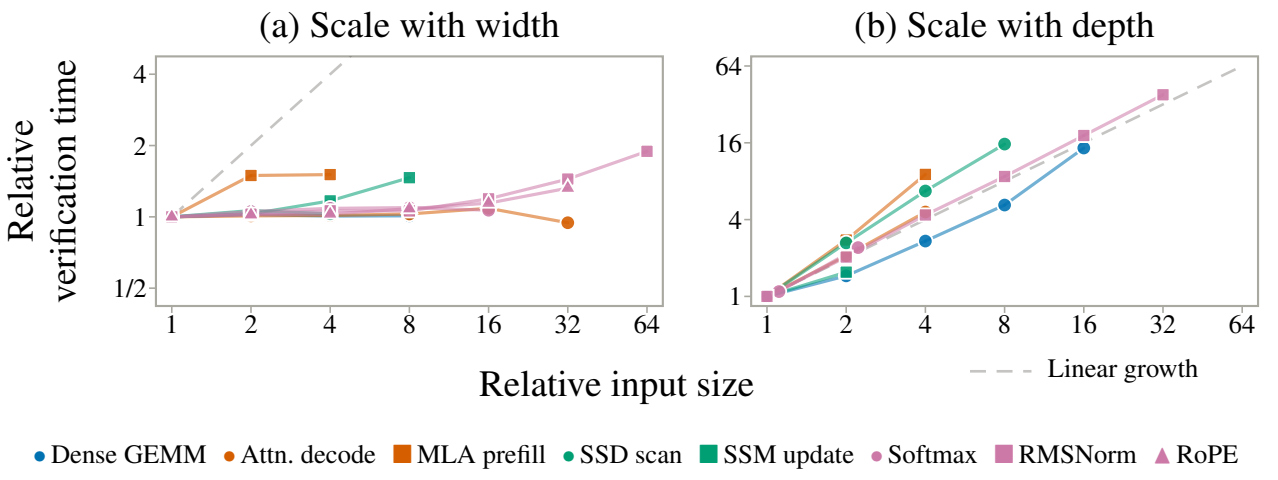

We examine how verification time scales with input size across different operators, including GEMM, attention, and normalization.
\autoref{fig:eval:size} varies \emph{width} (batch or output dimensions, such as GEMM's $M$) and \emph{depth} (reduction or sequence dimensions, such as $K$).

Increasing width by a factor of 8 changes verification time by less than 10\% for GEMM, attention decode, and softmax.
The additional outputs primarily enlarge existing template domains (\autoref{sec:scaling:templates}), which equivalence checking handles symbolically.
Increasing depth adds events and lengthens the accumulation chains compared during equivalence checking.
GEMM and attention decode exhibit approximately linear growth with depth, while MLA prefill and the state-space chunk scan grow more rapidly, with log--log slopes of 1.6 and 1.3, respectively.

\subsection{Case Studies}
\label{sec:eval:cases}

\input{figures/case-pytorch}

\paragraph{PyTorch: an unordered in-place update.}
In PyTorch~2.8.0, \sys{} reports a race in the backward kernel of \code{MultiMarginLoss}~\cite{paszke2019pytorch} (\autoref{fig:case-pytorch}).
Each stage assigns disjoint entries to threads, yet thread~0's store to the target gradient races with the in-place scaling by the entry's owner.
PyTorch~2.9.0 fixed the bug independently with a \code{\_\_syncthreads()} before the scaling.
The witness follows the path on which every class violates the margin.

\input{figures/case-cutlass}

\paragraph{CUTLASS: deallocation before a read completes.}
In CUTLASS's low-latency GQA example for Blackwell, each epilogue thread loads its accumulator from tensor memory and then arrives on the mbarrier that guards deallocation (\autoref{fig:case-cutlass}).
The arrival does not wait for the asynchronous load, and a register dependency orders the load only within its own thread.
Our fix adds \code{tcgen05.wait::ld}, as the kernel already does for another accumulator.
With CUDA~13.2, the compiler already emits this wait, so the generated code does not change.
The original kernel is correct only by a scheduling choice that the program does not guarantee.

\input{figures/case-flashmla}

\paragraph{FlashMLA: buffer reuse at loop exit.}
FlashMLA's Blackwell sparse-decoding kernel exchanges row maxima through a shared buffer and reuses the buffer after the loop (\autoref{fig:case-flashmla}).
A barrier at iteration entry protects reuse between iterations, but no barrier precedes the post-loop write.
On a B200, the kernel passed all 1,534 test cases, but a test-only delay before the partner warps' final read caused wrong outputs in 637.
Our fix adds a barrier before the reuse and passes all 1,534 cases under the same delay.

\input{figures/case-flashinfer}

\paragraph{FlashInfer: slot reuse across a warp shuffle.}
FlashInfer's fused add-RMSNorm kernel reduces the sum of squares of a row in 2 stages (\autoref{fig:case-flashinfer}).
In stage~2, every lane of warp~0 stores the total to \code{smem[0]}, the slot that lane~0 read before the shuffles.
PTX orders memory across lanes for \code{bar.warp.sync} but not for \code{shfl.sync} or a data dependency through it.
Our fix lets only lane~0 store each reduced value.

\subsection{Verification-Guided Synchronization Removal}

Beyond detecting correctness errors, we use \sys{} to identify synchronization that is redundant for a given kernel configuration.
We greedily attempt deletions in program order, considering barriers, mbarrier operations, asynchronous-copy waits, and \code{tcgen05} commits and waits.
Each deletion is retained if the modified kernel passes the synchronization, race, and lifecycle checks.

We evaluate 95 kernels that contain synchronization and pass these checks at the selected configurations.
We choose shapes that exercise the kernels' general execution patterns to reduce the likelihood of removing synchronization that is unused only in special cases.

We classify synchronization outside loops as \emph{static} and inside loops as \emph{dynamic}, excluding polling loops.
The search removes synchronization from 83 kernels, eliminating 37.7\% of executed points overall: 52.2\% of static points and 32.9\% of dynamic points.
Across these 83 kernels, the median runtime reduction on a B200 is 1.0\%; 53 run at least 0.5\% faster.

\begin{figure}[t]
\centering
\input{figures/fig-removal-kernels}
\caption{Synchronization removal for 83 kernels.
Top: fraction of executed points removed, split into static and dynamic points.
Bottom: B200 runtime reduction; line: median.}
\end{figure}

%% file: sections/confirmed-findings.tex
\begin{table*}[tp]
\centering
\fontsize{8.5}{9.5}\selectfont
\setlength{\tabcolsep}{4pt}
\renewcommand{\arraystretch}{1.0}
\caption{Findings and repair status by kernel.}
\label{tab:confirmed-findings}
\begin{tabular}{@{}>{\raggedright\arraybackslash}p{.26\textwidth}>{\raggedright\arraybackslash}p{.26\textwidth}>{\raggedright\arraybackslash}p{.24\textwidth}>{\raggedright\arraybackslash}p{\dimexpr.24\textwidth-6\tabcolsep\relax}@{}}
\toprule
Kernel source & Reported issue & Fix & Status \\
\midrule
CUTLASS 93: low-latency GQA &
Acc2 loads outlive TMEM allocation. &
Wait for TMEM loads before releasing TMEM. &
Confirmed by developers; fixed. \\
\addlinespace[2pt]
CUTLASS 77: FP16 FMHA forward &
Unordered TMEM statistics stores; load-warp phase mismatch. &
Wait for statistics stores before publication; synchronize the load warp after acquisition. &
Confirmed by developers; fixed. \\
\addlinespace[2pt]
CUTLASS 112: SSD &
Repeated TMEM free; readers still active. &
Wait for all readers; release TMEM from a single warp. &
Confirmed by developers; fixed. \\
\addlinespace[2pt]
FlashMLA: sparse decode, head64 &
Row-max buffer reused before partner reads finish. &
Add a warpgroup barrier before buffer reuse. &
Confirmed by developers; fixed. \\
\addlinespace[2pt]
FlashInfer: fused-add RMSNorm (BF16) &
Redundant stores; reduction-slot reuse. &
Restrict reduction-slot stores to \code{tx == 0}. &
Confirmed by developers; fixed. \\
\addlinespace[2pt]
FlashInfer: single decode, KV cache (BF16) &
Duplicate reduction-state and output stores. &
Restrict state stores to \code{tx == 0} and output stores to \code{tz == 0}. &
Confirmed by developers; fixed. \\
\addlinespace[2pt]
FlashInfer: merge-state (BF16) &
Duplicate log-sum-exp stores. &
Restrict log-sum-exp stores to \code{tx == 0}. &
Confirmed by developers; fixed. \\
\addlinespace[2pt]
vLLM: fused KDA decode &
Scratch overwritten before other lanes' reads. &
Synchronize the warp before lane~0 overwrites scratch. &
Confirmed by developers; fixed. \\
\addlinespace[2pt]
PyTorch 2.8: MultiMarginLoss backward &
Unordered gradient store and scaling. &
Add a block barrier before gradient scaling. &
Independently fixed in 2.9. \\
\addlinespace[2pt]
PyTorch 2.8: \code{kthvalue} &
Concurrent output writes. &
Select the index atomically; restrict output stores to thread~0. &
Independently fixed. \\
\addlinespace[2pt]
TensorRT-LLM: legacy KDA decode &
Scratch overwritten before other lanes' reads. &
Synchronize the warp before overwriting scratch. &
Independently fixed. \\
\addlinespace[2pt]
TileLang: pipelined \code{tcgen05} GEMM &
Shared operands overwritten during MMA. &
Wait for MMA completion before reusing shared operands. &
Confirmed by developers. \\
\addlinespace[2pt]
Triton 3.8: warp-specialized matmul &
Barrier invalidated before async release completes. &
Wait for the final asynchronous release before invalidating the barrier. &
Confirmed by developers. \\
\addlinespace[2pt]
FlashAttention: forward (10 variants) &
TMEM loads may outlive O-buffer release. &
Wait for TMEM loads before releasing the output buffer. &
Confirmed by developers. \\
\addlinespace[2pt]
FlashAttention: backward (3 variants) &
All warp lanes initialize the same mbarrier. &
Initialize the barrier from a single elected lane. &
Independently reported by other developers. \\
\bottomrule
\end{tabular}
\end{table*}

%% file: figures/fig-verification-time.tex
\definecolor{vtBar}{HTML}{1C5CAB}%
\definecolor{vtGrid}{HTML}{E1E0D9}%
\definecolor{vtAxis}{HTML}{A9A8A1}%
\definecolor{vtMuted}{HTML}{6E6D68}%
\begin{tikzpicture}[x=1in,y=1in,font=\scriptsize]
\draw[vtGrid,line width=0.3pt] (0,0.0000) -- (2.7500,0.0000);
\node[anchor=east,inner sep=1.5pt] at (0,0.0000) {0.01};
\draw[vtGrid,line width=0.3pt] (0,0.3462) -- (2.7500,0.3462);
\node[anchor=east,inner sep=1.5pt] at (0,0.3462) {1};
\draw[vtGrid,line width=0.3pt] (0,0.6923) -- (2.7500,0.6923);
\node[anchor=east,inner sep=1.5pt] at (0,0.6923) {100};
\fill[vtBar] (0.0020,0) rectangle (0.0125,0.0123);
\fill[vtBar] (0.0166,0) rectangle (0.0271,0.0410);
\fill[vtBar] (0.0311,0) rectangle (0.0416,0.0557);
\fill[vtBar] (0.0457,0) rectangle (0.0562,0.0842);
\fill[vtBar] (0.0602,0) rectangle (0.0707,0.0874);
\fill[vtBar] (0.0748,0) rectangle (0.0853,0.0922);
\fill[vtBar] (0.0893,0) rectangle (0.0998,0.1077);
\fill[vtBar] (0.1039,0) rectangle (0.1144,0.1151);
\fill[vtBar] (0.1184,0) rectangle (0.1289,0.1251);
\fill[vtBar] (0.1330,0) rectangle (0.1435,0.1348);
\fill[vtBar] (0.1475,0) rectangle (0.1580,0.1448);
\fill[vtBar] (0.1621,0) rectangle (0.1726,0.1551);
\fill[vtBar] (0.1766,0) rectangle (0.1871,0.2010);
\fill[vtBar] (0.1912,0) rectangle (0.2017,0.2118);
\fill[vtBar] (0.2057,0) rectangle (0.2162,0.2175);
\fill[vtBar] (0.2203,0) rectangle (0.2308,0.2202);
\fill[vtBar] (0.2348,0) rectangle (0.2453,0.2203);
\fill[vtBar] (0.2494,0) rectangle (0.2599,0.2214);
\fill[vtBar] (0.2639,0) rectangle (0.2744,0.2360);
\fill[vtBar] (0.2785,0) rectangle (0.2890,0.2418);
\fill[vtBar] (0.2930,0) rectangle (0.3035,0.2643);
\fill[vtBar] (0.3076,0) rectangle (0.3181,0.2663);
\fill[vtBar] (0.3221,0) rectangle (0.3326,0.2853);
\fill[vtBar] (0.3367,0) rectangle (0.3472,0.2862);
\fill[vtBar] (0.3512,0) rectangle (0.3617,0.2932);
\fill[vtBar] (0.3658,0) rectangle (0.3763,0.2944);
\fill[vtBar] (0.3803,0) rectangle (0.3908,0.2945);
\fill[vtBar] (0.3949,0) rectangle (0.4054,0.3000);
\fill[vtBar] (0.4094,0) rectangle (0.4199,0.3028);
\fill[vtBar] (0.4240,0) rectangle (0.4345,0.3038);
\fill[vtBar] (0.4385,0) rectangle (0.4490,0.3052);
\fill[vtBar] (0.4531,0) rectangle (0.4636,0.3105);
\fill[vtBar] (0.4676,0) rectangle (0.4781,0.3185);
\fill[vtBar] (0.4822,0) rectangle (0.4927,0.3290);
\fill[vtBar] (0.4967,0) rectangle (0.5072,0.3296);
\fill[vtBar] (0.5113,0) rectangle (0.5218,0.3358);
\fill[vtBar] (0.5258,0) rectangle (0.5363,0.3372);
\fill[vtBar] (0.5404,0) rectangle (0.5509,0.3411);
\fill[vtBar] (0.5549,0) rectangle (0.5654,0.3415);
\fill[vtBar] (0.5695,0) rectangle (0.5800,0.3422);
\fill[vtBar] (0.5840,0) rectangle (0.5945,0.3442);
\fill[vtBar] (0.5986,0) rectangle (0.6091,0.3463);
\fill[vtBar] (0.6131,0) rectangle (0.6236,0.3536);
\fill[vtBar] (0.6277,0) rectangle (0.6382,0.3539);
\fill[vtBar] (0.6422,0) rectangle (0.6527,0.3559);
\fill[vtBar] (0.6568,0) rectangle (0.6673,0.3561);
\fill[vtBar] (0.6713,0) rectangle (0.6818,0.3617);
\fill[vtBar] (0.6859,0) rectangle (0.6964,0.3633);
\fill[vtBar] (0.7004,0) rectangle (0.7109,0.3721);
\fill[vtBar] (0.7150,0) rectangle (0.7255,0.3745);
\fill[vtBar] (0.7296,0) rectangle (0.7400,0.3747);
\fill[vtBar] (0.7441,0) rectangle (0.7546,0.3775);
\fill[vtBar] (0.7587,0) rectangle (0.7691,0.3776);
\fill[vtBar] (0.7732,0) rectangle (0.7837,0.3805);
\fill[vtBar] (0.7878,0) rectangle (0.7982,0.3811);
\fill[vtBar] (0.8023,0) rectangle (0.8128,0.3830);
\fill[vtBar] (0.8169,0) rectangle (0.8273,0.3909);
\fill[vtBar] (0.8314,0) rectangle (0.8419,0.3958);
\fill[vtBar] (0.8460,0) rectangle (0.8564,0.3971);
\fill[vtBar] (0.8605,0) rectangle (0.8710,0.3989);
\fill[vtBar] (0.8751,0) rectangle (0.8855,0.3995);
\fill[vtBar] (0.8896,0) rectangle (0.9001,0.4009);
\fill[vtBar] (0.9042,0) rectangle (0.9146,0.4051);
\fill[vtBar] (0.9187,0) rectangle (0.9292,0.4055);
\fill[vtBar] (0.9333,0) rectangle (0.9437,0.4055);
\fill[vtBar] (0.9478,0) rectangle (0.9583,0.4067);
\fill[vtBar] (0.9624,0) rectangle (0.9728,0.4134);
\fill[vtBar] (0.9769,0) rectangle (0.9874,0.4140);
\fill[vtBar] (0.9915,0) rectangle (1.0019,0.4150);
\fill[vtBar] (1.0060,0) rectangle (1.0165,0.4178);
\fill[vtBar] (1.0206,0) rectangle (1.0310,0.4197);
\fill[vtBar] (1.0351,0) rectangle (1.0456,0.4203);
\fill[vtBar] (1.0497,0) rectangle (1.0601,0.4250);
\fill[vtBar] (1.0642,0) rectangle (1.0747,0.4256);
\fill[vtBar] (1.0788,0) rectangle (1.0892,0.4257);
\fill[vtBar] (1.0933,0) rectangle (1.1038,0.4276);
\fill[vtBar] (1.1079,0) rectangle (1.1183,0.4281);
\fill[vtBar] (1.1224,0) rectangle (1.1329,0.4310);
\fill[vtBar] (1.1370,0) rectangle (1.1474,0.4333);
\fill[vtBar] (1.1515,0) rectangle (1.1620,0.4368);
\fill[vtBar] (1.1661,0) rectangle (1.1765,0.4401);
\fill[vtBar] (1.1806,0) rectangle (1.1911,0.4402);
\fill[vtBar] (1.1952,0) rectangle (1.2056,0.4402);
\fill[vtBar] (1.2097,0) rectangle (1.2202,0.4409);
\fill[vtBar] (1.2243,0) rectangle (1.2347,0.4434);
\fill[vtBar] (1.2388,0) rectangle (1.2493,0.4437);
\fill[vtBar] (1.2534,0) rectangle (1.2638,0.4445);
\fill[vtBar] (1.2679,0) rectangle (1.2784,0.4471);
\fill[vtBar] (1.2825,0) rectangle (1.2929,0.4472);
\fill[vtBar] (1.2970,0) rectangle (1.3075,0.4495);
\fill[vtBar] (1.3116,0) rectangle (1.3220,0.4502);
\fill[vtBar] (1.3261,0) rectangle (1.3366,0.4587);
\fill[vtBar] (1.3407,0) rectangle (1.3511,0.4594);
\fill[vtBar] (1.3552,0) rectangle (1.3657,0.4615);
\fill[vtBar] (1.3698,0) rectangle (1.3802,0.4745);
\fill[vtBar] (1.3843,0) rectangle (1.3948,0.4757);
\fill[vtBar] (1.3989,0) rectangle (1.4093,0.4798);
\fill[vtBar] (1.4134,0) rectangle (1.4239,0.4853);
\fill[vtBar] (1.4280,0) rectangle (1.4384,0.4920);
\fill[vtBar] (1.4425,0) rectangle (1.4530,0.4937);
\fill[vtBar] (1.4571,0) rectangle (1.4675,0.4955);
\fill[vtBar] (1.4716,0) rectangle (1.4821,0.4980);
\fill[vtBar] (1.4862,0) rectangle (1.4966,0.4995);
\fill[vtBar] (1.5007,0) rectangle (1.5112,0.5127);
\fill[vtBar] (1.5153,0) rectangle (1.5257,0.5127);
\fill[vtBar] (1.5298,0) rectangle (1.5403,0.5160);
\fill[vtBar] (1.5444,0) rectangle (1.5548,0.5165);
\fill[vtBar] (1.5589,0) rectangle (1.5694,0.5171);
\fill[vtBar] (1.5735,0) rectangle (1.5839,0.5326);
\fill[vtBar] (1.5880,0) rectangle (1.5985,0.5343);
\fill[vtBar] (1.6026,0) rectangle (1.6130,0.5344);
\fill[vtBar] (1.6171,0) rectangle (1.6276,0.5421);
\fill[vtBar] (1.6317,0) rectangle (1.6421,0.5436);
\fill[vtBar] (1.6462,0) rectangle (1.6567,0.5445);
\fill[vtBar] (1.6608,0) rectangle (1.6712,0.5491);
\fill[vtBar] (1.6753,0) rectangle (1.6858,0.5529);
\fill[vtBar] (1.6899,0) rectangle (1.7003,0.5537);
\fill[vtBar] (1.7044,0) rectangle (1.7149,0.5568);
\fill[vtBar] (1.7190,0) rectangle (1.7294,0.5639);
\fill[vtBar] (1.7335,0) rectangle (1.7440,0.5719);
\fill[vtBar] (1.7481,0) rectangle (1.7585,0.5794);
\fill[vtBar] (1.7626,0) rectangle (1.7731,0.5823);
\fill[vtBar] (1.7772,0) rectangle (1.7876,0.5832);
\fill[vtBar] (1.7917,0) rectangle (1.8022,0.5841);
\fill[vtBar] (1.8063,0) rectangle (1.8167,0.5848);
\fill[vtBar] (1.8208,0) rectangle (1.8313,0.5900);
\fill[vtBar] (1.8354,0) rectangle (1.8458,0.5942);
\fill[vtBar] (1.8499,0) rectangle (1.8604,0.5947);
\fill[vtBar] (1.8645,0) rectangle (1.8749,0.5967);
\fill[vtBar] (1.8790,0) rectangle (1.8895,0.6000);
\fill[vtBar] (1.8936,0) rectangle (1.9040,0.6020);
\fill[vtBar] (1.9081,0) rectangle (1.9186,0.6099);
\fill[vtBar] (1.9227,0) rectangle (1.9331,0.6112);
\fill[vtBar] (1.9372,0) rectangle (1.9477,0.6144);
\fill[vtBar] (1.9518,0) rectangle (1.9622,0.6163);
\fill[vtBar] (1.9663,0) rectangle (1.9768,0.6179);
\fill[vtBar] (1.9809,0) rectangle (1.9913,0.6182);
\fill[vtBar] (1.9954,0) rectangle (2.0059,0.6182);
\fill[vtBar] (2.0100,0) rectangle (2.0204,0.6187);
\fill[vtBar] (2.0245,0) rectangle (2.0350,0.6194);
\fill[vtBar] (2.0391,0) rectangle (2.0496,0.6201);
\fill[vtBar] (2.0536,0) rectangle (2.0641,0.6211);
\fill[vtBar] (2.0682,0) rectangle (2.0787,0.6226);
\fill[vtBar] (2.0827,0) rectangle (2.0932,0.6232);
\fill[vtBar] (2.0973,0) rectangle (2.1078,0.6251);
\fill[vtBar] (2.1118,0) rectangle (2.1223,0.6278);
\fill[vtBar] (2.1264,0) rectangle (2.1369,0.6283);
\fill[vtBar] (2.1409,0) rectangle (2.1514,0.6292);
\fill[vtBar] (2.1555,0) rectangle (2.1660,0.6304);
\fill[vtBar] (2.1700,0) rectangle (2.1805,0.6330);
\fill[vtBar] (2.1846,0) rectangle (2.1951,0.6389);
\fill[vtBar] (2.1991,0) rectangle (2.2096,0.6397);
\fill[vtBar] (2.2137,0) rectangle (2.2242,0.6436);
\fill[vtBar] (2.2282,0) rectangle (2.2387,0.6526);
\fill[vtBar] (2.2428,0) rectangle (2.2533,0.6537);
\fill[vtBar] (2.2573,0) rectangle (2.2678,0.6575);
\fill[vtBar] (2.2719,0) rectangle (2.2824,0.6592);
\fill[vtBar] (2.2864,0) rectangle (2.2969,0.6615);
\fill[vtBar] (2.3010,0) rectangle (2.3115,0.6672);
\fill[vtBar] (2.3155,0) rectangle (2.3260,0.6700);
\fill[vtBar] (2.3301,0) rectangle (2.3406,0.6715);
\fill[vtBar] (2.3446,0) rectangle (2.3551,0.6765);
\fill[vtBar] (2.3592,0) rectangle (2.3697,0.6783);
\fill[vtBar] (2.3737,0) rectangle (2.3842,0.6791);
\fill[vtBar] (2.3883,0) rectangle (2.3988,0.6807);
\fill[vtBar] (2.4028,0) rectangle (2.4133,0.6814);
\fill[vtBar] (2.4174,0) rectangle (2.4279,0.6870);
\fill[vtBar] (2.4319,0) rectangle (2.4424,0.6922);
\fill[vtBar] (2.4465,0) rectangle (2.4570,0.6932);
\fill[vtBar] (2.4610,0) rectangle (2.4715,0.6947);
\fill[vtBar] (2.4756,0) rectangle (2.4861,0.6956);
\fill[vtBar] (2.4901,0) rectangle (2.5006,0.6965);
\fill[vtBar] (2.5047,0) rectangle (2.5152,0.7006);
\fill[vtBar] (2.5192,0) rectangle (2.5297,0.7119);
\fill[vtBar] (2.5338,0) rectangle (2.5443,0.7124);
\fill[vtBar] (2.5483,0) rectangle (2.5588,0.7149);
\fill[vtBar] (2.5629,0) rectangle (2.5734,0.7157);
\fill[vtBar] (2.5774,0) rectangle (2.5879,0.7317);
\fill[vtBar] (2.5920,0) rectangle (2.6025,0.7338);
\fill[vtBar] (2.6065,0) rectangle (2.6170,0.7464);
\fill[vtBar] (2.6211,0) rectangle (2.6316,0.7543);
\fill[vtBar] (2.6356,0) rectangle (2.6461,0.7641);
\fill[vtBar] (2.6502,0) rectangle (2.6607,0.7655);
\fill[vtBar] (2.6647,0) rectangle (2.6752,0.7715);
\fill[vtBar] (2.6793,0) rectangle (2.6898,0.7957);
\fill[vtBar] (2.6938,0) rectangle (2.7043,0.8519);
\fill[vtBar] (2.7084,0) rectangle (2.7189,0.8602);
\fill[vtBar] (2.7229,0) rectangle (2.7334,0.8721);
\fill[vtBar] (2.7375,0) rectangle (2.7480,0.8798);
\draw[vtAxis,line width=0.4pt] (0,0) -- (2.7500,0);
\draw[vtMuted,line width=0.4pt,densely dashed] (0,0.4745) -- (2.7500,0.4745);
\node[anchor=south west,inner sep=1pt,text=vtMuted] at (0.02,0.4745) {median 5.5\,s};
\node[anchor=north] at (1.3750,-0.02) {189 configurations, sorted by analysis time};
\node[rotate=90,anchor=south] at (-0.24,0.4500) {Time (s)};
\end{tikzpicture}

%% file: figures/fig-scaling-size.tex
\definecolor{fgemm}{HTML}{0072B2}
\definecolor{fattn}{HTML}{D55E00}
\definecolor{fssm}{HTML}{009E73}
\definecolor{frow}{HTML}{CC79A7}
\definecolor{scAxis}{HTML}{A9A8A1}
\definecolor{scMuted}{HTML}{6E6D68}
\begin{tikzpicture}[x=1in,y=1in,font=\scriptsize]
\begin{scope}[shift={(0,0)}]
\draw[scAxis,line width=0.3pt] (-0.018,0.0464286) -- (0,0.0464286);
\node[anchor=east,inner sep=1.5pt,font=\tiny] at (0,0.0464286) {1/2};
\draw[scAxis,line width=0.3pt] (-0.018,0.2321429) -- (0,0.2321429);
\node[anchor=east,inner sep=1.5pt,font=\tiny] at (0,0.2321429) {1};
\draw[scAxis,line width=0.3pt] (-0.018,0.4178571) -- (0,0.4178571);
\node[anchor=east,inner sep=1.5pt,font=\tiny] at (0,0.4178571) {2};
\draw[scAxis,line width=0.3pt] (-0.018,0.6035714) -- (0,0.6035714);
\node[anchor=east,inner sep=1.5pt,font=\tiny] at (0,0.6035714) {4};
\draw[scAxis,line width=0.3pt] (0.04125,0) -- (0.04125,-0.018);
\node[anchor=north,inner sep=1.5pt,font=\tiny] at (0.04125,0) {1};
\draw[scAxis,line width=0.3pt] (0.2475,0) -- (0.2475,-0.018);
\node[anchor=north,inner sep=1.5pt,font=\tiny] at (0.2475,0) {2};
\draw[scAxis,line width=0.3pt] (0.45375,0) -- (0.45375,-0.018);
\node[anchor=north,inner sep=1.5pt,font=\tiny] at (0.45375,0) {4};
\draw[scAxis,line width=0.3pt] (0.66,0) -- (0.66,-0.018);
\node[anchor=north,inner sep=1.5pt,font=\tiny] at (0.66,0) {8};
\draw[scAxis,line width=0.3pt] (0.86625,0) -- (0.86625,-0.018);
\node[anchor=north,inner sep=1.5pt,font=\tiny] at (0.86625,0) {16};
\draw[scAxis,line width=0.3pt] (1.0725,0) -- (1.0725,-0.018);
\node[anchor=north,inner sep=1.5pt,font=\tiny] at (1.0725,0) {32};
\draw[scAxis,line width=0.3pt] (1.27875,0) -- (1.27875,-0.018);
\node[anchor=north,inner sep=1.5pt,font=\tiny] at (1.27875,0) {64};
\draw[scAxis,line width=0.4pt] (0,0) rectangle (1.32,0.65);
\begin{scope}\clip (0,0) rectangle (1.32,0.65);
\draw[scMuted,draw opacity=0.4,line width=0.5pt,densely dashed] (0.04125,0.2321429) -- (1.27875,1.3464286);
\draw[fgemm,line width=0.6pt,draw opacity=0.55,line cap=round] (0.04125,0.2321429) -- (0.2475,0.238175);
\draw[fgemm,line width=0.6pt,draw opacity=0.55,line cap=round] (0.2475,0.238175) -- (0.45375,0.2342336);
\draw[fgemm,line width=0.6pt,draw opacity=0.55,line cap=round] (0.45375,0.2342336) -- (0.66,0.2348134);
\draw[fattn,line width=0.6pt,draw opacity=0.55,line cap=round] (0.04125,0.2321429) -- (0.2475,0.2346519);
\draw[fattn,line width=0.6pt,draw opacity=0.55,line cap=round] (0.2475,0.2346519) -- (0.45375,0.2357453);
\draw[fattn,line width=0.6pt,draw opacity=0.55,line cap=round] (0.45375,0.2357453) -- (0.66,0.2396749);
\draw[fattn,line width=0.6pt,draw opacity=0.55,line cap=round] (0.66,0.2396749) -- (0.86625,0.2544522);
\draw[fattn,line width=0.6pt,draw opacity=0.55,line cap=round] (0.86625,0.2544522) -- (1.0725,0.2168143);
\draw[fattn,line width=0.6pt,draw opacity=0.55,line cap=round] (0.04125,0.2321429) -- (0.2475,0.3398536);
\draw[fattn,line width=0.6pt,draw opacity=0.55,line cap=round] (0.2475,0.3398536) -- (0.45375,0.3421665);
\draw[fssm,line width=0.6pt,draw opacity=0.55,line cap=round] (0.04125,0.2321429) -- (0.2475,0.2484899);
\draw[fssm,line width=0.6pt,draw opacity=0.55,line cap=round] (0.2475,0.2484899) -- (0.45375,0.2402231);
\draw[fssm,line width=0.6pt,draw opacity=0.55,line cap=round] (0.04125,0.2321429) -- (0.2475,0.2396967);
\draw[fssm,line width=0.6pt,draw opacity=0.55,line cap=round] (0.2475,0.2396967) -- (0.45375,0.2737619);
\draw[fssm,line width=0.6pt,draw opacity=0.55,line cap=round] (0.45375,0.2737619) -- (0.66,0.3341536);
\draw[frow,line width=0.6pt,draw opacity=0.55,line cap=round] (0.04125,0.2321429) -- (0.2475,0.2460736);
\draw[frow,line width=0.6pt,draw opacity=0.55,line cap=round] (0.2475,0.2460736) -- (0.45375,0.2543078);
\draw[frow,line width=0.6pt,draw opacity=0.55,line cap=round] (0.45375,0.2543078) -- (0.66,0.2561851);
\draw[frow,line width=0.6pt,draw opacity=0.55,line cap=round] (0.66,0.2561851) -- (0.86625,0.2493949);
\draw[frow,line width=0.6pt,draw opacity=0.55,line cap=round] (0.04125,0.2321429) -- (0.2475,0.2391213);
\draw[frow,line width=0.6pt,draw opacity=0.55,line cap=round] (0.2475,0.2391213) -- (0.45375,0.2478549);
\draw[frow,line width=0.6pt,draw opacity=0.55,line cap=round] (0.45375,0.2478549) -- (0.66,0.2476428);
\draw[frow,line width=0.6pt,draw opacity=0.55,line cap=round] (0.66,0.2476428) -- (0.86625,0.2791185);
\draw[frow,line width=0.6pt,draw opacity=0.55,line cap=round] (0.86625,0.2791185) -- (1.0725,0.3305343);
\draw[frow,line width=0.6pt,draw opacity=0.55,line cap=round] (1.0725,0.3305343) -- (1.27875,0.4028396);
\draw[frow,line width=0.6pt,draw opacity=0.55,line cap=round] (0.04125,0.2321429) -- (0.2475,0.2375317);
\draw[frow,line width=0.6pt,draw opacity=0.55,line cap=round] (0.2475,0.2375317) -- (0.45375,0.2395081);
\draw[frow,line width=0.6pt,draw opacity=0.55,line cap=round] (0.45375,0.2395081) -- (0.66,0.2533669);
\draw[frow,line width=0.6pt,draw opacity=0.55,line cap=round] (0.66,0.2533669) -- (0.86625,0.2678864);
\draw[frow,line width=0.6pt,draw opacity=0.55,line cap=round] (0.86625,0.2678864) -- (1.0725,0.3068758);
\path[draw=white,line width=0.35pt,fill=fgemm,fill opacity=0.9] (0.04125,0.2321429) circle (0.0190);
\path[draw=white,line width=0.35pt,fill=fgemm,fill opacity=0.9] (0.2475,0.238175) circle (0.0190);
\path[draw=white,line width=0.35pt,fill=fgemm,fill opacity=0.9] (0.45375,0.2342336) circle (0.0190);
\path[draw=white,line width=0.35pt,fill=fgemm,fill opacity=0.9] (0.66,0.2348134) circle (0.0190);
\path[draw=white,line width=0.35pt,fill=fattn,fill opacity=0.9] (0.04125,0.2321429) circle (0.0190);
\path[draw=white,line width=0.35pt,fill=fattn,fill opacity=0.9] (0.2475,0.2346519) circle (0.0190);
\path[draw=white,line width=0.35pt,fill=fattn,fill opacity=0.9] (0.45375,0.2357453) circle (0.0190);
\path[draw=white,line width=0.35pt,fill=fattn,fill opacity=0.9] (0.66,0.2396749) circle (0.0190);
\path[draw=white,line width=0.35pt,fill=fattn,fill opacity=0.9] (0.86625,0.2544522) circle (0.0190);
\path[draw=white,line width=0.35pt,fill=fattn,fill opacity=0.9] (1.0725,0.2168143) circle (0.0190);
\path[draw=white,line width=0.35pt,fill=fattn,fill opacity=0.9] (0.02415,0.2150429) rectangle (0.05835,0.2492429);
\path[draw=white,line width=0.35pt,fill=fattn,fill opacity=0.9] (0.2304,0.3227536) rectangle (0.2646,0.3569536);
\path[draw=white,line width=0.35pt,fill=fattn,fill opacity=0.9] (0.43665,0.3250665) rectangle (0.47085,0.3592665);
\path[draw=white,line width=0.35pt,fill=fssm,fill opacity=0.9] (0.04125,0.2321429) circle (0.0190);
\path[draw=white,line width=0.35pt,fill=fssm,fill opacity=0.9] (0.2475,0.2484899) circle (0.0190);
\path[draw=white,line width=0.35pt,fill=fssm,fill opacity=0.9] (0.45375,0.2402231) circle (0.0190);
\path[draw=white,line width=0.35pt,fill=fssm,fill opacity=0.9] (0.02415,0.2150429) rectangle (0.05835,0.2492429);
\path[draw=white,line width=0.35pt,fill=fssm,fill opacity=0.9] (0.2304,0.2225967) rectangle (0.2646,0.2567967);
\path[draw=white,line width=0.35pt,fill=fssm,fill opacity=0.9] (0.43665,0.2566619) rectangle (0.47085,0.2908619);
\path[draw=white,line width=0.35pt,fill=fssm,fill opacity=0.9] (0.6429,0.3170536) rectangle (0.6771,0.3512536);
\path[draw=white,line width=0.35pt,fill=frow,fill opacity=0.9] (0.04125,0.2321429) circle (0.0190);
\path[draw=white,line width=0.35pt,fill=frow,fill opacity=0.9] (0.2475,0.2460736) circle (0.0190);
\path[draw=white,line width=0.35pt,fill=frow,fill opacity=0.9] (0.45375,0.2543078) circle (0.0190);
\path[draw=white,line width=0.35pt,fill=frow,fill opacity=0.9] (0.66,0.2561851) circle (0.0190);
\path[draw=white,line width=0.35pt,fill=frow,fill opacity=0.9] (0.86625,0.2493949) circle (0.0190);
\path[draw=white,line width=0.35pt,fill=frow,fill opacity=0.9] (0.02415,0.2150429) rectangle (0.05835,0.2492429);
\path[draw=white,line width=0.35pt,fill=frow,fill opacity=0.9] (0.2304,0.2220213) rectangle (0.2646,0.2562213);
\path[draw=white,line width=0.35pt,fill=frow,fill opacity=0.9] (0.43665,0.2307549) rectangle (0.47085,0.2649549);
\path[draw=white,line width=0.35pt,fill=frow,fill opacity=0.9] (0.6429,0.2305428) rectangle (0.6771,0.2647428);
\path[draw=white,line width=0.35pt,fill=frow,fill opacity=0.9] (0.84915,0.2620185) rectangle (0.88335,0.2962185);
\path[draw=white,line width=0.35pt,fill=frow,fill opacity=0.9] (1.0554,0.3134343) rectangle (1.0896,0.3476343);
\path[draw=white,line width=0.35pt,fill=frow,fill opacity=0.9] (1.26165,0.3857396) rectangle (1.29585,0.4199396);
\path[draw=white,line width=0.35pt,fill=frow,fill opacity=0.9] (0.04125,0.2558929) -- (0.0205875,0.2202679) -- (0.0619125,0.2202679) -- cycle;
\path[draw=white,line width=0.35pt,fill=frow,fill opacity=0.9] (0.2475,0.2612817) -- (0.2268375,0.2256567) -- (0.2681625,0.2256567) -- cycle;
\path[draw=white,line width=0.35pt,fill=frow,fill opacity=0.9] (0.45375,0.2632581) -- (0.4330875,0.2276331) -- (0.4744125,0.2276331) -- cycle;
\path[draw=white,line width=0.35pt,fill=frow,fill opacity=0.9] (0.66,0.2771169) -- (0.6393375,0.2414919) -- (0.6806625,0.2414919) -- cycle;
\path[draw=white,line width=0.35pt,fill=frow,fill opacity=0.9] (0.86625,0.2916364) -- (0.8455875,0.2560114) -- (0.8869125,0.2560114) -- cycle;
\path[draw=white,line width=0.35pt,fill=frow,fill opacity=0.9] (1.0725,0.3306258) -- (1.0518375,0.2950008) -- (1.0931625,0.2950008) -- cycle;
\end{scope}
\node[anchor=base,inner sep=0pt] at (0.66,0.70) {(a) Scale with width};
\end{scope}
\begin{scope}[shift={(1.55,0)}]
\draw[scAxis,line width=0.3pt] (-0.018,0.025) -- (0,0.025);
\node[anchor=east,inner sep=1.5pt,font=\tiny] at (0,0.025) {1};
\draw[scAxis,line width=0.3pt] (-0.018,0.225) -- (0,0.225);
\node[anchor=east,inner sep=1.5pt,font=\tiny] at (0,0.225) {4};
\draw[scAxis,line width=0.3pt] (-0.018,0.425) -- (0,0.425);
\node[anchor=east,inner sep=1.5pt,font=\tiny] at (0,0.425) {16};
\draw[scAxis,line width=0.3pt] (-0.018,0.625) -- (0,0.625);
\node[anchor=east,inner sep=1.5pt,font=\tiny] at (0,0.625) {64};
\draw[scAxis,line width=0.3pt] (0.04125,0) -- (0.04125,-0.018);
\node[anchor=north,inner sep=1.5pt,font=\tiny] at (0.04125,0) {1};
\draw[scAxis,line width=0.3pt] (0.2475,0) -- (0.2475,-0.018);
\node[anchor=north,inner sep=1.5pt,font=\tiny] at (0.2475,0) {2};
\draw[scAxis,line width=0.3pt] (0.45375,0) -- (0.45375,-0.018);
\node[anchor=north,inner sep=1.5pt,font=\tiny] at (0.45375,0) {4};
\draw[scAxis,line width=0.3pt] (0.66,0) -- (0.66,-0.018);
\node[anchor=north,inner sep=1.5pt,font=\tiny] at (0.66,0) {8};
\draw[scAxis,line width=0.3pt] (0.86625,0) -- (0.86625,-0.018);
\node[anchor=north,inner sep=1.5pt,font=\tiny] at (0.86625,0) {16};
\draw[scAxis,line width=0.3pt] (1.0725,0) -- (1.0725,-0.018);
\node[anchor=north,inner sep=1.5pt,font=\tiny] at (1.0725,0) {32};
\draw[scAxis,line width=0.3pt] (1.27875,0) -- (1.27875,-0.018);
\node[anchor=north,inner sep=1.5pt,font=\tiny] at (1.27875,0) {64};
\draw[scAxis,line width=0.4pt] (0,0) rectangle (1.32,0.65);
\begin{scope}\clip (0,0) rectangle (1.32,0.65);
\draw[scMuted,draw opacity=0.4,line width=0.5pt,densely dashed] (0.04125,0.025) -- (1.27875,0.625);
\draw[fgemm,line width=0.6pt,draw opacity=0.55,line cap=round] (0.04125,0.025) -- (0.2475,0.0784567);
\draw[fgemm,line width=0.6pt,draw opacity=0.55,line cap=round] (0.2475,0.0784567) -- (0.45375,0.1690783);
\draw[fgemm,line width=0.6pt,draw opacity=0.55,line cap=round] (0.45375,0.1690783) -- (0.66,0.2629126);
\draw[fgemm,line width=0.6pt,draw opacity=0.55,line cap=round] (0.66,0.2629126) -- (0.86625,0.4114172);
\draw[fattn,line width=0.6pt,draw opacity=0.55,line cap=round] (0.04125,0.025) -- (0.2475,0.130769);
\draw[fattn,line width=0.6pt,draw opacity=0.55,line cap=round] (0.2475,0.130769) -- (0.45375,0.2446931);
\draw[fattn,line width=0.6pt,draw opacity=0.55,line cap=round] (0.04125,0.025) -- (0.2475,0.1717809);
\draw[fattn,line width=0.6pt,draw opacity=0.55,line cap=round] (0.2475,0.1717809) -- (0.45375,0.3426738);
\draw[fssm,line width=0.6pt,draw opacity=0.55,line cap=round] (0.04125,0.025) -- (0.2475,0.1641786);
\draw[fssm,line width=0.6pt,draw opacity=0.55,line cap=round] (0.2475,0.1641786) -- (0.45375,0.2994775);
\draw[fssm,line width=0.6pt,draw opacity=0.55,line cap=round] (0.45375,0.2994775) -- (0.66,0.4216482);
\draw[fssm,line width=0.6pt,draw opacity=0.55,line cap=round] (0.04125,0.025) -- (0.2475,0.0881134);
\draw[frow,line width=0.6pt,draw opacity=0.55,line cap=round] (0.04125,0.025) -- (0.0723425,0.0375461);
\draw[frow,line width=0.6pt,draw opacity=0.55,line cap=round] (0.0723425,0.0375461) -- (0.2785925,0.1521208);
\draw[frow,line width=0.6pt,draw opacity=0.55,line cap=round] (0.04125,0.025) -- (0.2475,0.1274927);
\draw[frow,line width=0.6pt,draw opacity=0.55,line cap=round] (0.2475,0.1274927) -- (0.45375,0.2370933);
\draw[frow,line width=0.6pt,draw opacity=0.55,line cap=round] (0.45375,0.2370933) -- (0.66,0.3372339);
\draw[frow,line width=0.6pt,draw opacity=0.55,line cap=round] (0.66,0.3372339) -- (0.86625,0.4439606);
\draw[frow,line width=0.6pt,draw opacity=0.55,line cap=round] (0.86625,0.4439606) -- (1.0725,0.5500562);
\path[draw=white,line width=0.35pt,fill=fgemm,fill opacity=0.9] (0.04125,0.025) circle (0.0190);
\path[draw=white,line width=0.35pt,fill=fgemm,fill opacity=0.9] (0.2475,0.0784567) circle (0.0190);
\path[draw=white,line width=0.35pt,fill=fgemm,fill opacity=0.9] (0.45375,0.1690783) circle (0.0190);
\path[draw=white,line width=0.35pt,fill=fgemm,fill opacity=0.9] (0.66,0.2629126) circle (0.0190);
\path[draw=white,line width=0.35pt,fill=fgemm,fill opacity=0.9] (0.86625,0.4114172) circle (0.0190);
\path[draw=white,line width=0.35pt,fill=fattn,fill opacity=0.9] (0.04125,0.025) circle (0.0190);
\path[draw=white,line width=0.35pt,fill=fattn,fill opacity=0.9] (0.2475,0.130769) circle (0.0190);
\path[draw=white,line width=0.35pt,fill=fattn,fill opacity=0.9] (0.45375,0.2446931) circle (0.0190);
\path[draw=white,line width=0.35pt,fill=fattn,fill opacity=0.9] (0.02415,0.0079) rectangle (0.05835,0.0421);
\path[draw=white,line width=0.35pt,fill=fattn,fill opacity=0.9] (0.2304,0.1546809) rectangle (0.2646,0.1888809);
\path[draw=white,line width=0.35pt,fill=fattn,fill opacity=0.9] (0.43665,0.3255738) rectangle (0.47085,0.3597738);
\path[draw=white,line width=0.35pt,fill=fssm,fill opacity=0.9] (0.04125,0.025) circle (0.0190);
\path[draw=white,line width=0.35pt,fill=fssm,fill opacity=0.9] (0.2475,0.1641786) circle (0.0190);
\path[draw=white,line width=0.35pt,fill=fssm,fill opacity=0.9] (0.45375,0.2994775) circle (0.0190);
\path[draw=white,line width=0.35pt,fill=fssm,fill opacity=0.9] (0.66,0.4216482) circle (0.0190);
\path[draw=white,line width=0.35pt,fill=fssm,fill opacity=0.9] (0.02415,0.0079) rectangle (0.05835,0.0421);
\path[draw=white,line width=0.35pt,fill=fssm,fill opacity=0.9] (0.2304,0.0710134) rectangle (0.2646,0.1052134);
\path[draw=white,line width=0.35pt,fill=frow,fill opacity=0.9] (0.04125,0.025) circle (0.0190);
\path[draw=white,line width=0.35pt,fill=frow,fill opacity=0.9] (0.0723425,0.0375461) circle (0.0190);
\path[draw=white,line width=0.35pt,fill=frow,fill opacity=0.9] (0.2785925,0.1521208) circle (0.0190);
\path[draw=white,line width=0.35pt,fill=frow,fill opacity=0.9] (0.02415,0.0079) rectangle (0.05835,0.0421);
\path[draw=white,line width=0.35pt,fill=frow,fill opacity=0.9] (0.2304,0.1103927) rectangle (0.2646,0.1445927);
\path[draw=white,line width=0.35pt,fill=frow,fill opacity=0.9] (0.43665,0.2199933) rectangle (0.47085,0.2541933);
\path[draw=white,line width=0.35pt,fill=frow,fill opacity=0.9] (0.6429,0.3201339) rectangle (0.6771,0.3543339);
\path[draw=white,line width=0.35pt,fill=frow,fill opacity=0.9] (0.84915,0.4268606) rectangle (0.88335,0.4610606);
\path[draw=white,line width=0.35pt,fill=frow,fill opacity=0.9] (1.0554,0.5329562) rectangle (1.0896,0.5671562);
\end{scope}
\node[anchor=base,inner sep=0pt] at (0.66,0.70) {(b) Scale with depth};
\end{scope}
\node[rotate=90,anchor=south,align=center] at (-0.16,0.325) {Relative\\verification time};
\node[anchor=north] at (1.20,-0.12) {Relative input size};
\draw[scMuted,draw opacity=0.4,line width=0.5pt,densely dashed] (2.12,-0.17) -- (2.28,-0.17);
\node[anchor=base west,inner sep=0pt,font=\tiny] at (2.32,-0.186) {Linear growth};
\node[anchor=north,inner sep=0pt,font=\tiny] at (1.25,-0.36) {%
  \textcolor{fgemm}{$\bullet$}\,Dense~GEMM\hspace{3pt}%
  \textcolor{fattn}{$\bullet$}\,Attn.~decode\hspace{3pt}%
  \textcolor{fattn}{$\blacksquare$}\,MLA~prefill\hspace{3pt}%
  \textcolor{fssm}{$\bullet$}\,SSD~scan\hspace{3pt}%
  \textcolor{fssm}{$\blacksquare$}\,SSM~update\hspace{3pt}%
  \textcolor{frow}{$\bullet$}\,Softmax\hspace{3pt}%
  \textcolor{frow}{$\blacksquare$}\,RMSNorm\hspace{3pt}%
  \textcolor{frow}{$\blacktriangle$}\,RoPE};
\end{tikzpicture}

%% file: figures/case-pytorch.tex
\begin{figure}[t]
\lstset{style=kernel,numbers=none,xleftmargin=2pt,literate={!=}{{$\neq$}}1}
\begin{lstlisting}
  buffer[t] = 0
  for i = t, t + 128, ... with i != target:
    gradInput[i] = h(i)
    buffer[t] -= h(i)
  syncthreads()
  if t == 0: gradInput[target] = sum(buffer)
(*@\textcolor{swGood}{+ syncthreads()}@*)
  for i = t, t + 128, ...: gradInput[i] *= gradOutput
\end{lstlisting}
\centering
\begin{tikzpicture}[x=1cm,y=-0.43cm,
  ev/.style={swbox,minimum width=#1,minimum height=3.4mm,inner ysep=0.6pt},ev/.default=2.6cm,
  fix/.style={dashed,draw=swGood,text=swGood},
  swdep/.style={-{Stealth[length=1.3mm]},line width=0.5pt,draw=black!30},
  race/.style={draw=swBad,line width=0.9pt,rounded corners=2pt,inner sep=0.6pt}]
  \node[swrole] at (0,-0.85) {thread 0};
  \node[swrole] at (3.2,-0.85) {thread 5};
  \node[ev] (b0) at (0,0) {store \code{buffer[0]}};
  \node[ev] (b5) at (3.2,0) {store \code{buffer[5]}};
  \node[ev=6.1cm] (s1) at (1.6,0.9) {\code{syncthreads}: one event of all 128 threads};
  \node[ev] (r0) at (0,1.8) {load \code{buffer[0:128]}};
  \node[ev] (w0) at (0,2.7) {store \code{gradInput[5]}};
  \node[ev=6.1cm,fix] (s2) at (1.6,3.6) {+ \code{syncthreads} (PyTorch 2.9.0)};
  \node[ev] (l5) at (3.2,4.5) {load \code{gradInput[5]}};
  \node[ev] (w5) at (3.2,5.4) {store \code{gradInput[5]}};
  \draw[swdep] (b0) -- (b0 |- s1.north);
  \draw[swdep] (b5) -- (b5 |- s1.north);
  \draw[swdep] (s1.south -| r0) -- (r0);
  \draw[swdep] (r0) -- (w0);
  \draw[swdep,rounded corners=2pt] ([xshift=-1.5mm]s1.south east) |- (4.9,1.45) |- (l5.east);
  \draw[swdep] (l5) -- (w5);
  \node[race,fit=(w0)] {};
  \node[race,fit=(l5)] {};
\end{tikzpicture}
\caption{PyTorch~2.8.0 \code{MultiMarginLoss} backward and its witness (target class~5).
In Figures~\ref{fig:case-pytorch}--\ref{fig:case-flashinfer}, red marks the unordered events and + the fix, here from PyTorch~2.9.0.}
\label{fig:case-pytorch}
\end{figure}

%% file: figures/case-cutlass.tex
\begin{figure}[t]
\lstset{style=kernel,numbers=none,xleftmargin=2pt}
\begin{minipage}[t]{0.5\columnwidth}
\begin{lstlisting}
  // epilogue thread
  r = tcgen05.ld(acc2)
(*@\textcolor{swGood}{+ tcgen05.wait::ld()}@*)
  if lane < 16:
    send r to reduction CTA
  arrive(tmem_bar)
\end{lstlisting}
\end{minipage}\hfill
\begin{minipage}[t]{0.48\columnwidth}
\begin{lstlisting}[xleftmargin=0pt]
// MMA warp
arrive(tmem_bar)
wait(tmem_bar, 1)
tcgen05.dealloc(acc1, acc2)
\end{lstlisting}
\end{minipage}
\centering
\begin{tikzpicture}[x=1cm,y=-0.43cm,
  ev/.style={swbox,minimum width=#1,minimum height=3.4mm,inner ysep=0.6pt},ev/.default=2.3cm,
  hw/.style={ev=#1,fill=black!7},hw/.default=2.3cm,
  endn/.style={swphase,minimum width=1.6cm,minimum height=3.4mm,inner ysep=0.6pt},
  fix/.style={dashed,draw=swGood,text=swGood},
  fixdep/.style={-{Stealth[length=1.3mm]},line width=0.5pt,draw=swGood,dashed},
  swdep/.style={-{Stealth[length=1.3mm]},line width=0.5pt,draw=black!30},
  swhb/.style={-{Stealth[length=1.4mm]},line width=0.7pt,draw=swHB},
  race/.style={draw=swBad,line width=0.9pt,rounded corners=2pt,inner sep=0.6pt}]
  \node[swrole] at (0,-0.9) {epilogue thread};
  \node[swrole] at (2.75,-0.9) {asynchronous};
  \node[swrole] at (5.5,-0.9) {MMA warp};
  \node[ev] (li) at (0,0) {tcgen05.ld (issue)};
  \node[hw] (lx) at (2.75,0) {tcgen05.ld (execute)};
  \node[ev,fix] (lw) at (0,1) {+ wait::ld (return)};
  \node[ev] (ea) at (0,2) {arrive \code{tmem\_bar}};
  \node[endn] (en) at (2.75,3) {$\cmp{\barname{tmem\_bar}}{1}$};
  \node[ev] (ma) at (5.5,0) {arrive \code{tmem\_bar}};
  \node[ev] (mw) at (5.5,1) {wait \code{tmem\_bar} (issue)};
  \node[ev] (mr) at (5.5,3) {wait \code{tmem\_bar} (return)};
  \node[ev] (md) at (5.5,4) {tcgen05.dealloc};
  \draw[swdep] (li.east) -- (lx.west);
  \draw[swdep,rounded corners=2pt] (li.west) -| (-1.4,2) -- (ea.west);
  \draw[fixdep] (lx.west) -- (lw.north east);
  \draw[fixdep] (lw) -- (ea);
  \draw[swhb] (ea.east) -- (en.north west);
  \draw[swhb,rounded corners=2pt] (ma.west) -| (4.12,2.3) -- (en.north east);
  \draw[swdep] (ma) -- (mw);
  \draw[swdep] (mw) -- (mr);
  \draw[swhb] (en.east) -- (mr.west);
  \draw[swdep] (mr) -- (md);
  \node[race,fit=(lx)] {};
  \node[race,fit=(md)] {};
\end{tikzpicture}
\caption{CUTLASS low-latency GQA epilogue for Blackwell and its lifetime witness.
Shaded events are asynchronous; blue edges are synchronization.}
\label{fig:case-cutlass}
\end{figure}

%% file: figures/case-flashmla.tex
\begin{figure}[t]
\lstset{style=kernel,numbers=none,xleftmargin=2pt,literate={^}{{$\oplus$}}1}
\begin{lstlisting}
  for each block of keys:
    wg0_sync(); ...
    max_buf[t] = cur_max
    wg0_sync()
    cur_max = max(cur_max, max_buf[t ^ 64]); ...
(*@\textcolor{swGood}{+ wg0\_sync()}@*)
  max_buf[t] = li
  wg0_sync()
  li += max_buf[t ^ 64]
\end{lstlisting}
\centering
\begin{tikzpicture}[x=1cm,y=-0.36cm,
  ev/.style={swbox,minimum width=#1,minimum height=3.4mm,inner ysep=0.6pt},ev/.default=2.6cm,
  fix/.style={dashed,draw=swGood,text=swGood},
  swdep/.style={-{Stealth[length=1.3mm]},line width=0.5pt,draw=black!30},
  race/.style={draw=swBad,line width=0.9pt,rounded corners=2pt,inner sep=0.6pt}]
  \node[swrole] at (0,-1.0) {thread 0};
  \node[swrole] at (3.2,-1.0) {thread 64};
  \node[ev] (a0) at (0,0) {store \code{max\_buf[0]} (max)};
  \node[ev] (b0) at (3.2,0) {store \code{max\_buf[64]} (max)};
  \node[ev=6.1cm] (s1) at (1.6,1.2) {\code{wg0\_sync} of the last iteration};
  \node[ev] (a1) at (0,2.4) {load \code{max\_buf[64]}};
  \node[ev] (b1) at (3.2,2.4) {load \code{max\_buf[0]}};
  \node[ev=6.1cm,fix] (fx) at (1.6,3.6) {+ \code{wg0\_sync} (our fix)};
  \node[ev] (a2) at (0,4.8) {store \code{max\_buf[0]} (li)};
  \node[ev] (b2) at (3.2,4.8) {store \code{max\_buf[64]} (li)};
  \node[ev=6.1cm] (s2) at (1.6,6.0) {\code{wg0\_sync}};
  \draw[swdep] (a0) -- (a0 |- s1.north);
  \draw[swdep] (b0) -- (b0 |- s1.north);
  \draw[swdep] (s1.south -| a1) -- (a1);
  \draw[swdep] (s1.south -| b1) -- (b1);
  \draw[swdep,rounded corners=2pt] (a1.west) -| (-1.6,4.8) -- (a2.west);
  \draw[swdep,rounded corners=2pt] (b1.east) -| (4.8,4.8) -- (b2.east);
  \draw[swdep] (a2) -- (a2 |- s2.north);
  \draw[swdep] (b2) -- (b2 |- s2.north);
  \node[race,fit=(b1)] {};
  \node[race,fit=(a2)] {};
\end{tikzpicture}
\caption{FlashMLA sparse-decoding row-maximum exchange for threads~0 and~64 of one row.}
\label{fig:case-flashmla}
\end{figure}

%% file: figures/case-flashinfer.tex
\begin{figure}[t]
\lstset{style=kernel,numbers=none,xleftmargin=2pt}
\begin{lstlisting}
  smem[warp] = sum of squares of the warp
  syncthreads()
  if warp == 0:
    s = smem[lane] if lane < num_warps else 0
    for k = 16, 8, 4, 2, 1: s += shfl_xor(s, k)
(*@\textcolor{swGood}{+\ \ \ if lane == 0:}@*)
      smem[0] = s
  syncthreads()
  r = rsqrt(smem[0] / d + eps)
\end{lstlisting}
\centering
\begin{tikzpicture}[x=1cm,y=-0.4cm,
  ev/.style={swbox,minimum width=#1,minimum height=3.4mm,inner ysep=0.6pt},ev/.default=2.6cm,
  swdep/.style={-{Stealth[length=1.3mm]},line width=0.5pt,draw=black!30},
  exdep/.style={-{Stealth[length=1.3mm]},line width=0.6pt,draw=black!60,densely dotted},
  race/.style={draw=swBad,line width=0.9pt,rounded corners=2pt,inner sep=0.6pt}]
  \node[swrole] at (0,-0.9) {lane 0};
  \node[swrole] at (3.2,-0.9) {lane 1};
  \node[ev=6.1cm] (s0) at (1.6,0) {\code{syncthreads}};
  \node[ev] (r0) at (0,1.05) {load \code{smem[0]}};
  \node[ev] (r1) at (3.2,1.05) {load \code{smem[1]}};
  \node[ev=6.1cm] (sh) at (1.6,2.35) {\code{shfl\_xor} (five events)};
  \node[ev] (w0) at (0,3.65) {store \code{smem[0]}};
  \node[ev] (w1) at (3.2,3.65) {store \code{smem[0]}};
  \node[ev=6.1cm] (s1) at (1.6,4.7) {\code{syncthreads}};
  \draw[swdep] (s0.south -| r0) -- (r0);
  \draw[swdep] (s0.south -| r1) -- (r1);
  \draw[exdep] (r0) -- (r0 |- sh.north);
  \draw[exdep] (r1) -- (r1 |- sh.north);
  \draw[exdep] (sh.south -| w0) -- (w0);
  \draw[exdep] (sh.south -| w1) -- (w1);
  \draw[swdep,rounded corners=2pt] (r0.west) -| (-1.6,3.65) -- (w0.west);
  \draw[swdep,rounded corners=2pt] (r1.east) -| (4.8,3.65) -- (w1.east);
  \draw[swdep] (w0) -- (w0 |- s1.north);
  \draw[swdep] (w1) -- (w1 |- s1.north);
  \node[race,fit=(r0)] {};
  \node[race,fit=(w1)] {};
\end{tikzpicture}
\caption{FlashInfer fused add-RMSNorm reduction for lanes~0 and~1 of warp~0.
Dotted edges order execution only.}
\label{fig:case-flashinfer}
\end{figure}

%% file: figures/fig-removal-kernels.tex
\definecolor{rmOut}{HTML}{1C5CAB}%
\definecolor{rmIn}{HTML}{86B6EF}%
\definecolor{rmBar}{HTML}{52514E}%
\definecolor{rmGrid}{HTML}{E1E0D9}%
\definecolor{rmAxis}{HTML}{A9A8A1}%
\definecolor{rmMuted}{HTML}{6E6D68}%
\begin{tikzpicture}[x=1in,y=1in,font=\scriptsize]
\draw[rmAxis,line width=0.4pt] (0,0.62) -- (2.46,0.62);
\node[anchor=east,inner sep=1.5pt] at (0,0.62) {0};
\draw[rmGrid,line width=0.3pt] (0,0.93) -- (2.46,0.93);
\node[anchor=east,inner sep=1.5pt] at (0,0.93) {50};
\draw[rmGrid,line width=0.3pt] (0,1.24) -- (2.46,1.24);
\node[anchor=east,inner sep=1.5pt] at (0,1.24) {100};
\draw[rmAxis,line width=0.4pt] (0,0) -- (2.46,0);
\node[anchor=east,inner sep=1.5pt] at (0,0) {0};
\draw[rmGrid,line width=0.3pt] (0,0.1438) -- (2.46,0.1438);
\node[anchor=east,inner sep=1.5pt] at (0,0.1438) {5};
\draw[rmGrid,line width=0.3pt] (0,0.2875) -- (2.46,0.2875);
\node[anchor=east,inner sep=1.5pt] at (0,0.2875) {10};
\draw[rmGrid,line width=0.3pt] (0,0.4313) -- (2.46,0.4313);
\node[anchor=east,inner sep=1.5pt] at (0,0.4313) {15};
\node[rotate=90,anchor=south,align=center] at (-0.2,0.93) {Removed (\%)};
\node[rotate=90,anchor=south,align=center] at (-0.2,0.23) {Runtime\\reduction (\%)};
\fill[rmOut] (0.0047,0.62) rectangle (0.0249,1.24);
\fill[rmBar] (0.0047,0) rectangle (0.0249,0.0049);
\fill[rmOut] (0.0344,0.62) rectangle (0.0545,0.9374);
\fill[rmIn] (0.0344,0.9444) rectangle (0.0545,1.0333);
\fill[rmBar] (0.0344,0) rectangle (0.0545,0.1903);
\fill[rmOut] (0.064,0.62) rectangle (0.0842,0.9523);
\fill[rmIn] (0.064,0.9593) rectangle (0.0842,1.0171);
\fill[rmBar] (0.064,0) rectangle (0.0842,0.1276);
\fill[rmOut] (0.0937,0.62) rectangle (0.1138,0.7891);
\fill[rmIn] (0.0937,0.7961) rectangle (0.1138,1.0145);
\fill[rmBar] (0.0937,0) rectangle (0.1138,0.1975);
\fill[rmOut] (0.1233,0.62) rectangle (0.1435,0.8611);
\fill[rmIn] (0.1233,0.8681) rectangle (0.1435,1.0058);
\fill[rmBar] (0.1233,0) rectangle (0.1435,0.0667);
\fill[rmOut] (0.1529,0.62) rectangle (0.1731,0.8896);
\fill[rmIn] (0.1529,0.8966) rectangle (0.1731,0.9884);
\fill[rmBar] (0.1529,0) rectangle (0.1731,0.0489);
\fill[rmOut] (0.1826,0.62) rectangle (0.2027,0.775);
\fill[rmIn] (0.1826,0.782) rectangle (0.2027,0.9743);
\fill[rmBar] (0.1826,0) rectangle (0.2027,0.1593);
\fill[rmOut] (0.2122,0.62) rectangle (0.2324,0.8108);
\fill[rmIn] (0.2122,0.8178) rectangle (0.2324,0.9634);
\fill[rmBar] (0.2122,0) rectangle (0.2324,0.0558);
\fill[rmOut] (0.2419,0.62) rectangle (0.262,0.807);
\fill[rmIn] (0.2419,0.814) rectangle (0.262,0.9546);
\fill[rmBar] (0.2419,0) rectangle (0.262,0.272);
\fill[rmOut] (0.2715,0.62) rectangle (0.2916,0.744);
\fill[rmIn] (0.2715,0.751) rectangle (0.2916,0.9507);
\fill[rmBar] (0.2715,0) rectangle (0.2916,0.0635);
\fill[rmOut] (0.3011,0.62) rectangle (0.3213,0.8314);
\fill[rmIn] (0.3011,0.8384) rectangle (0.3213,0.9441);
\fill[rmBar] (0.3011,0) rectangle (0.3213,0.13);
\fill[rmOut] (0.3308,0.62) rectangle (0.3509,0.8314);
\fill[rmIn] (0.3308,0.8384) rectangle (0.3509,0.9441);
\fill[rmBar] (0.3308,0) rectangle (0.3509,-0.008);
\fill[rmOut] (0.3604,0.62) rectangle (0.3806,0.75);
\fill[rmIn] (0.3604,0.757) rectangle (0.3806,0.94);
\fill[rmBar] (0.3604,0) rectangle (0.3806,0.0664);
\fill[rmOut] (0.39,0.62) rectangle (0.4102,0.6502);
\fill[rmIn] (0.39,0.6572) rectangle (0.4102,0.9376);
\fill[rmBar] (0.39,0) rectangle (0.4102,0.0908);
\fill[rmOut] (0.4197,0.62) rectangle (0.4398,0.93);
\fill[rmBar] (0.4197,0) rectangle (0.4398,-0.0066);
\fill[rmOut] (0.4493,0.62) rectangle (0.4695,0.93);
\fill[rmBar] (0.4493,0) rectangle (0.4695,0.0167);
\fill[rmOut] (0.479,0.62) rectangle (0.4991,0.93);
\fill[rmBar] (0.479,0) rectangle (0.4991,0.0253);
\fill[rmOut] (0.5086,0.62) rectangle (0.5288,0.7251);
\fill[rmIn] (0.5086,0.7321) rectangle (0.5288,0.9247);
\fill[rmBar] (0.5086,0) rectangle (0.5288,0.0615);
\fill[rmOut] (0.5382,0.62) rectangle (0.5584,0.6535);
\fill[rmIn] (0.5382,0.6605) rectangle (0.5584,0.9216);
\fill[rmBar] (0.5382,0) rectangle (0.5584,0.0126);
\fill[rmOut] (0.5679,0.62) rectangle (0.588,0.7822);
\fill[rmIn] (0.5679,0.7892) rectangle (0.588,0.9213);
\fill[rmBar] (0.5679,0) rectangle (0.588,0.3321);
\fill[rmOut] (0.5975,0.62) rectangle (0.6177,0.8193);
\fill[rmIn] (0.5975,0.8263) rectangle (0.6177,0.9079);
\fill[rmBar] (0.5975,0) rectangle (0.6177,0.182);
\fill[rmOut] (0.6272,0.62) rectangle (0.6473,0.6291);
\fill[rmIn] (0.6272,0.6361) rectangle (0.6473,0.9026);
\fill[rmBar] (0.6272,0) rectangle (0.6473,-0.0092);
\fill[rmOut] (0.6568,0.62) rectangle (0.6769,0.6282);
\fill[rmIn] (0.6568,0.6352) rectangle (0.6769,0.8974);
\fill[rmBar] (0.6568,0) rectangle (0.6769,0.0276);
\fill[rmOut] (0.6864,0.62) rectangle (0.7066,0.6772);
\fill[rmIn] (0.6864,0.6842) rectangle (0.7066,0.8966);
\fill[rmBar] (0.6864,0) rectangle (0.7066,0.0017);
\fill[rmOut] (0.7161,0.62) rectangle (0.7362,0.7278);
\fill[rmIn] (0.7161,0.7348) rectangle (0.7362,0.8963);
\fill[rmBar] (0.7161,0) rectangle (0.7362,0.0871);
\fill[rmOut] (0.7457,0.62) rectangle (0.7659,0.6889);
\fill[rmIn] (0.7457,0.6959) rectangle (0.7659,0.8956);
\fill[rmBar] (0.7457,0) rectangle (0.7659,0.1144);
\fill[rmOut] (0.7753,0.62) rectangle (0.7955,0.6889);
\fill[rmIn] (0.7753,0.6959) rectangle (0.7955,0.8956);
\fill[rmBar] (0.7753,0) rectangle (0.7955,0.0788);
\fill[rmOut] (0.805,0.62) rectangle (0.8251,0.64);
\fill[rmIn] (0.805,0.647) rectangle (0.8251,0.8933);
\fill[rmBar] (0.805,0) rectangle (0.8251,0.0546);
\fill[rmIn] (0.8346,0.62) rectangle (0.8548,0.8907);
\fill[rmBar] (0.8346,0) rectangle (0.8548,0.0558);
\fill[rmOut] (0.8643,0.62) rectangle (0.8844,0.638);
\fill[rmIn] (0.8643,0.645) rectangle (0.8844,0.8896);
\fill[rmBar] (0.8643,0) rectangle (0.8844,0.0299);
\fill[rmOut] (0.8939,0.62) rectangle (0.9141,0.6402);
\fill[rmIn] (0.8939,0.6472) rectangle (0.9141,0.8896);
\fill[rmBar] (0.8939,0) rectangle (0.9141,0.0431);
\fill[rmOut] (0.9235,0.62) rectangle (0.9437,0.7971);
\fill[rmIn] (0.9235,0.8041) rectangle (0.9437,0.8857);
\fill[rmBar] (0.9235,0) rectangle (0.9437,-0.0043);
\fill[rmOut] (0.9532,0.62) rectangle (0.9733,0.7251);
\fill[rmIn] (0.9532,0.7321) rectangle (0.9733,0.8827);
\fill[rmBar] (0.9532,0) rectangle (0.9733,0.1811);
\fill[rmOut] (0.9828,0.62) rectangle (1.003,0.707);
\fill[rmIn] (0.9828,0.714) rectangle (1.003,0.8702);
\fill[rmBar] (0.9828,0) rectangle (1.003,0.0773);
\fill[rmOut] (1.0125,0.62) rectangle (1.0326,0.868);
\fill[rmBar] (1.0125,0) rectangle (1.0326,0.176);
\fill[rmOut] (1.0421,0.62) rectangle (1.0622,0.8525);
\fill[rmBar] (1.0421,0) rectangle (1.0622,0.0046);
\fill[rmOut] (1.0717,0.62) rectangle (1.0919,0.6328);
\fill[rmIn] (1.0717,0.6398) rectangle (1.0919,0.8501);
\fill[rmBar] (1.0717,0) rectangle (1.0919,-0.0221);
\fill[rmOut] (1.1014,0.62) rectangle (1.1215,0.6328);
\fill[rmIn] (1.1014,0.6398) rectangle (1.1215,0.8501);
\fill[rmBar] (1.1014,0) rectangle (1.1215,-0.0267);
\fill[rmOut] (1.131,0.62) rectangle (1.1512,0.6344);
\fill[rmIn] (1.131,0.6414) rectangle (1.1512,0.8435);
\fill[rmBar] (1.131,0) rectangle (1.1512,0.0069);
\fill[rmOut] (1.1606,0.62) rectangle (1.1808,0.6315);
\fill[rmIn] (1.1606,0.6385) rectangle (1.1808,0.8381);
\fill[rmBar] (1.1606,0) rectangle (1.1808,0.228);
\fill[rmOut] (1.1903,0.62) rectangle (1.2104,0.6315);
\fill[rmIn] (1.1903,0.6385) rectangle (1.2104,0.8381);
\fill[rmBar] (1.1903,0) rectangle (1.2104,0.2217);
\fill[rmOut] (1.2199,0.62) rectangle (1.2401,0.6677);
\fill[rmIn] (1.2199,0.6747) rectangle (1.2401,0.8298);
\fill[rmBar] (1.2199,0) rectangle (1.2401,0.0451);
\fill[rmOut] (1.2496,0.62) rectangle (1.2697,0.6677);
\fill[rmIn] (1.2496,0.6747) rectangle (1.2697,0.8298);
\fill[rmBar] (1.2496,0) rectangle (1.2697,0.0273);
\fill[rmOut] (1.2792,0.62) rectangle (1.2994,0.8267);
\fill[rmBar] (1.2792,0) rectangle (1.2994,0.004);
\fill[rmOut] (1.3088,0.62) rectangle (1.329,0.8267);
\fill[rmBar] (1.3088,0) rectangle (1.329,0.0184);
\fill[rmOut] (1.3385,0.62) rectangle (1.3586,0.8267);
\fill[rmBar] (1.3385,0) rectangle (1.3586,0.0032);
\fill[rmIn] (1.3681,0.62) rectangle (1.3883,0.8267);
\fill[rmBar] (1.3681,0) rectangle (1.3883,0.0196);
\fill[rmOut] (1.3978,0.62) rectangle (1.4179,0.6308);
\fill[rmIn] (1.3978,0.6378) rectangle (1.4179,0.8249);
\fill[rmBar] (1.3978,0) rectangle (1.4179,0.1785);
\fill[rmOut] (1.4274,0.62) rectangle (1.4475,0.6272);
\fill[rmIn] (1.4274,0.6342) rectangle (1.4475,0.8219);
\fill[rmBar] (1.4274,0) rectangle (1.4475,0.0118);
\fill[rmOut] (1.457,0.62) rectangle (1.4772,0.6451);
\fill[rmIn] (1.457,0.6521) rectangle (1.4772,0.8211);
\fill[rmBar] (1.457,0) rectangle (1.4772,0.05);
\fill[rmOut] (1.4867,0.62) rectangle (1.5068,0.802);
\fill[rmIn] (1.4867,0.809) rectangle (1.5068,0.8191);
\fill[rmBar] (1.4867,0) rectangle (1.5068,0.0931);
\fill[rmOut] (1.5163,0.62) rectangle (1.5365,0.6466);
\fill[rmIn] (1.5163,0.6536) rectangle (1.5365,0.8149);
\fill[rmBar] (1.5163,0) rectangle (1.5365,-0.0052);
\fill[rmOut] (1.5459,0.62) rectangle (1.5661,0.6656);
\fill[rmIn] (1.5459,0.6726) rectangle (1.5661,0.8115);
\fill[rmBar] (1.5459,0) rectangle (1.5661,0.0224);
\fill[rmOut] (1.5756,0.62) rectangle (1.5957,0.6717);
\fill[rmIn] (1.5756,0.6787) rectangle (1.5957,0.806);
\fill[rmBar] (1.5756,0) rectangle (1.5957,0.0244);
\fill[rmOut] (1.6052,0.62) rectangle (1.6254,0.6677);
\fill[rmIn] (1.6052,0.6747) rectangle (1.6254,0.7988);
\fill[rmBar] (1.6052,0) rectangle (1.6254,0.1469);
\fill[rmOut] (1.6349,0.62) rectangle (1.655,0.6665);
\fill[rmIn] (1.6349,0.6735) rectangle (1.655,0.7982);
\fill[rmBar] (1.6349,0) rectangle (1.655,0.0914);
\fill[rmOut] (1.6645,0.62) rectangle (1.6847,0.6961);
\fill[rmIn] (1.6645,0.7031) rectangle (1.6847,0.794);
\fill[rmBar] (1.6645,0) rectangle (1.6847,0.1052);
\fill[rmOut] (1.6941,0.62) rectangle (1.7143,0.7891);
\fill[rmBar] (1.6941,0) rectangle (1.7143,0.0043);
\fill[rmOut] (1.7238,0.62) rectangle (1.7439,0.651);
\fill[rmIn] (1.7238,0.658) rectangle (1.7439,0.7853);
\fill[rmBar] (1.7238,0) rectangle (1.7439,0.0549);
\fill[rmOut] (1.7534,0.62) rectangle (1.7736,0.651);
\fill[rmIn] (1.7534,0.658) rectangle (1.7736,0.7853);
\fill[rmBar] (1.7534,0) rectangle (1.7736,0.0293);
\fill[rmOut] (1.7831,0.62) rectangle (1.8032,0.7027);
\fill[rmIn] (1.7831,0.7097) rectangle (1.8032,0.7853);
\fill[rmBar] (1.7831,0) rectangle (1.8032,0.4433);
\fill[rmOut] (1.8127,0.62) rectangle (1.8328,0.7027);
\fill[rmIn] (1.8127,0.7097) rectangle (1.8328,0.7853);
\fill[rmBar] (1.8127,0) rectangle (1.8328,0.443);
\fill[rmOut] (1.8423,0.62) rectangle (1.8625,0.6261);
\fill[rmIn] (1.8423,0.6331) rectangle (1.8625,0.7796);
\fill[rmBar] (1.8423,0) rectangle (1.8625,-0.0049);
\fill[rmOut] (1.872,0.62) rectangle (1.8921,0.6269);
\fill[rmIn] (1.872,0.6339) rectangle (1.8921,0.7784);
\fill[rmBar] (1.872,0) rectangle (1.8921,0.0034);
\fill[rmOut] (1.9016,0.62) rectangle (1.9218,0.6848);
\fill[rmIn] (1.9016,0.6918) rectangle (1.9218,0.7773);
\fill[rmBar] (1.9016,0) rectangle (1.9218,0.0983);
\fill[rmOut] (1.9312,0.62) rectangle (1.9514,0.775);
\fill[rmBar] (1.9312,0) rectangle (1.9514,0.0029);
\fill[rmOut] (1.9609,0.62) rectangle (1.981,0.775);
\fill[rmBar] (1.9609,0) rectangle (1.981,0.0198);
\fill[rmOut] (1.9905,0.62) rectangle (2.0107,0.775);
\fill[rmBar] (1.9905,0) rectangle (2.0107,0.0029);
\fill[rmOut] (2.0202,0.62) rectangle (2.0403,0.6258);
\fill[rmIn] (2.0202,0.6328) rectangle (2.0403,0.7721);
\fill[rmBar] (2.0202,0) rectangle (2.0403,-0.0023);
\fill[rmOut] (2.0498,0.62) rectangle (2.07,0.6247);
\fill[rmIn] (2.0498,0.6317) rectangle (2.07,0.7656);
\fill[rmBar] (2.0498,0) rectangle (2.07,-0.015);
\fill[rmOut] (2.0794,0.62) rectangle (2.0996,0.6486);
\fill[rmIn] (2.0794,0.6556) rectangle (2.0996,0.7631);
\fill[rmBar] (2.0794,0) rectangle (2.0996,0.0259);
\fill[rmOut] (2.1091,0.62) rectangle (2.1292,0.7578);
\fill[rmBar] (2.1091,0) rectangle (2.1292,0.0115);
\fill[rmOut] (2.1387,0.62) rectangle (2.1589,0.6455);
\fill[rmIn] (2.1387,0.6525) rectangle (2.1589,0.7559);
\fill[rmBar] (2.1387,0) rectangle (2.1589,0.222);
\fill[rmIn] (2.1684,0.62) rectangle (2.1885,0.7529);
\fill[rmBar] (2.1684,0) rectangle (2.1885,0.0463);
\fill[rmOut] (2.198,0.62) rectangle (2.2181,0.6307);
\fill[rmIn] (2.198,0.6377) rectangle (2.2181,0.7376);
\fill[rmBar] (2.198,0) rectangle (2.2181,-0.0043);
\fill[rmOut] (2.2276,0.62) rectangle (2.2478,0.7348);
\fill[rmBar] (2.2276,0) rectangle (2.2478,0.0086);
\fill[rmIn] (2.2573,0.62) rectangle (2.2774,0.7327);
\fill[rmBar] (2.2573,0) rectangle (2.2774,0.0152);
\fill[rmOut] (2.2869,0.62) rectangle (2.3071,0.6397);
\fill[rmIn] (2.2869,0.6467) rectangle (2.3071,0.7283);
\fill[rmBar] (2.2869,0) rectangle (2.3071,0.0058);
\fill[rmOut] (2.3165,0.62) rectangle (2.3367,0.7233);
\fill[rmBar] (2.3165,0) rectangle (2.3367,0.0515);
\fill[rmOut] (2.3462,0.62) rectangle (2.3663,0.6975);
\fill[rmBar] (2.3462,0) rectangle (2.3663,-0.0003);
\fill[rmOut] (2.3758,0.62) rectangle (2.396,0.6975);
\fill[rmBar] (2.3758,0) rectangle (2.396,0.0009);
\fill[rmOut] (2.4055,0.62) rectangle (2.4256,0.6421);
\fill[rmBar] (2.4055,0) rectangle (2.4256,-0.0003);
\fill[rmOut] (2.4351,0.62) rectangle (2.4553,0.6355);
\fill[rmBar] (2.4351,0) rectangle (2.4553,-0.0009);
\draw[rmMuted,line width=0.5pt] (0,0.0293) -- (2.5,0.0293);
\node[anchor=west,inner sep=1pt,text=rmMuted,align=left] at (2.5,0.0293) {median\\1.0\%};
\node[anchor=north] at (1.23,-0.035) {83 kernels with a removal, sorted by removed share};
\fill[rmOut] (-0.3,1.35) rectangle (-0.2,1.41);
\node[anchor=west,inner sep=2pt] at (-0.2,1.38) {Removed outside loops};
\fill[rmIn] (1.02,1.35) rectangle (1.12,1.41);
\node[anchor=west,inner sep=2pt] at (1.12,1.38) {Removed inside loops};
\end{tikzpicture}

%% file: sections/related.tex
\section{Related Work}

\paragraph{GPU formal verification.}
GPU verifiers use symbolic execution, test amplification, and thread abstraction to check data races and barrier safety~\cite{li2010pug,li2012gklee,betts2012gpuverify,leung2012amplification,cogumbreiro2021faial}.
Proof-oriented languages also support race-freedom and functional-correctness proofs within their source-language semantics~\cite{martinez2026kuiper}.
TensorLean~\cite{tensorlean2026} targets PTX-level proofs of race freedom, termination, and functional correctness against tensor-level specifications.
TxVerify~\cite{hou2026txverify} uses agent-assisted Lean proofs to establish safety, progress, termination, and functional correctness of PTX kernels under a shared semantic core and explicit kernel-specific assumptions.
ProofWright~\cite{chatterjee2026proofwright} uses LLM-generated annotations and proofs with VerCors and Rocq to verify CUDA safety and functional correctness.
WEFT~\cite{sharma2015weft} establishes well-synchronization for named-barrier protocols in warp-specialized kernels.
\sys{} extends this reasoning to mbarriers whose transaction counters can increase within a phase.
Work on GPU weak-memory semantics~\cite{alglave2015gpu,lustig2019ptx} addresses memory-visibility effects, an orthogonal concern to the asynchronous-completion and barrier protocols studied here.

\paragraph{Equivalence checking.}
Translation validation checks individual compilations rather than the compiler itself~\cite{pnueli1998tv,necula2000tv,lopes2021alive2}.
Volta~\cite{driscoll2025volta,costa2026aicompiler} checks PTX equivalence and race freedom; Racer~\cite{vu2026racer} also models asynchronous-copy completion.
RESOLVE~\cite{vedadigargary2026resolve} combines concurrency testing with proofs of reduced-concurrency kernels.
\sys{} checks byte-counted barrier protocols and resource lifetimes in asynchronous kernels before comparing their output expressions with a reference over real arithmetic.
Its algebraic checks build on identity testing for expressions with exponentiation~\cite{li2026identity}.

\paragraph{Kernel generation.}
ML compilers and superoptimizers generate efficient kernels through compiler transformations and schedule search~\cite{chen2018tvm,tillet2019triton,zheng2020ansor,wu2025mirage}.
LLM-based systems also generate and refine kernels using execution and performance feedback~\cite{dai2026cudaagent,saba2026cutegen,costa2026aicompiler}.
Correctness-preserving optimization has been studied for tensor-graph, expression, and scheduling rewrites~\cite{yang2021tensat,newcomb2020halide,liu2022atl}.
Exo-GPU~\cite{akeley2026exogpu} verifies sequential--parallel equivalence for asynchronous tensor-core programs.
These guarantees concern programs and transformations within the respective representations.
\sys{} complements kernel-generation methods by checking synchronization and computation in emitted PTX, independently of the source language, compiler, or generating agent.

%% file: sections/proofs-sync.tex
\section{Execution Model and Synchronization Proofs}

\subsection{Execution Model and Event Construction}
\label{app:sync-model}

We formulate the execution model based on our interpretation of the PTX ISA documentation~\cite{nvidia2026ptx}.

\paragraph{State and instruction families.}
Specialization fixes the finite per-thread instruction sequences, barrier identities, counts, and parity arguments.
Memory addresses and computed values retain their runtime dataflow.
For each admitted state wait~$w$, extraction resolves its token-producing arrival~$s(w)$ and retains $s(w)\to\issue(w)$.
The token's generation is captured at runtime; its producing occurrence is fixed.
A configuration contains per-thread program counters and registers, byte-addressed memory, pending asynchronous work and completion histories, barrier states, wait targets, and resource lifetimes.
A step advances one thread, performs an asynchronous effect, or completes a collective whose required participants have arrived.
Threads have independent program counters.

The following families share rules parameterized by operand types, address spaces, layouts, and participant sets~\cite{nvidia2026ptx}.
An admitted variant must determine these parameters and every effect it performs; numerical verification additionally requires its value semantics.

\begin{center}
\footnotesize
\renewcommand{\arraystretch}{1.12}
\begin{tabularx}{\linewidth}{@{}>{\raggedright\arraybackslash}p{.24\linewidth}@{\hspace{.8em}}>{\raggedright\arraybackslash}X@{}}
\toprule
Family & Execution rule \\
\midrule
Register and control operations &
Evaluate the declared function of register operands, including arithmetic, conversions, predicates, and address calculations; advance the thread's program counter along its specialized path. \\
Memory and matrix-fragment operations &
Read or write the bytes selected by the address, width, and layout; matrix loads/stores and register MMAs additionally use their prescribed lane-to-element mappings and participation rules. \\
Asynchronous copy and tensor operations &
Issue pending work, then perform its reads, computation, and writes under the rules below.
This covers classic/bulk copies, TMA, asynchronous cluster stores, and supported \code{tcgen05} MMA, load/store, copy, and shift variants. \\
Collectives &
Match the required threads at the same instance.
Warp exchanges, reductions, and elections compute per-thread results; warp/CTA barriers release participants after arrival; cluster barriers separate nonblocking arrival from wait. \\
Resources and descriptors &
TMEM allocation publishes a resource, deallocation ends its lifetime, and permit operations update allocation rights.
Tensor-map replacement and copy modify descriptor memory used by transfers. \\
\bottomrule
\end{tabularx}
\end{center}

An access reads or writes storage, including thread-local registers involved in asynchronous work, at the step prescribed by the instruction.
A read obtains the current bytes, and a write takes effect immediately in this model (\autoref{sec:background:gpu}).
Widths, zero filling, sparse metadata, scaling, and tile layouts determine the accessed bytes and value functions.
An MMA reads its accumulator only when accumulation is enabled.
Supported bulk reductions perform an atomic read--modify--write per destination element, retaining the operation and scope; this gives no atomicity to the entire transfer.
General \code{atom}/\code{red} instructions are outside the admitted subset.

\paragraph{Asynchronous work and completion.}
Issue creates pending work with the instruction's operand-use timing and lets the issuing thread continue.
Operands captured at issue retain those values; later accesses occur at their prescribed effects.
Effects execute after issue and their causal prerequisites; a destination write uses the values obtained by its source reads.
Reuse of source or destination storage is checked against outstanding accesses.
Only the specified pipeline dependencies constrain different asynchronous operations; source order alone does not serialize them.
CTA groups, multicast masks, and layouts determine the actual source and destination owners and notification targets.

Completion tracking uses the following per-thread histories, whose membership is fixed by instruction order independently of completion timing.

\begin{center}
\footnotesize
\renewcommand{\arraystretch}{1.12}
\begin{tabularx}{\linewidth}{@{}>{\raggedright\arraybackslash}p{.25\linewidth}@{\hspace{.8em}}>{\raggedright\arraybackslash}X@{}}
\toprule
Mechanism & Tracked work and completion action \\
\midrule
Classic/bulk groups &
Commit groups the previously uncommitted operations, with separate classic and bulk queues.
Groups complete in commit order; their members need not be ordered.
\code{wait\_group N} waits for all but the newest $N$ groups; bulk \code{.read} waits only for source and tensor-map reads.
Classic \code{wait\_all} commits and waits for all groups. \\
Byte notification &
TMA/bulk loads and \code{st.async} perform \code{complete\_tx} on each target barrier after its destination writes, with that target's byte count. \\
Classic-copy arrival &
\code{cp.async.mbarrier.arrive} notifies after the prior copies complete; without \code{.noinc}, issue first adds one pending arrival, matched by the delayed arrival. \\
Tensor-core arrival &
\code{tcgen05.commit} tracks the cumulative prefix of prior MMA/copy/shift operations of the same thread and CTA group.
After both commit issue and that work complete, it produces one arrival per target barrier. \\
Tensor-memory wait &
\code{tcgen05.wait::ld/st} waits for all prior matching loads/stores of its thread, with the instruction's warp-participation requirements. \\
\bottomrule
\end{tabularx}
\end{center}

Notifications remain separately schedulable counter updates.
An analysis-level abstraction may group memory effects into one execute step, as in the running example, provided it preserves the analyzed conflicts, completion conditions, and read/write values.

\paragraph{Mbarriers.}
For each initialized barrier~$B$, $\beta(B)=(g,a,x)$ records its generation and pending arrival/byte counts; initially it is uninitialized.
Initialization with $n_B>0$ sets $(0,n_B,0)$.
An update or wait issue before initialization is an error.
Arrival, byte announcement, combined arrival/announcement, and byte completion have increments $(-r,0)$, $(0,b)$, $(-r,b)$, and $(0,-b)$, where $r$ is the arrival count and $b$ the byte count.
The classic-copy reservation above has increment $(1,0)$.
Each increment is one atomic update:
from $(g,a,x)$ compute $(a',x')=(a,x)+(\Delta a,\Delta x)$; an invalid counter state is an error, and $(a',x')=(0,0)$ completes the phase and resets to $(g+1,n_B,0)$ atomically.
Otherwise the new state is $(g,a',x')$.
The condition $(a',x')\lexge(0,0)$ means $a'>0$ or ($a'=0$ and $x'\ge0$); bytes may be negative while arrivals remain~\cite{nvidia2026ptx}.
Write $V_B$ for the permitted pre-reset counter states: this lower bound together with the selected instruction variant's counter bounds.
An arrival marked \code{.noComplete} additionally produces an error if its update completes the phase.

\paragraph{Named counted barriers.}
A named slot starts at generation~0 with its count unconfigured.
For PTX named barriers, a registration represents one warp's arrival after its required non-exited lanes rendezvous; counts and contributions use the same units.
An arrival carries a declared count~$n$; the first arrival of a phase sets its pending count to~$n$, and subsequent arrivals with a different count produce an error.
Registration subtracts its arrival contribution; reaching zero completes the phase, advances the generation, and clears the configuration~\cite{sharma2015weft}.
A blocking registration also captures its generation in the same atomic step.
Counts may differ between phases, and the physical slot is retained across reuse.
Neither phase membership nor a fixed rendezvous is inferred from one trace during extraction.
For named slots, $V_B$ consists of the permitted nonnegative arrival counts with byte count zero.

\paragraph{Waits.}
Write $q=\issue(w)$ for entry and $r=\ret(w)$ for return.
The three forms retain distinct semantics:
\begin{itemize}[leftmargin=1.5em,itemsep=2pt]
\item A \emph{parity wait} records $\theta(w)=g$ if its argument~$p$ equals $g\bmod2$, and $\theta(w)=g-1$ otherwise, at entry.
This is a ghost assignment; a probe tests the current parity, returning when $p\ne g\bmod2$.
\item A \emph{state wait} evaluates its token operand at entry.
A token denotes the barrier lifetime and generation~$h$ captured before the update of~$s(w)$.
The wait records $\theta(w)=h$.
A probe is valid only for the same barrier lifetime with $g\in\{h,h+1\}$, and succeeds at $g=h+1$.
Invalid state tokens and expired probes are errors.
\item A \emph{named wait} has its blocking registration as~$q$ and records that registration's generation.
Its return is enabled once that generation completes, even if later generations have completed.
\end{itemize}
Target~$-1$ occurs only for an initially ready parity wait.
A pure retry loop retains entry and successful return; its argument stays fixed and failed probes affect only retry control.
One-shot probes whose results affect other computation require their own events.
Fixed-member warp and cluster collectives retain their prescribed participation and instance rules; a split arrival retains its nonblocking continuation.

\paragraph{Event-graph construction.}
From these rules, retain issue, storage-access, notification, barrier, wait-return, collective, and lifetime steps as the finite event set~$E$.
Local computations become value expressions; accesses that may interact with outstanding asynchronous work remain explicit, including those of synchronous instructions.
Each retained access belongs to its instruction-prescribed step, and each captured value precedes its uses.
A structurally fixed collective joins its required arrivals in a shared step; split arrival/wait retains the separate continuation points.
Named registration follows the warp-local rendezvous, retaining the slot's dynamic phase membership.
The graph~$G$ contains the following fixed dependencies:
\begin{enumerate}[leftmargin=1.5em,itemsep=1pt]
\item A thread's next instruction follows async issue, or completion of a synchronous instruction or blocking wait.
Every wait has $\issue(w)\to\ret(w)$; a pure mbarrier-wait issue has no other outgoing edge in~$G$.
\item Each asynchronous effect follows issue and the causal prerequisites specified by its instruction; each notification follows its tracked effects and its own issue.
\item Group and completion waits follow the operations selected by their histories; collectives follow the required arrivals and precede released continuations.
Group-completion steps join their commit, members, and any required earlier group completion; this does not order all memory effects across groups.
\end{enumerate}
Specified pipeline constraints are also retained; no edges are added from a chosen scheduling order.
Completion waits order the corresponding outstanding accesses; a storage read alone supplies no completion edge from a pending writer.
Lifecycle obligations are checked against the resulting order rather than inserted as edges.

\paragraph{Event execution.}
The event configuration contains $(D,\beta,\theta)$ and the storage and value state, where $D\subseteq E$ contains the executed events.
Initially $D=\varnothing$, mbarriers are uninitialized, named slots are unconfigured, and wait targets are unset.
An event~$e$ other than a barrier-wait return can be attempted when $e\notin D$ and $\operatorname{pred}_G(e)\subseteq D$, where $\operatorname{pred}_G(e)$ contains its direct predecessors.
Its dynamic checks either succeed or produce an error; they do not silently block the attempt.
Barrier-wait returns additionally use their typed conditions above; an invalid pending state probe can produce an error.
Updates and barrier-wait issues are \emph{barrier operations}; all events except initializations, updates, and barrier-wait issues/returns are \emph{ordinary} for synchronization.
For a named blocking registration, the update and wait issue are the same event.
Ordinary events leave $\beta$ and $\theta$ unchanged; the other events follow the initialization, update, issue, and return rules above.
A successful step adds~$e$ to~$D$.
An error terminates execution and records the failing event.

Write $\initof(B)$ for the initialization event.

\paragraph{Executions.}
A \emph{trace}~$\pi$ records finitely many steps from the initial configuration, possibly ending in an error; $\evs(\pi)$ is its event set.
For the fixed launch configuration, traces range over the admitted initial inputs; every check uses this same domain.

\begin{definition}[Complete traces and safety]
A trace~$\pi$ is \emph{complete} if $\evs(\pi)=E$.
It is \emph{safe} if no barrier error occurs.
A safe incomplete trace is \emph{stuck} if no event can extend it.
\end{definition}

\begin{definition}[Phase assignment]
In a safe trace~$\pi$, $\phaseof_\pi(e)$ maps an update of~$B$ to $\ph{B}{g}$, where $g$ is its generation at execution, and maps a wait issue $\issue(w)$ to $\ph{B}{\theta(w)}$.
The latter is defined even if the wait has not returned, and may be $\ph{B}{-1}$.
\end{definition}

Well-synchronization alone permits deadlock: \code{init(B,1); wait(B,0)} always targets phase~0 but never completes.
\autoref{thm:well-synchronization} therefore requires a safe complete reference trace.

\begin{lemma}[Instruction--event correspondence]
\label{lem:instruction-event-correspondence}
Fix an admissible input~$x$.
Let $\mathcal T_{\rm inst}(P,x)$ and $\mathcal T_{\rm evt}(P,x)$ be the event sequences of the instruction-family model above and its extracted event model, respectively.
Asynchronous instructions and pure retries use the issue/effect/return steps specified above.
Then
\[
  \rho\bigl(\mathcal T_{\rm inst}(P,x)\bigr)
  =\mathcal T_{\rm evt}(P,x),
\]
where $\rho$ erases steps represented solely by value expressions and failed valid probes, and identifies matched collective steps.
It preserves retained storage accesses, values, barrier updates, wait targets, and errors.
The statement uses ungrouped effects.
\end{lemma}
\begin{proof}
Relate configurations with equal retained storage, barrier states, wait targets, pending work, completion histories, and available operand values.
The initial configurations are related.
By construction, retained steps have identical enabling conditions and updates, including errors; erased computations evaluate retained expressions from captured values, and failed valid probes stutter.
Conversely, every enabled event is realized by inserting the required local computations and collective arrivals.
Both directions preserve the relation and recorded effects.
Induction on finite traces proves both inclusions.
\end{proof}

\subsection{Characterization of Well-Synchronization}

Fix a safe complete reference trace~$\tau$ and its candidate graph~$H$.
Write $C=\contr{B}{k}$ for a reference phase's contributors.
Let $\mathcal I(C)$ be its ideals under~$\reach$, and let $n_{B,k}$ be its initial count: $n_B$ for an mbarrier, or the count of its first named registration.
Set
\[
  \stateof(I)=(n_{B,k},0)+\sum_{u\in I}\labelof(u)
  \qquad (I\in\mathcal I(C)).
\]
A phase is \emph{closed} when it completes in~$\tau$, and \emph{open} otherwise.
An empty unconfigured named phase has no counter obligation.
Let $\operatorname{Lin}(H)$ be the linear extensions of reachability restricted to program events~$E$, and $\operatorname{Top}(H)$ their prefixes.
Completion vertices only record resets.
For $S\subseteq E$, write $\mathord{\downarrow}_H S=S\cup\{e\in E\mid \exists f\in S:\ e\reach f\}$.
The augmented reference trace witnesses that $H$ is acyclic.

\subsubsection{Checks}

The full, typed form of the four checks in \autoref{alg:certify} is:
\begin{enumerate}[leftmargin=2.5em,itemsep=3pt,label=(C\arabic*)]
\item \emph{Initialization.}
For every mbarrier update or wait issue~$e$ on~$B$, $\initof(B)\reach e$.
\item \emph{Separation.}
For every $k>0$,
\[
  \forall d\in\contr{B}{k-1},\
  \forall u\in\contr{B}{k}:\quad d\reach u.
\]
\item \emph{Counters.}
For every phase and $I\in\mathcal I(C)$, $\stateof(I)\in V_B$.
\item \emph{Closed phases.}
For every closed phase,
\[
  \forall I\in\mathcal I(C),\ I\subsetneq C:
  \quad \stateof(I)\ne(0,0).
\]
Every \code{.noComplete} update $u\in C$ has a successor $v\in C$ with $u\reach v$.
\item \emph{Open phases.}
For every open phase and $I\in\mathcal I(C)$, $\stateof(I)\ne(0,0)$.
\item \emph{Parity waits.}
For target $\ph{B}{k}$ and $q=\issue(w)$,
\[
\begin{aligned}
 k>0&\ \Longrightarrow\
   \forall d\in\contr{B}{k-1}:\ d\reach q,\\
 \ph{B}{k+1}\text{ closed}
   &\ \Longrightarrow\
   \exists d\in\contr{B}{k+1}:\ q\reach d.
\end{aligned}
\]
\item \emph{State waits.}
For target $\ph{B}{k}$,
\[
 \ph{B}{k+1}\text{ closed}
 \ \Longrightarrow\
 \exists d\in\contr{B}{k+1}:\ \ret(w)\reach d.
\]
\end{enumerate}

A state wait has $\phaseof_\tau(s(w))=\phaseof_\tau(\issue(w))$ and $s(w)\to\issue(w)$ in~$G$.
These follow from resolved token provenance and safety of~$\tau$.
Named waits capture their registration's phase.
Safety of~$\tau$ also gives count agreement within each named phase and $\stateof(C)=(0,0)$ exactly for closed phases.

The displayed pseudocode is the parity-wait specialization without \code{.noComplete} updates and with counter upper bounds already established.
For these cases, (C3)--(C5) reduce to its proper-ideal test: the full-set state is fixed by~$\tau$.
The full check uses the typed conditions above; all are reachability or finite ideal tests.

\subsubsection{Soundness}

Assume (C1)--(C7).
We prove every trace safe and phase-consistent with~$\tau$.

\paragraph{Induction hypothesis.}
For a trace prefix~$\pi$, put $D_\pi=\evs(\pi)$ and insert completion markers at actual resets to obtain~$\widehat\pi$.
Induct on event attempts, maintaining:
\begin{enumerate}[leftmargin=2.5em,itemsep=2pt,label=(I\arabic*)]
\item $\phaseof_\pi=\phaseof_\tau$ on executed barrier operations.
\item For each reference closed phase,
\[
 \ph{B}{k}\text{ completed}
 \iff \contr{B}{k}\subseteq D_\pi.
\]
No reference open phase completes.
For an initialized mbarrier or configured named phase, the current counters equal $\stateof(\contr{B}{g}\cap D_\pi)$.
A named slot is unconfigured until its first registration.
\item $\widehat\pi$ is ordered and downward-closed in~$H$.
\item $\pi$ is safe.
Every pending mbarrier wait targeting $\ph{B}{k}$ has $g\in\{k,k+1\}$.
\end{enumerate}

\paragraph{Base.}
For $\pi=\epsilon$, no operation or phase has occurred.
Initial counts are positive, and every closed phase has a nonempty contributor set.
Thus (I1)--(I4) hold.

\paragraph{Step.}
Let $e$ be the next attempted event.
For an event other than a barrier-wait return, every incoming edge in~$H$ is already in~$G$.
Its execution condition and (I3) therefore give
\[
  d\reach e\ \Longrightarrow\ d\in\evs(\widehat\pi).
  \tag{*}
\]

\emph{Ordinary event or initialization.}
An ordinary event leaves barrier state unchanged.
Initialization sets $(g,a,x)=(0,n_B,0)$ with $n_B>0$.
Together with~(*), these preserve (I1)--(I4).

\emph{Update $u\in C=\contr{B}{k}$.}
By (C1), the mbarrier is initialized.
By (C2), (*) and (I2), $g\ge k$.
If $g>k$, (I2) would imply $u\in D_\pi$ or completion of an open phase.
Hence $g=k$.

The set $I=C\cap(D_\pi\cup\{u\})$ is an ideal by~(*) and (I3).
By (I1)--(I2), its post-update, pre-reset counters are $\stateof(I)$, which belongs to~$V_B$ by (C3).
Named registrations use the same $n_{B,k}$ as in~$\tau$.
Thus the update is counter-safe and preserves (I1).
A state-producing arrival captures~$k$; a blocking named arrival also assigns its wait target~$k$.

By (C4)--(C5) and the reference full-set sums,
\[
  \stateof(I)=(0,0)
  \iff \ph{B}{k}\text{ is closed and }I=C.
\]
The last contributor is not \code{.noComplete}, by (C4).
This proves (I2); inserting the completion marker after its contributors preserves (I3).

For a pending parity wait with return~$r$, (C6) gives $r\reach d$ for a contributor of the next closed phase: every path from its pure issue passes through~$r$.
For a state wait, (C7) gives this directly.
By (I3), $d$ has not occurred, so (I2) excludes completion of that next phase.
An open next phase cannot complete either.
This preserves (I4).

\emph{Pure wait issue $q=\issue(w)$ targeting $\ph{B}{k}$.}
By (C1), $B$ is initialized.
For a parity wait with $k\ge0$, (C6), (*) and (I2) give $g\ge k$.
The upper condition of (C6), or noncompletion of an open next phase, gives $g\le k+1$.
Consequently parity $k\bmod2$ selects~$k$.
For $k=-1$, the same upper condition gives $g=0$ and parity~1 selects~$-1$.

For a state wait, $s(w)\to q$ and (I1) give a token of generation~$k$ and $g\ge k$.
For a closed next phase, (C7), (I3) and $q\to\ret(w)$ give $g\le k+1$; for an open next phase, (I2) gives the same bound.
Thus its token is valid and selects~$k$.
In both cases, (*) preserves (I3), counters are unchanged, and (I1)--(I4) hold.

\emph{Wait return $r=\ret(w)$ targeting $\ph{B}{k}$.}
For mbarrier waits, (I4) makes success equivalent to completion of~$k$.
For named waits this is the ticket condition.
If $k\ge0$, (I2) supplies every contributor and the completion marker, satisfying the added edge $\cmp{B}{k}\to r$.
For $k=-1$, no such edge exists.
The $G$-predecessors also occurred, so (I3) holds.
No counters or assignments change, preserving the other invariants.

A failed valid probe stutters.
Invariant (I4) excludes an expired-state error; (C1) and (C3)--(C4) exclude the remaining barrier errors.
The induction proves well-synchronization.

\subsubsection{Graph--Execution Correspondence}

\begin{lemma}[Pending-wait bound]
\label{lem:retry-correspondence}
If $P$ is well-synchronized, every pending mbarrier wait targeting $\ph{B}{k}$ satisfies $g\le k+1$.
Its success condition is therefore equivalent to completion of~$k$.
\end{lemma}
\begin{proof}
For a state wait, $g\ge k+2$ enables an expired-token error, contradicting safety.
For a parity wait, suppose a prefix before return reaches $g\ge k+2$.
Remove its pure issue~$q$ and failed probes and append~$q$ to that prefix.
This preserves executability: $q$ changes only its private target and its sole outgoing edge in~$G$ leads to the absent return.
The new target is $g$ or $g-1$, both greater than~$k$, contradicting phase stability.
The argument also covers $k=-1$.
\end{proof}

\begin{lemma}[Graph--execution correspondence]
\label{lem:graph-execution-correspondence}
If $P$ is well-synchronized, then, for every admissible input~$x$,
\[
  \mathcal T_{\rm evt}(P,x)=\operatorname{Top}(H).
\]
Moreover, a reference closed phase completes exactly when all its contributors occur; reference open phases never complete.
\end{lemma}
\begin{proof}
\emph{Extension.}
Fix an execution prefix~$\pi$ and a linear extension~$\sigma$ of~$H$.
Append the subsequence of~$\sigma$ omitting events already in~$\pi$.
At each step, all $G$-predecessors have occurred.
Every event other than a barrier-wait return can therefore be attempted; universal barrier safety excludes a barrier error and stability fixes its phase.
At a return targeting $\ph{B}{k}$, $k\ge0$, all reference contributors have occurred in phase~$k$.
Their sum is zero, so that phase has completed.
The pending-wait bound enables mbarrier returns; named tickets remain enabled after completion.
Target~$-1$ is initially ready.
Thus every prefix extends to a complete execution.

\emph{Completion boundaries.}
If a closed phase completes while some $u\in C$ is absent, the extension executes~$u$ in a later generation, violating stability.
If every contributor occurs, their zero sum completes the phase.
The same argument forbids early completion of an open phase; its nonzero full-set sum forbids completion at its last contributor.

\emph{Equality.}
Execution respects~$G$.
Completion boundaries and the pending-wait bound supply all added contributor-to-completion-to-return edges, giving $\mathcal T_{\rm evt}(P,x)\subseteq\operatorname{Top}(H)$.
Conversely, extend a topological prefix to a linear extension and apply the extension construction from~$\epsilon$.
This proves the reverse inclusion.
\end{proof}

By synchronization soundness, acceptance therefore guarantees that every execution prefix respects~$\hb$ and extends to a complete execution.
No reachable incomplete state is stuck.

\subsubsection{Completeness}

Assume $P$ is well-synchronized.
By \autoref{lem:graph-execution-correspondence}, every linear extension of~$H$ is a safe complete execution with the reference phase assignment.
For distinct events, the finite-poset fact
\[
  e\nreach f
  \ \Longrightarrow\
  \exists\sigma\in\operatorname{Lin}(H):\ f<_\sigma e
  \tag{**}
\]
provides the following witnesses.

\emph{Initialization (C1).}
If $\initof(B)\nreach e$, (**) puts the use before initialization, contradicting safety.

\emph{Separation (C2).}
If $d\nreach u$ for $d\in\contr{B}{k-1}$ and $u\in\contr{B}{k}$, (**) puts $u$ before~$d$.
Phase~$k-1$ is then incomplete, contradicting the assignment of~$u$ to~$k$.

\emph{Ideals (C3)--(C5).}
For $I\in\mathcal I(C)$, its downward closure $D=\mathord{\downarrow}_H I$ satisfies $D\cap C=I$.
An extension starting with~$D$ therefore realizes precisely~$I$ within the phase.
Every nonempty proper ideal of a closed phase is a pre-reset state, so safety gives $\stateof(I)\in V_B$, and completion boundaries give $\stateof(I)\ne(0,0)$.
The same argument applies to all nonempty ideals of an open phase.
The empty ideal has state $(n_{B,k},0)$; full closed-phase states are zero by~$\tau$.
Finally, a maximal contributor can be last among~$C$ in an extension, so a closed phase cannot have a maximal \code{.noComplete} update.

\emph{Parity waits (C6).}
If some $d\in\contr{B}{k-1}$ does not reach~$q$, (**) issues~$q$ before~$d$, with $g<k$ and target at most~$g$.
If $q$ reaches no member of $\contr{B}{k+1}$, then $q\notin\mathord{\downarrow}_H\contr{B}{k+1}$, so that downward closure can precede~$q$.
When that phase is closed, this yields $g\ge k+2$ and target at least $g-1>k$.
Both contradict stability; the upper argument includes $k=-1$.

\emph{State waits (C7).}
If $r=\ret(w)$ reaches no member of the closed phase $\contr{B}{k+1}$, an extension places that phase before~$r$.
If issue follows the phase, its token is already expired; otherwise the pending wait exceeds the bound of \autoref{lem:retry-correspondence}.
Both contradict safety.

Thus (C1)--(C7) are necessary and sufficient, proving the full typed characterization underlying \autoref{thm:well-synchronization}.

\subsection{Minimum-Ideal Reduction}
\label{app:min-ideal}

\begin{lemma}[Minimum-ideal reduction]
A minimum cut computes the lexicographically least ideal state; one cut per maximal contributor computes the least proper-ideal state.
Affine counter bounds use the same reduction.
\end{lemma}
\begin{proof}
\emph{Scalarization.}
Write $\labelof(e)=(\Delta a_e,\Delta x_e)$ and set
\[
\begin{aligned}
 T&=\sum_{e\in C}|\Delta x_e|,& M&=T+1,\\
 w_e&=M\Delta a_e+\Delta x_e,&
 F(I)&=Mn_{B,k}+\sum_{e\in I}w_e.
\end{aligned}
\]
Integer arrival counts and byte-sum differences bounded by~$T$ make $F$ preserve lexicographic order and comparison with zero.

\emph{Closure cut.}
On vertices $\{s,t\}\cup C$, use source~$s$, sink~$t$, and capacities
\[
\begin{array}{lll}
 s\to e&\text{capacity }-w_e& (w_e<0),\\
 e\to t&\text{capacity } w_e& (w_e>0),\\
 f\to e&\text{capacity } L& (e\reach f),
\end{array}
\]
where $L>\sum_{e\in C}|w_e|$.
The empty ideal costs less than~$L$, so minimum cuts select ideals on the source side, whose capacities are
\[
  \operatorname{cap}(I)
  =\sum_{w_e<0}(-w_e)+\sum_{e\in I}w_e.
\]
Thus minimum cut minimizes~$F$.

\emph{Proper ideals.}
Every proper ideal omits a maximal contributor, so for nonempty~$C$,
\[
 \min_{\substack{I\in\mathcal I(C)\\I\subsetneq C}}F(I)
 =\min_{m\in\operatorname{Max}(C)}
   \min_{\substack{I\in\mathcal I(C)\\m\notin I}}F(I).
\]
Adding $m\to t$ with capacity~$L$ enforces each exclusion; empty~$C$ has no proper ideals.

After excluding lexicographically negative states, a zero ideal exists exactly when the corresponding minimum is zero.
Affine counter bounds use the same closure construction on their slack.
\end{proof}

%% file: sections/proofs-memory.tex
\section{Memory Safety and Equivalence Proofs}

\subsection{Memory Safety}
\label{app:memory-assumptions}
\label{app:memory-races}

Fix the finite event set~$E$ and a nonempty admissible-input domain~$X$ under the instruction model of \autoref{app:sync-model}.
Each $x\in X$ fixes all initial bytes, including unspecified scratch contents; allocation identities are exact, and distinct pointer parameters refer to disjoint allocations.
Events have deterministic, defined semantics, account for all their accesses, and use their specified internal read/write order.
Each access operand of an event~$e$ is computed from the input, values produced by events $d\hb e$, or earlier reads in~$e$'s prescribed internal order.
These operands determine its addresses, widths, and layouts.
By \autoref{lem:instruction-event-correspondence} and \autoref{lem:graph-execution-correspondence}, after synchronization checking passes, event executions are exactly the topological prefixes of~$\hb$.
Completion vertices contribute paths to~$\hb$ but are not memory-performing events in~$E$.

An access of width~$w$ at $(\alpha,o)$ touches $\{(\alpha,o+b):0\le b<w\}$; a tile touches the union of its element accesses.
Here $\alpha$ identifies an allocation or a thread-local register; register identities include the owning thread and are disjoint from memory allocations.
The overlap query ranges over admissible inputs~$x\in X$ and complete executions of~$P$ on~$x$.
A disjoint verdict excludes overlap throughout this domain; an overlap verdict certifies an input and execution with overlapping accesses; unknown leaves the pair unresolved.
A resource respects its lifetime when $\mathit{alloc}(r)\hb e\hb\mathit{dealloc}(r)$ for every use~$e$ other than its boundaries; reallocations have fresh identities, and external boundaries follow the scope assumptions.

\paragraph{Consequences of the checks.}
After synchronization checking passes, a race-free verdict from \autoref{alg:races} guarantees that every distinct conflicting pair is ordered by~$\hb$ on every admissible input and complete execution.
A report race$(e,f)$ supplies an admissible input~$x$ and complete execution~$\pi$ in which $e,f$ conflict and neither $e\hb f$ nor $f\hb e$.
The lifetime inequalities hold in every linear extension, so every use lies between allocation and deallocation whenever the lifetime check passes.
\autoref{lem:instruction-event-correspondence} transfers these guarantees to the instruction model.

\subsection{Determinate Reads}

A read byte's supplier is its last preceding write, possibly within the same event, or its initial-memory cell.

\begin{proof}[Read determinacy (\autoref{sec:races:determinate})]
Fix $x\in X$ and induct on~$\hb$, and within each event~$e$ on its prescribed access order, proving that all operands, accesses, read suppliers, and read/write values agree across complete executions.

\emph{Operands.}
The next access uses the input, preceding events' values, or earlier reads in~$e$, so the induction hypotheses fix its operands and access geometry.

\emph{Suppliers.}
For a byte~$b$ read before any internal write to it, every event~$f\ne e$ writing~$b$ in any execution conflicts with~$e$ and is therefore comparable with it by \autoref{app:memory-races}.
If $e\hb f$, that write cannot supply the read; if $f\hb e$, the induction hypothesis fixes its access and value in every execution.
Consequently $W_b(e)=\{f\mid f\hb e,\ b\in\mathit{writes}(f)\}$ is fixed and totally ordered, with supplier
\[
  \mathit{supplier}(e,b)=
  \begin{cases}
    \max_{\hb}W_b(e),&W_b(e)\ne\varnothing,\\
    \mathit{initial}(b),&W_b(e)=\varnothing.
  \end{cases}
\]

\emph{Values.}
A read of a byte already written within~$e$ obtains the last such write's value, fixed by the inner induction hypothesis.
Deterministic computation fixes the resulting values and writes, closing both inductions; the same chain argument gives a fixed final write and value for every byte.
\end{proof}

\subsection{Memory Elimination and Equivalence}
\label{app:memory-equivalence}

Use the declared scalar semantics of \autoref{sec:reference:egraph}, with matching byte selection, encoding, decoding, and operator preconditions.
Every accepted identity must establish its side conditions, such as nonzero divisors, on the original admissible-input domain.
The input correspondence identifies both programs' initial data, and the output correspondence covers every observable output.

\begin{theorem}[Soundness of functional verification]
\label{thm:equivalence}
Let $P$ satisfy the conditions of \autoref{app:memory-assumptions}, and let $R$ be its reference under the scalar semantics and input/output correspondences specified above.
If synchronization, race, and lifetime checking accept $P$, and functional verification (\autoref{sec:reference}) accepts $(P,R)$, then, for every $x\in X$, every complete execution of $P$ on $x$ produces the same observable outputs as $R$ on the corresponding input.
\end{theorem}

\begin{proof}[Proof of \autoref{thm:equivalence}]
Fix corresponding admissible inputs~$x$ and any complete kernel execution.
By the read determinacy established in \autoref{sec:races:determinate}, its outputs equal those in the elimination order.

\emph{Symbolic execution.}
Let $M_j,S_j$ be the concrete and symbolic states of retained storage after $j$ events in that order.
Induct on~$j$ with the invariant that every produced value node evaluates to its concrete value and, for every retained storage byte~$(\alpha,b)$,
\begin{equation}
  \operatorname{eval}_x(S_j(\alpha,b))=M_j(\alpha,b).
\end{equation}
The initial correspondence establishes the invariant.
Each retained read obtains the same bytes and decoded value; each write updates the same bytes with equal values, preserving untouched storage.
Scalar and tile computations preserve values by their declared semantics and internal access order; synchronization-only steps leave storage unchanged.
Thus the output expressions evaluate to the kernel's outputs; faithful reference lowering gives the corresponding reference values.

\emph{Proof steps.}
Induct over the proof steps, using previously established equalities only.
For coverage, if $\phi:S\to D$ covers every index exactly once and $e_s=f_{\phi(s)}$, then
\[
  \sum_{s\in S}e_s=\sum_{d\in D}f_d.
\]
Coefficient equality of the finite generating functions certifies exact-once coverage.
Normalization preserves real semantics~\cite{li2026identity}, and congruence handles opaque operators.
An inductive proof checks $I(l,s_l)$, $I(k,s_k)\Rightarrow I(k+1,s_{k+1})$ for $l\le k<n$, and that $I(n,s_n)$ entails the desired equality; finite induction proves it.

The accepted output equalities establish \autoref{thm:equivalence}.
\end{proof}